\documentclass[11pt,a4paper]{article}
\usepackage{jcappub}

\usepackage{adjustbox}

\newcommand\dosingle[1]{#1}  \newcommand\dodouble[1]{ }

\newcommand\nice[1]{#1}    \newcommand\subm[1]{}   
\subm{ \renewcommand\dosingle[1]{ }   \renewcommand\dodouble[1]{#1} }

\dodouble{ \documentclass[referee]{aa} }

\usepackage[T1]{fontenc}
\usepackage{aecompl}

\usepackage{graphicx}

\usepackage{amsfonts}
\usepackage{amsmath}

\usepackage{csquotes}

\usepackage{rotating}

\usepackage[varg]{txfonts}
\usepackage{fixltx2e} 

\newcommand\mystamp[1]{#1}

\newcommand\mystamppreamble{
  \usepackage{eso-pic}
  \usepackage{color}
  \definecolor{redstamp}{rgb}{0.99,0.80,0.90} 
  \usepackage{datetime}
  \usepackage[normalem]{ulem}
}

\mystamp{\mystamppreamble}

\usepackage{url}

\usepackage{etoolbox}

\newtoggle{postrefAA} \newtoggle{postrefBB} \newtoggle{postrefCC}

\newtoggle{prerefAA} \newtoggle{prerefBB} \newtoggle{prerefCC}
\newtoggle{prerefDD} \newtoggle{prerefEE} \newtoggle{prerefFF} \newtoggle{prerefGG}

\togglefalse{postrefAA} 

\togglefalse{postrefBB} 

\togglefalse{postrefCC}

\togglefalse{prerefAA}

\togglefalse{prerefBB} \togglefalse{prerefCC}
\togglefalse{prerefDD} \togglefalse{prerefEE} \togglefalse{prerefFF} \togglefalse{prerefGG}

\iftoggle{prerefAA}{
  \newcommand\prerefereeAAchanges[1]{{\bf \large \color{myred} #1}}     \usepackage{color}  \definecolor{myred}{rgb}{0.7,0.0,0.2}
}{
  \newcommand\prerefereeAAchanges[1]{#1}    
}

\iftoggle{prerefBB}{
       \usepackage{color}  \definecolor{myred}{rgb}{0.7,0.0,0.2}
}{
      
}

\iftoggle{prerefCC}{
       \usepackage{color}  \definecolor{myred}{rgb}{0.7,0.0,0.2}
}{
      
}

\iftoggle{prerefDD}{
       \usepackage{color}  \definecolor{myred}{rgb}{0.7,0.0,0.2}
}{
      
}

\iftoggle{prerefEE}{
       \usepackage{color}  \definecolor{myred}{rgb}{0.7,0.0,0.2}
}{
      
}

\iftoggle{prerefFF}{
       \usepackage{color}  \definecolor{myred}{rgb}{0.7,0.0,0.2}
}{
      
}

\iftoggle{prerefGG}{
       \usepackage{color}  \definecolor{myred}{rgb}{0.7,0.0,0.2}
}{
      
}

\iftoggle{postrefAA}{
  \newcommand\postrefereeAAchanges[1]{{\bf \large \color{myred} #1}}     \usepackage{color}  \definecolor{myred}{rgb}{0.7,0.0,0.2}
}{
  \newcommand\postrefereeAAchanges[1]{#1}    
}

\iftoggle{postrefBB}{
  \newcommand\postrefereeBBchanges[1]{{\bf \large \color{myred} #1}} \newcommand\postrefereeBBstart{ \bf \large \color{myred} }  \newcommand\postrefereeBBstop{ \rm \color{black} }  \usepackage{color}  \definecolor{myred}{rgb}{0.7,0.0,0.2}
}{
  \newcommand\postrefereeBBchanges[1]{#1}  \newcommand\postrefereeBBstart{  }  \newcommand\postrefereeBBstop{ }
}

\iftoggle{postrefCC}{
       \usepackage{color}  \definecolor{myred}{rgb}{0.7,0.0,0.2}
}{
      
}

\usepackage{color}
\definecolor{mygreen}{rgb}{0.11, 0.65, 0.02}

 \usepackage{hyperref}

\nice{

\providecommand{\url}[1]{\href{#1}{#1}}
\providecommand{\newblock}{}
}

\providecommand{\adsurl}[1]{}

\newcommand\BIBAABST{JHEP_arXiv}

\newcommand\SSS{Sect.~}
\newcommand\SSSS{Sections~}

\providecommand\apj{ApJ}                 
                 
\providecommand\apjs{ApJSupp}                 
                 
\providecommand\aap{A\&A}            
\providecommand\mnras{MNRAS}

\providecommand\PRL{Physical Review Letters}

\providecommand\PRL{PRL}
\providecommand\prd{PRD}

\providecommand\jcap{JCAP}

\providecommand\physrep{Phys.Rep.}

\providecommand\grg{Gen.~Rel.~Grav.}
\providecommand\annalesBruxelles{Ann. de la Soc. Sc. de Brux.}
\providecommand\annrevnucpartphys{Ann.~Rev.~Nucl.~Part.~Sci.}
\providecommand\SPIEConfSeries{SPIE Conf. Ser. }

\providecommand\pnas{Proc.~Nat.~Acad.~Sci.}
\providecommand\cqg{Class.~Quant.~Gra.}

\providecommand\nucphysbprocsupp{Nucl.~Phys.~B~Proc.~Supp.}

\providecommand\ijmpd{Int. J. Mod. Phys. D}

\providecommand\mdash{---}

\nice{  }

\newcommand\gtapprox{\,\lower.6ex\hbox{$\buildrel >\over \sim$} \, }
\newcommand\ltapprox{\,\lower.6ex\hbox{$\buildrel <\over \sim$} \, }
\newcommand\propapprox{\,\lower.6ex\hbox{$\buildrel \propto\over \sim$} \, }

\newcommand\arcs{\ifmmode {'' }\else $'' $\fi}     
\newcommand\arcm{\ifmmode {' }\else $' $\fi}       
\newcommand\ddeg{\ifmmode^\circ\else$^\circ$\fi}    

\newcommand\diffd{\mathrm{d}}

\newcommand\frtoday{Le\space\number\day\space\ifcase\month\or
  janvier\or f\'evrier\or mars\or avril\or mai\or juin\or
  juillet\or ao\^ut\or septembre\or octobre\or novembre\or
d\'ecembre\fi\space \number\year}
\def\frdutoday{du\space\number\day\space\ifcase\month\or
  janvier\or f\'evrier\or mars\or avril\or mai\or juin\or
  juillet\or ao\^ut\or septembre\or octobre\or novembre\or
d\'ecembre\fi\space \number\year}

\newcommand\todayISO{\number\year-\ifnum\month<10 0\fi\number\month-\ifnum\day<10 0\fi\number\day}

\newcommand{\CA}{{\cal A}}
\newcommand{\CD}{{\cal D}}

\newcommand{\CI}{{\cal I}}
\newcommand{\CJ}{{\cal J\mathop{\vphantom{\rule{0ex}{1.9ex}}}}}
\newcommand{\CQ}{{\cal Q}}
\newcommand{\CR}{{\cal R}}

\newcommand{\average}[1]{\left\langle #1 \right\rangle_\CD}

\newcommand{\initaverageriem}[1]{\left\langle #1 \right\rangle_{\CI}}
\newcommand{\Gdet}{\initial{G}}

\newcommand\initinvII{\initaverageriem{\initial{\invII}}}

\newcommand\initavinvII{\initaverageriem{\initial{\invII}}}
\newcommand\initavinvIII{\initaverageriem{\initial{\invIII}}}

\newcommand{\initial}[1]{{{#1}_{\mathbf i}}}

\newcommand\closedopen[2]{\ensuremath{[#1,#2)}}

\newcommand\Omm{\Omega_{\mathrm{m}}}
\newcommand\Ommzero{\Omega_{\mathrm{m0}}}

\newcommand\OmLam{\Omega_{\Lambda}} 
\newcommand\OmLamzero{\Omega_{\Lambda0}} 
\newcommand\Omk{\Omega_{\mathrm{k}}}

\newcommand\OmmD{\Omega_{\mathrm{m}}^{\CD}}

\newcommand\OmRD{\Omega_{\CR}^{\CD}}
\newcommand\OmkD{\Omega_{k}^{\CD}}
\newcommand\OmRDPS{\Delta_{\CR}^{\CD}}

\newcommand\OmQD{\Omega_{\CQ}^{\CD}}
\newcommand\OmQDNewt{\Omega_{\CQ}^{\CD}}

\newcommand\aeff{a_{\mathrm{eff}}}

\newcommand\tcollapse{t_{\mathrm{coll}}}

\newcommand\aturnaround{a_{\mathrm{turn}}}
\newcommand\tturnaround{t_{\mathrm{turn}}}
\newcommand\Hturnaround{H_{\mathrm{turn}}}
\newcommand\NnearTA{N_{\mathrm{turn}}}

\newcommand{\invI}{{\mathrm{I}}}
\newcommand{\invII}{{\mathrm{II}}}
\newcommand{\invIII}{{\mathrm{III}}}

\newcommand\hzeroeff{h_{\mathrm{eff}}}

\newcommand\Heff{H_{\mathrm{eff}}}

\newcommand\LCDM{$\Lambda$CDM}

\newcommand\RAMSES{{\sc ramses}}

\newcommand\RAMSESscalav{{\sc ramses-scalav}}
\newcommand\DTFE{{\sc dtfe}}

\newcommand\inhomog{{\sc inhomog}}

\newcommand\RCeff{L_{\CR}^{\CD}}

\newcommand\Lbox{L_{\mathrm{box}}}
\newcommand\LD{L_{\CD}}

\newcommand\LDTFE{L_{\mathrm{DTFE}}}
\newcommand\LN{L_N}

\newcommand\Toptobottom{From top to bottom}
\newcommand\toptobottom{from top to bottom}

\newtheorem{proposition}{\protect\postrefereeAAchanges{Proposition}}

\usepackage{color}
\definecolor{mygreen}{rgb}{0.11, 0.65, 0.02}
\definecolor{myred}{rgb}{0.9,0,0}

\title{Does spatial flatness forbid the turnaround epoch of collapsing structures?}

\newcommand\TCfAaddress{Institute of Astronomy,  
  Faculty of Physics, Astronomy and Informatics,
  Nicolaus Copernicus University,  
  Grudziadzka 5,
  87-100 Toru\'n, Poland}
\newcommand\CRALaddress{Univ Lyon, Ens de Lyon, Univ Lyon1, CNRS, Centre de Recherche Astrophysique de
  Lyon UMR5574, F--69007, Lyon, France}
\newcommand\NCBJaddress{Department of Fundamental Research,
  National Centre for Nuclear Research,
  Pasteura 7, 02-093 Warszawa, Poland}

\author[1,2]{Boudewijn F. Roukema,}
\author[3,2]{Jan J. Ostrowski}
\affiliation[1]{\TCfAaddress}
\affiliation[2]{\CRALaddress}
\affiliation[3]{\NCBJaddress}

\emailAdd{boud {\em at} astro.uni.torun.pl, Jan.Jakub.Ostrowski {\em at} ncbj.gov.pl}

\date{\frtoday}

\abstract{
    {Cosmological observational analysis frequently assumes that the
      Universe is spatially flat.}
    {We aim to non-perturbatively check the conditions under which a
      flat or nearly flat expanding dust
      universe, including the $\Lambda$-cold-dark-matter ({\LCDM})
      model if interpreted as strictly flat, forbids the gravitational
      collapse of structure. We quantify spatial curvature at turnaround.}
    {We use the Hamiltonian constraint to determine the pointwise
      conditions required for an overdensity to reach its turnaround
      epoch in an exactly flat spatial domain. We illustrate
      this with a plane-symmetric, exact,
      cosmological solution of the Einstein equation, extending
      earlier work.  More generally, for a
      standard initial power spectrum, we use the relativistic Zel'dovich approximation
      implemented in {\sc{}inhomog} to numerically estimate how much
      positive spatial curvature is required to allow turnaround
      at typical epochs/length scales
      in almost-Einstein--de~Sitter (EdS)/{\LCDM} models
      with inhomogeneous curvature.}
    {We find that gravitational collapse in a spatially exactly flat,
      irrotational, expanding, dust universe is relativistically
      forbidden pointwise.  In the spatially flat plane-symmetric
      model considered here, pancake collapse is excluded both pointwise and in
      averaged domains.  In an almost-EdS/{\LCDM} model, the per-domain average curvature in collapsing
      domains almost always becomes strongly positive prior to
      turnaround, with the expansion-normalised curvature functional
      reaching $\OmRD\sim-5$. We show analytically that a special
      case gives $\OmRD=-5$ exactly (if normalised using the EdS
      expansion rate) at turnaround.}
    {An interpretation of {\LCDM} as literally 3-Ricci flat would
      forbid structure formation.  \protect\postrefereeAAchanges{The
        difference between relativistic cosmology and a strictly flat
        {\LCDM} model is fundamental in principle, but we find that the
        geometrical effect is weak.}}
}

\keywords{
    Cosmology: theory --
cosmological parameters --
large-scale structure of Universe --
dark matter}

\begin{document}

\mystamp{}

\maketitle

\newcommand\FIGWIDTHONE{0.95\columnwidth}
\newcommand\FIGWIDTHTWO{1.05\columnwidth}

\newcommand\tOmRDTA{
\begin{table}
  \caption{Zero point $\OmRD(\initaverageriem{\invII}=0)$ estimated
    using Theil--Sen robust linear fit of $\OmRD$ against
    $\initaverageriem{\invII}$ at turnaround ($|H_{\CD}| <
    1$~km/s/Mpc), with robust estimate of the uncertainty (1.4826
    times the median absolute deviation from 100 bootstrap Theil--Sen
    fits, corresponding to 68\% probability for a Gaussian
    distribution).
    \label{t-OmRDTA}}
  $\begin{array}{c ccc}
    \hline\hline
    \LD & 2.5 & 10 & 40     \rule{0ex}{2.5ex}\\ 
    \mathrm{Mpc}/\hzeroeff
    \\
    \hline
    \rule{0ex}{2.5ex}
     \mbox{EdS}
 &     -3.71 \pm      0.05
 &     -4.77 \pm      0.04
 &     -5.37 \pm      0.01
 \\
 \mbox{{\LCDM}}
 &     -4.24 \pm      0.08
 &     -4.24 \pm      0.03
 & \mbox{--}
 \\
     \hline
  \end{array}$
\end{table}
}

\newcommand\tOmRDTAplanesym{
\begin{table}
  \caption{
    Zero point of $\OmRDPS(\initaverageriem{\invII}=0)$, defined in
    Eq.~\protect\eqref{e-OmRDTA-planesym},
    near turnaround, estimated
    by robust statistics, as in Table~\protect\ref{t-OmRDTA}.
    The special case initial conditions should yield $\OmRDPS = -6$.
    \label{t-OmRDTA-planesym}}
  $\begin{array}{c ccc}
    \hline\hline
    \LD & 2.5 & 10 & 40    \rule{0ex}{2.5ex}\\ 
    \mathrm{Mpc}/\hzeroeff
    \\
    \hline
    \rule{0ex}{2.5ex}
     \mbox{EdS}
 &     -6.50 \pm      0.02
 &     -6.17 \pm      0.02
 &     -5.99 \pm      0.01
 \\
 \mbox{{\LCDM}}
 &     -6.08 \pm      0.05
 &     -5.85 \pm      0.01
 & \mbox{--}
 \\
     \hline
  \end{array}$
\end{table}
}

\newcommand\tOmRDTAstats{
\begin{table}
  \caption{Statistics of $\OmRD(\initaverageriem{\invII}=0)$
    near turnaround (defined as $|H_{\CD}| < 1$~km/s/Mpc),
    mean $\mu(\OmRD)$, standard deviation $\sigma(\OmRD)$,
    median $\mu'(\OmRD)$, robust dispersion estimate
    (1.4826 times the median absolute deviation) $\sigma'(\OmRD)$.
    \label{t-OmRDTA-stats}}
  $\begin{array}{c rrr}
    \hline\hline
    \LD & 2.5 & 10 & 40     \rule{0ex}{2.5ex}\\ 
    \mathrm{Mpc}/\hzeroeff
    \\
    \hline
    \rule{0ex}{2.5ex}
     & \multicolumn{3}{c}{\mbox{EdS}} \\
 \mu(\OmRD) &     -3.95 &     -4.84 &     -5.48 \\
 \sigma(\OmRD) &      0.92 &      0.96 &      0.36 \rule{0ex}{2.5ex} \\
 \mu'(\OmRD) &     -3.73 &     -4.71 &     -5.38 \rule{0ex}{2.5ex} \\
 \sigma'(\OmRD) &      0.76 &      0.58 &      0.20 \rule{0ex}{2.5ex} \\
 & \multicolumn{3}{c}{\mbox{{\LCDM}}} \\
 \mu(\OmRD) &     -4.30 &     -4.26 & \mbox{--} \\
 \sigma(\OmRD) &      0.70 &      0.32 & \mbox{--} \rule{0ex}{2.5ex} \\
 \mu'(\OmRD) &     -4.25 &     -4.24 & \mbox{--} \rule{0ex}{2.5ex} \\
 \sigma'(\OmRD) &      0.57 &      0.37 & \mbox{--} \rule{0ex}{2.5ex} \\
     \hline
  \end{array}$
\end{table}
}

\newcommand\tnonnegOmRDTA{
\begin{table}
  \caption{Fraction $\NnearTA(\OmRD \ge 0)\,/\,\NnearTA$ of
    near-turnaround domains
    that have negative (or zero) curvature;
    as can be inferred from
    Figs~\protect\ref{f-II-OmRD-EdS}--\protect\ref{f-OmmOmQ-LCDM},
    it is numerically unlikely for a domain to reach turnaround
    unless its curvature is positive.
    \label{t-nonnegOmRDTA}}
  $\begin{array}{c ccc}
    \hline\hline
    \LD & 2.5 & 10 & 40     \rule{0ex}{2.5ex}\\ 
    \mathrm{Mpc}/\hzeroeff
    \\
    \hline
    \rule{0ex}{2.5ex}
     \mbox{EdS}
 & 0/140 
 & 0/357 
 & 0/266 
 \\
 \mbox{{\LCDM}}
 & 0/230 
 & 0/277 
 & \mbox{--} 
 \\
     \hline
  \end{array}$
\end{table}
}

\newcommand\tversions{
\begin{table}
  \caption{Software version numbers, {\sc git} commit hashes, and {\sc git} repositories.
    \label{t-versions}}
  \begin{tabular}{lll}
    \hline\hline
    package, URL \rule{0ex}{2.5ex} 
    & version
    & {\sc git} commit hash
    \\
    \hline
    {\sc mpgrafic}\rule{0ex}{2.5ex} 
    & 0.3.18
    &  19103c7
     \\
    \multicolumn{3}{l}{\url{https://bitbucket.org/broukema/mpgrafic}}
    \\
    \DTFE{}  & 1.1.1.Q
    & 8efe489
     \\
    \multicolumn{3}{l}{\url{https://bitbucket.org/broukema/dtfe}}
    \\
    \inhomog{} &
    0.1.10
    & 876c7e5
     \\
    \multicolumn{3}{l}{\url{https://bitbucket.org/broukema/inhomog}}
    \\
    \RAMSES{} &
    ramses-use-mpif08
    & 7b64713
     \\
    \multicolumn{3}{l}{\url{https://bitbucket.org/broukema/ramses-use-mpif08}}
    \\
    \RAMSESscalav{} &
    --
    & df379ab
     \\
    \multicolumn{3}{l}{\url{https://bitbucket.org/broukema/ramses-scalav}}
    \\
    \hline
  \end{tabular} \\
\end{table}
}  

\newcommand\tversionsnbody{
\begin{table}
  \caption{\protect\postrefereeBBchanges{Software {\sc git} commit hash
    that differs from Table~\protect\ref{t-versions}; used for Fig.~\protect\ref{f-OmRDEdS-nbody}.}
    \label{t-versions-nbody}}
  \begin{tabular}{lll}
    \hline\hline
    package, URL \rule{0ex}{2.5ex} 
    & {\sc git} commit hash
    \\
    \hline
    \RAMSESscalav{} \rule{0ex}{2.5ex} 
    & 14bfebd
     \\
    \multicolumn{3}{l}{\url{https://bitbucket.org/broukema/ramses-scalav}}
    \\
    \hline
  \end{tabular} \\
\end{table}
}

\section{Introduction} \label{s-intro}

We investigate the degree to which \prerefereeAAchanges{the} spatial flatness
\prerefereeAAchanges{of a compact spatial domain of the Universe}
would relativistically prevent
\prerefereeAAchanges{the domain from gravitationally collapsing}.
Cosmological observations are usually analysed
\prerefereeAAchanges{by modelling
  the Universe as} a Friedmann--Lema\^{\i}tre--Robertson--Walker
(FLRW) \citep{Fried23,Lemaitre31ell,Rob35} model with a flat comoving
spatial section
(e.g.
\postrefereeAAchanges{\cite{Cole05BAO,Eisenstein05,
    RyanYunRatra19,
    Bennett13WMAP9}; and
  \cite{Planck2015cosmparam} apart from
  the gravitational lensing analyses therein;}
and references therein).
\prerefereeAAchanges{However, this use of flatness for calculations coexists with the
  contradictory characteristic of spatially inhomogeneous curvature.
  The seeds of structure formation
  are frequently modelled as early epoch curvature pertubations
  associated with density pertubations, in
  linear perturbation theory \citep[e.g.][and references
  therein]{Durrer96HelvPertTheory},
  and allowed to grow}
into non-linear gravitationally
collapsed (or \prerefereeAAchanges{nearly} emptied) structures by the current epoch, based on
the hypothesis that the perturbed solutions are
good approximations to exact
solutions of the Einstein equation for as long as the perturbations remain
weak. Empirically, this mathematical
hypothesis seems to be {\em a posteriori} reasonable.
However, the Einstein equation imposes constraints that are still
being explored and not yet completely understood.
Non-perturbative, exact, nearly-FLRW cosmological solutions
of the Einstein equation
\citep[for a review, see][]{Krasinski06book} can, as we show here,
provide \prerefereeAAchanges{stronger} constraints on structure formation
\prerefereeAAchanges{than those provided by linear perturbation theory}.
It is
\postrefereeAAchanges{important to improve our understanding of}
the distinction between the
{\LCDM} ($\Lambda$ cold dark matter) model,
interpreted literally as having flat FLRW spatial sections
\prerefereeAAchanges{with non-linear curvature inhomogeneities that by
  coincidence cancel exactly to a uniform, flat average background,
versus} strictly general-relativistic models that account for structure
\prerefereeAAchanges{formation with as few linearisations as possible.}
\prerefereeAAchanges{Better understanding of inhomogeneous
  curvature at the scale of a few tens of megaparsecs can
  potentially contribute to the understanding of large-scale
  average scalar curvature.}
This is especially important in the context of the upcoming generation
of major extragalactic surveys including those of ground-based photometric projects such
as LSST (Large Synoptic Survey Telescope; \cite{TysonLSST03}), spectroscopic projects
such as 4MOST (4-metre Multi-Object Spectroscopic Telescope;
\postrefereeAAchanges{\cite{deJong12VISTA4MOST,Richard19CRS4MOST}}),
DESI (Dark Energy Spectroscopic Instrument; \cite{Levi13DESI}) and
eBOSS (extended Baryon Oscillation Spectroscopic Survey; \cite{Zhao15eBOSSpredict}),
and the space-based projects Euclid \citep{EuclidScienceBook2010}
and COREmfive \citep{DelabrouilleCORE18,deZotti15full,deZotti17full}.

As a step towards this clarification,
the primary aim of this work is to non-perturbatively check the conditions under which a
flat or nearly flat expanding dust (general-relativistic)
universe forbids the gravitational collapse of structure.
By \enquote*{flat}, we refer to the spatial curvature for a
flow-orthogonal spacetime foliation, as detailed below in
\SSS\ref{s-method-GR-fluid}. This foliation is defined by a physical criterion
and is thus
gauge independent (see \SSS\ref{s-gauge-weak-dependence} for more discussion).

\prerefereeAAchanges{While Newtonian
  gravity in standard $N$-body simulations can be interpreted as
  requiring positive spatial curvature for gravitational collapse, this
  relativistic interpretation only follows
  from a quite restricted line of reasoning.
  This \enquote*{Newtonian limit} argument follows from a Newtonian slicing (a Newtonian-gauge restriction)
  of a relativistic cosmological spacetime
  with a spatial section that is topologically $\mathbb{E}^3$ or
  $\mathbb{S}^1\times \mathbb{S}^1\times \mathbb{S}^1$
  and that has a line element restricted to
  \begin{align}
    \diffd s^2 &= a^2 \left\{ -(1+2\psi) \, \diffd \tau^2 + (1-2\varphi) \left[ \diffd x^2 + \diffd y^2 + \diffd z^2 \right]\right\} \,,
  \end{align}
  where $a = a(\tau)$ is a flat FLRW background scale factor depending on conformal time $\tau$,
  and $\psi$ and $\varphi$ are scalar fields on the spacetime.
  The 3-Ricci scalar curvature is
  \begin{align}
    {}^3R = {{8\,\varphi\,
        \left(
        \varphi_{,x\,x}
        +\,\varphi_{,y\,y}
        +\,\varphi_{,z\,z}
        \right)
        -4\,\left(
        \varphi_{,x\,x}
        +\,\varphi_{,y\,y}
        +\,\varphi_{,z\,z}
        \right)
        -6\,\left(
        \varphi_{,x}^2
        + \,\varphi_{,y}^2
        +\,\varphi_{,z}^2
        \right)}
      \over{a^2\,\left(2\,\varphi -1\right)^3}} \,,
    \label{e-Newtonian-gauge}
  \end{align}
  where commas indicate partial derivatives.
  For linear perturbation theory with a slowly varying potential $\varphi$ and weak second derivatives of
  the potential, that is, for
  $|\varphi| \ll 1,\;
  |\varphi_{,x^i}| \ll 1,\;
  |\varphi_{,x^ix^i}| \ll 1 \;\forall x^i \in \{x,y,z\}$,
  Eq.~ \eqref{e-Newtonian-gauge} can be approximated as
  \begin{align}
    {}^3R &\approx 4\,a^{-2}\, \left( \varphi_{,x\,x} +  \varphi_{,y\,y} +  \varphi_{,z\,z} \right)
    \nonumber \\
    &= 4\,a^{-2}\, \nabla_{\mathbb{E}^3}^2 \varphi
    \nonumber \\
    &= 16 \pi \,G\, \delta\rho\,,
  \end{align}
  for an overdensity $\delta\rho$,
  where the final line follows from the Newtonian expanding universe
  form of Poisson's equation.
  Thus, since gravitational collapse will usually require a positive overdensity, this should
  be associated with positive spatial curvature in the Newtonian restriction
  if the potential $\varphi$ and its first and second derivatives are weak enough.
  (For wider discussion,
  see especially Eq.~(4.5) of \cite{KodamaSasaki84}; or \cite{Durrer96HelvPertTheory}
  and references therein.) Here, we do not make these restrictions.}

\prerefereeAAchanges{Beyond this interpretive, approximate, Newtonian-gauge--restriction sense,
  the Newtonian formulae and vector space structure
  of standard $N$-body simulations
  are not modified to take into account spatial curvature.
  Thus, the
  \enquote*{Newtonian cosmology} that directly enters
  into calculations that are coded in the non-linear evolution of
  standard $N$-body simulations
  \citep[e.g.][]{BaglaPaddy97,Teyssier02,Springel05GADGET2,GuilletTeyssier11},
  is, in principle, relativistically inaccurate by not taking into account curvature
  beyond this intepretive sense.}
  Here, by
  Newtonian cosmology we mean work such as that of
  \cite{BuchertEhlers97} and the terminology of
  Ellis (\cite{Ellis1971RelCos}; \cite{Ellis1971RelCosRepub}, p.~611), who observes that spatial
  curvature is \enquote*{essentially general-relativistic in character} and
  that the 3-Ricci tensor \enquote*{has no Newtonian analogue}, and of
  Ellis et al.\/ \cite[][p.~149]{EllisMaartensMacCall12}, who note that in Newtonian
  cosmology, the Friedmann equation curvature constant is \enquote*{a constant
  of integration, with no relation to spatial curvature}.

\postrefereeAAchanges{In standard $N$-body simulations,} spatial flatness
is built into the simulations via
two-point flat-space Newtonian gravitational attraction for \prerefereeAAchanges{the \enquote*{particle-particle}
mode} of calculation, Fourier analysis for the \enquote*{mesh} mode of calculating gravitational
potentials, vector arithmetic
\prerefereeAAchanges{(in the universal covering space)}
that is \postrefereeAAchanges{undefined} in a space that is not a vector space,
\postrefereeAAchanges{and distance calculations that use a flat FLRW metric
  (not \prerefereeAAchanges{an inhomogeneous} Newtonian-gauge metric).}
Using these methods is consistent with the assumptions of Newtonian cosmology
\postrefereeAAchanges{(in the sense defined above),}
reliant on the notion of an absolute spacetime, which provides a flat
embedding for cosmological fluid trajectories. In Newtonian
cosmology, one can change from Eulerian coordinates, associated
with the absolute spacetime, to Lagrangian coordinates, which trace fluid
evolution. This allows associating the relativistic notions of
extrinsic and intrinsic spatial curvature to properties of
fluid trajectories in Newtonian cosmology. The fact
that the Eulerian to Lagrangian coordinate
transformation is invertible prior to shell crossing
sometimes leads to the belief that
the extrinsic and intrinsic curvatures are merely artefacts of
a specific coordinate system.
In particular, in the
context of cosmological perturbation theory,
a diffeomorphism is used to map
between the curved manifold
and a flat background
with pulled-back fields,
giving the misleading impression that scalar
quantities associated with curvature are not invariant.

\postrefereeAAchanges{To the best of our knowledge, it has not been shown prior to this work that}
positive intrinsic spatial curvature
in the fluid rest frame is
unavoidable if a collapsing domain is to reach the turnaround epoch,
\prerefereeAAchanges{apart from the approximate, Newtonian-gauge--restriction argument
  above that imposes several restrictive assumptions.}
This is yet another example of how
extrinsic and intrinsic curvatures are real physical quantities
that are not removable by any coordinate transformation.

With the aim of developing new tests for observational cosmology,
we also wish to estimate approximate values of positive
spatial curvature that are required
at characteristic mass scales and epochs in
the almost-Einstein--de~Sitter (EdS) and almost-{\LCDM} models in order to allow overdensities
to reach their turnaround epochs.
We use the term \enquote*{almost-} to indicate that inhomogeneous
spatial curvature is allowed in the models.
The initial spatial curvature at any point or region
is typically weak.
However, in contrast to the evolution of the spatially constant curvature of a non-flat FLRW
model, which is static in FLRW comoving units but
weakens in physical inverse square length units as the Universe expands,
generic spatial curvature evolution tends to strengthen initially
weak curvature.

Moreover, we show that the positive curvature at turnaround must,
at least in a particular EdS case, occur at a particular critical value.
The motivation for considering the EdS case is that a
cosmic microwave background (CMB) normalised EdS model together
with structure formation may provide the extra 16\% expansion
(scale factor value) needed for the combined model to be observationally
viable without dark energy \citep*[][Eq.~(13)]{RMBO16Hbg1}.

Following the standard FLRW and scalar averaging conventions, the
expansion-rate--normalised curvature functional adopted here
($\OmRD$, see Eq.~\eqref{e-defn-OmRD}) has the opposite
sign to that of the scalar curvature itself. In the scalar averaging context, the
\enquote*{Omegas} are referred to as functionals rather than parameters, because they
depend on fields on spatial hypersurfaces.
\postrefereeAAchanges{We set the initial values
of the average scale factor in a domain $\CD$,
of the effective (globally averaged) scale factor,
and of $a(t)$,
to be equal, that is, we set
$a_{\CD}(\initial{t})$
and $\aeff(\initial{t})$ to be equal to
$a(\initial{t})$, respectively,}
where this value is normalised to the CMB value using a {\LCDM} proxy
that reaches a unity scale factor at the current epoch,
as detailed in \SSS\ref{s-method-QZA}.

In \SSS\ref{s-method} we present our method, including definitions and
terminology in
\SSS\ref{s-method-GR-fluid},
the Hamiltonian constraint in \SSS\ref{s-method-Hamiltonian-constraint}, and
the definition of an illustrative,
one--free-function, plane-symmetric case in \SSS\ref{s-method-P-only-case}.
In \SSS\ref{s-method-EdS-LCDM} we describe how we analytically investigate a special case of
initial conditions in an almost-EdS model (\SSS\ref{s-method-EdS-II-III-zero}),
and in the more general case, for standard cosmological $N$-body simulation initial conditions,
for the EdS and {\LCDM} cases (\SSS\ref{s-method-QZA}), where we also include the effects
of scalar averaging.
Results are given in \SSS\ref{s-results}.
In \SSS\ref{s-results-Hamiltonian-constraint} we examine the pointwise (unaveraged)
Hamiltonian constraint, which relates,
in particular, the expansion rate, density and curvature.
In \SSS\ref{s-results-P-only-case}
we consider the plane-symmetric, exact, flat, expanding universe example.
In \SSS\ref{s-results-EdS-LCDM} we present
the results in a special analytical almost-EdS case that yields a critical value
(\SSS\ref{s-results-EdS-II-III-zero})
and more general numerical results in the almost-EdS and
almost-{\LCDM} cases (\SSS\ref{s-results-EdS-LCDM-subsub}).
We discuss the results in \SSS\ref{s-discuss}
\postrefereeAAchanges{and applications to numerical general relativity}
in \SSS\ref{s-OmRD-HCD-relation},
and conclude in \SSS\ref{s-conclu}.

\section{Method} \label{s-method}

\subsection{Universe model} \label{s-method-GR-fluid}

As in \cite{Buch00scalav} and following \cite{Ehlers61,Ehlers61en}, we generalise beyond the FLRW model,
allowing initial inhomogeneities to, {\em a priori}, gravitationally
collapse, expand and become voids, or form more complex structures that
together yield the cosmic web, as follows.
We adopt the notation of \cite*{BuchRZA2}
for an irrotational dust universe foliated by flow-orthogonal spatial sections
that are labelled by a time coordinate $t$ defined by the fluid proper time.
This notation differs slightly from that of \cite{Buch00scalav}.
We adopt Roman indices to indicate spatial coordinates,
we use the Einstein summation convention, and an overdot ($\dot{\ }$) indicates
derivatives with respect to $t$.
An FLRW \enquote*{reference model} (not necessarily
an average model; we thus avoid the ambiguous
term \enquote*{background model}) is used here, with scale factor $a(t)$.
The extrinsic curvature $K^i_j$ is used to define
the expansion tensor
$\Theta^i_j := -K^i_j$, the
expansion scalar (without removal of a reference model Hubble--Lema\^{\i}tre flow;
\cite{Hubble1929,Lemaitre31ell,Lemaitre31elltrans})
$\Theta := \Theta^i_i \equiv - K^i_i$,
and the peculiar-expansion tensor
$\theta^i_j : = \Theta^i_j - H \,\delta^i_j$
(for the Kronecker delta $\delta^i_j$),
where a reference model expansion rate $H := \dot{a}/a$ is subtracted; the
shear tensor is defined $\sigma^i_j := \Theta^i_j - \frac{1}{3} \Theta \, \delta^i_j$
and the shear scalar
$\sigma^2 := \frac{1}{2} \sigma^i_j \, \sigma^j_i$;
the spatial scalar curvature is
$\CR := \CR^i_i$; density is $\rho$; the gravitational constant is $G$; and
an optional cosmological constant is $\Lambda$.

The FLRW models
solve the Einstein equation in a way in which $a^2(t)\, \CR $
is spatially constant (curvature is set to be homogeneous)
and setting $\theta^i_j = 0$, or equivalently, $\sigma^i_j = 0$ and $\Theta = 3H$.
The EdS and {\LCDM} models are part of the FLRW subcase in which
$\CR = 0$.
In sections \SSSS\ref{s-method-EdS-LCDM}
and \ref{s-results-EdS-LCDM}
we consider models that are almost-FLRW at early times.

\subsection{Hamiltonian constraint} \label{s-method-Hamiltonian-constraint}

The time--time component of the Einstein equation gives the
Hamiltonian constraint
\citep*{EllisBruniHwang90vort,
TsagasChallMaart08,
Magni2012masters,
BuchRas12},
which is presented in an elegant form
in \cite[][Eq.~(4a)]{Buch00scalav}:
\begin{align}
  \frac{1}{3}\Theta^2 &= 8\pi G \rho + \sigma^2 - \frac{1}{2} \CR + \Lambda \,.
  \label{e-Buch00dust-4a}
\end{align}
In \SSS\ref{s-results-Hamiltonian-constraint}, we use this equation
to determine the conditions
required for an initially weak overdensity
expanding with the rest of the Universe
to decelerate and reach its turnaround epoch at
points that are exactly flat spatially.
The above equation generalises the
Friedmann equation of the FLRW model, which can be
written with terms matching, respectively, those of
Eq.~\eqref{e-Buch00dust-4a},
\begin{align}
  3\,H^2 &= 3\,H^2 \Omm + 0 + 3\,H^2 \Omk + 3\,H^2 \OmLam\,,
  \label{e-Friedmann-FLRW}
\end{align}
with matter density parameter $\Omm$,
curvature parameter $\Omk$, and dark energy parameter $\OmLam$,
all of which are given here as time-dependent values,
not current-epoch constants.

\subsection{Plane-symmetric subcase} \label{s-method-P-only-case}
We illustrate the blocking of gravitational collapse (inability to reach the turnaround
epoch) with an
exact non-perturbative solution of the Einstein equation.
This is a subcase of the plane-symmetric case,
which has been considered by several authors
(\cite*{Ellis67planecase,
  Kras08szekplane,
  DidioVD12wallbackreact};
\cite{AdamekDDK14szekplane}). The existence of
this exact but reasonably simple
inhomogeneous cosmology solution (a Szekeres model; \cite{SzekeresP75})
potentially offers a powerful method of
calibrating relativistic cosmology software.
Here,
we examine the subcase in the form that has an
\enquote*{exact perturbation} $P(w,t)$ in a single direction $w$ in a universe whose spatial
section is either the infinite Euclidean space $\mathbb{E}^3$ or
the 3-torus $\mathbb{T}^3$.
By \enquote*{exact perturbation}, we mean that the perturbation $P$
can be studied without Taylor expansions in $P$ and without
dropping any Taylor or non-Taylor high-order terms in $P$.

The metric is expressed here as
\citep[][Eq.~(66), \SSS{}V.A]{BuchRZA2},
\begin{align}
  \diffd s^2 = &-\diffd t^2
  + a(t)^2
  \left[ \diffd x^2 + \diffd y^2
    + \left(1 + P(w,t)\rule{0ex}{2ex} \right)^2 \diffd w^2 \right],
  \label{e-plane-collapse-RZA}
\end{align}
with flow-orthogonal spatial sections,
where $P$ is by definition a plane-symmetric function, which we
require to be smooth.
Inserting this metric into the Einstein equation shows that
the scale factor $a(t)$ and the associated expansion rate
$H(t) := \dot{a}/a$ must be the standard flat FLRW solutions
(for arbitrary $\Lambda$).
In other words, $a(t)$ has to be the scale factor solution
for what we can consider to be
a flat \enquote*{reference model} universe.
The coordinates $x,y,w$ are comoving with the fluid.
We assume as an initial condition at an
early epoch that $|P(w,\initial{t})| \ll 1$.
Situations in which the line element in
the inhomogeneous direction $w$ is compressed to zero, that is,
in which $\lim_{t \rightarrow \tcollapse^+} P = -1$ at an epoch $\tcollapse$,
can be expected to represent gravitational \enquote*{pancake} collapse.
In
\SSS\ref{s-results-P-only-case}, we consider whether an initial
overdensity defined on a $w$ plane of the function $P$ can decelerate
sufficiently to reach its turnaround epoch, and extend the EdS reference model case
considered by \cite{BuchRZA2} to also include the case of
a {\LCDM} reference model.

To extend this analysis from pointwise collapse to averaged
collapse within a comoving spatial domain $\CD$
using the spatial metric volume element $\diffd \mu_{\mathbf{g}}$,
we need the scalar averaging operator $\average{.}$, defined
for a scalar field $\CA$ as
\begin{align}
  \average{\CA} := \frac{\int_{\CD} \CA \,\diffd \mu_{\mathbf{g}}}{\int_{\CD} \diffd \mu_{\mathbf{g}}}
  \label{e-defn-D-average}
\end{align}
(see also Eq.~\eqref{e-Riem-vs-Lag-average} below).
Applying this to Eq.~\eqref{e-Buch00dust-4a} and shifting
$\average{\Theta^2}/3$ to the right-hand side
yields
\begin{align}
  \frac{1}{3}\average{\Theta}^2 &= 8\pi G \average{\rho}
  + \average{\sigma^2}
  - \frac{1}{3}\average{\left(\Theta - \average{\Theta}\right)^2}
  - \frac{1}{2} \average{\CR} + \Lambda
  \label{e-Buch00dust-4a-averaged}
\end{align}
\citep[][Eqs~(10), (11), and references therein]{BuchRZA2}.

\subsection{Almost-Einstein--de~Sitter and almost-{\LCDM} models} \label{s-method-EdS-LCDM}

Since a realistic cosmological model must allow gravitational collapse
of density perturbations, the results below
(\SSS\ref{s-results-Hamiltonian-constraint},
\SSS\ref{s-results-P-only-case})
imply that positive spatial curvature has to be allowed
in such a model in order to allow pointwise turnaround to be reached.
The scalar-averaged behaviour in a spatial domain, as represented in
Eq.~\eqref{e-Buch00dust-4a-averaged}, has an extra term in the right-hand
side. Since $\average{\left(\Theta - \average{\Theta}\right)^2}$ is necessarily
non-negative, it could, in principle, provide an alternative physical driver
than positive curvature for allowing gravitational collapse through to the
turnaround epoch, provided that it is sufficiently bigger than the averaged
shear scalar $\average{\sigma^2}$.
To study this, as shown in
\cite*{BKS00}, \cite*{BuchRZA1}, and \cite[][and references therein]{BuchRZA2}
and
implemented with free-licensed software
as shown in \cite{Roukema17silvir},
the relativistic Zel'dovich approximation \postrefereeAAchanges{(RZA)}
can be used to model gravitational collapse and void formation
in the cosmological context much faster than using $N$-body
simulations, without any need to assume spherically
symmetric collapse, and permitting positive (and non-positive)
spatial scalar curvature.
\postrefereeAAchanges{This is done by integrating the averaged Raychaudhuri equation, where the averaging corresponds to
  a Lagrangian spatial domain $\CD$,
  starting with standard cosmological
  inhomogeneous initial conditions.
  This equation yields $\ddot{a}_{\CD}$,
  the second time derivative
  of the per-domain average scale factor $a_{\CD}$,
  and depends, in particular, on the temporal evolution
  of the kinematical backreaction functional.
The kinematical backreaction functional,} defined
\citep[][Eq.~(11)]{BuchRZA2}
\begin{align}
  \CQ_{\CD} &:=
  \frac{2}{3} \average{\left(\Theta - \average{\Theta}\right)^2}
  - 2 \average{\sigma^2} \,,
  \label{e-defn-Q}
\end{align}
combines the
\postrefereeAAchanges{expansion variance term,}
$\average{\left(\Theta - \average{\Theta}\right)^2}$,
and the shear scalar.
\postrefereeAAchanges{By definition, $\CQ_\CD$ only makes
  sense as an average quantity (it would be zero in the limit
  at a spatial point).}

\postrefereeAAchanges{More specifically,
  we use the average Raychaudhuri
  equation
  as expressed in Eq.~(25) of \cite{Roukema17silvir},
  and use Eqs~(30), (31)
  of \cite{Roukema17silvir} \citep[][Eq.~(50)]{BuchRZA2}
  to approximate the kinematical backreaction
  temporal evolution $\CQ_\CD(t)$.}
The relation of the evolution of non-collapsing domains to
collapsed (virialised) domains, a major theme of
\cite{Roukema17silvir}, is beyond the scope of the present
paper, and will be revisited in future work.

The Raychaudhuri and the RZA kinematical backreaction
($\CQ_\CD$) evolution
equations are algebraically identical in Newtonian and
relativistic cosmological models,
as shown in \cite{BuchRZA2}, although they have
different interpretations and, in principle, physically different
meanings. For numerical purposes, it is usually assumed that there are
no significant differences between the initial conditions for {\LCDM}
$N$-body (Newtonian)
simulations and those for relativistic cosmology simulations.  For
this reason, this method can be conveniently referred to as the QZA method,
where Q represents the kinematical backreaction term and
ZA stands for the relativistic Zel'dovich approximation that to some
degree can be interpreted as Newtonian.
The main practical difference between the Newtonian and relativistic
cases, as shown below in
\SSS\ref{s-results-Hamiltonian-constraint} and illustrated in
\SSS\ref{s-results-P-only-case},
is not directly in QZA itself,
but is instead
an effect of the Hamiltonian constraint, which can induce
at least one fundamental difference between the relativistic and
Newtonian cases.
The Raychaudhuri equation and the RZA
kinematical backreaction evolution equation,
as used in the QZA, do not,
in practice, distinguish the two cases directly, but because
of the Hamiltonian constraint,
not all
perturbation modes are allowed: a relativistic constraint can prevent
growing modes.
This is how flatness can forbid structure formation.

The case considered here can be compared to the Newtonian setting in terms
of the Hamiltonian constraint, which in the absence of spatial curvature
relates the initial density to the shear and expansion scalars. In this
context, a physical situation allowed in the Newtonian formulation
of a setting with a particular (plane) symmetry is forbidden in the
analogous general relativistic solution due to the difference in the
corresponding versions of the Hamiltonian constraint.

In this part of our work,
we make the reasonable hypotheses that strictly zero curvature is
not required in the cosmological model, and that
the slowing down to the turnaround epoch is
not blocked by the Hamiltonian constraint,
since positive curvature is allowed, and
in an averaged domain,
kinematical backreaction may be positive due to its
non-negative expansion variance component,
$\average{\left(\Theta - \average{\Theta}\right)^2}$.
In \SSS\ref{s-method-EdS-II-III-zero} we investigate the behaviour of
the average (spatial) per-domain
expansion-rate--normalised scalar curvature functional
\begin{align}
  \OmRD := - \frac{\average{\CR}}{6 \, {\Heff}^2}
  \label{e-defn-OmRD}
\end{align}
(defined as in \cite[][Eqs~(10), (11)]{WiegBuch10},
where $\Heff$ is the global volume-weighted
mean expansion rate;
see Eq.~\eqref{e-Riem-vs-Lag-average} here for the
definition of averaging)
in a special case of initial conditions in a spatial
domain. In \SSS\ref{s-method-QZA}, we simulate the general case, using
standard cosmological $N$-body initial conditions.

\subsubsection{Almost-EdS special case} \label{s-method-EdS-II-III-zero}

In addition to the definitions of \SSS\ref{s-method-GR-fluid},
we need to use several elements of the relativistic Zel'dovich approximation
apparatus, with the use of a reference model (in principle, reference-model--free
calculations are possible).
The FLRW parameters of the reference model, in this case EdS,
include the scale factor $a(t)$ and the Hubble--Lema\^{\i}tre parameter $H := \dot{a}/{a}$.
The peculiar volume deformation
$\CJ(t,\mathbf{X})$ at foliation time $t$ and Lagrangian position $\mathbf{X}$ is defined
as in \cite[][Eq.~(42)]{BuchRZA2}
\begin{align}
  \CJ(t,\mathbf{X}) &:= 1 + \xi(t) \initial{\invI} + \xi^2(t) \initial{\invII} +  \xi^3(t) \initial{\invIII} \,,
  \label{e-CJ-defn}
\end{align}
where $\xi(t)$ is a normalised, zero-pointed linear perturbation theory growth function defined in
\cite[][Eqs~(32), (33)]{BuchRZA2},
${\invI}, {\invII}, {\invIII}$ are \postrefereeAAchanges{defined}
as the principal scalar invariants of the peculiar-expansion
tensor $\theta^i_j$ (\SSS\ref{s-method-GR-fluid}),
and $\initial{\invI}, \initial{\invII}, \initial{\invIII}$
are their values on the initial hypersurface.
For the EdS reference model, writing $\initial{a}$ as the initial scale factor,
we write the normalised growth function as
\begin{align}
  \xi = \frac{a - \initial{a}}{\initial{a}} \,.
  \label{e-xi-EdS}
\end{align}

Averages of a scalar functional $\CA$ can be either spatial metric
(\enquote*{Riemannian}) averages
$\average{\CA}$ or Lagrangian averages $\initaverageriem{\CA}$
\citep[][Eqs~(1)--(8)]{BuchRZA2}. In particular,
from \cite[][Eqs~(2), (3)]{BuchRZA2} we need the spatial (Riemannian)
volume $V_{\CD}$ and average scale factor $a_{\CD}$ on a
Lagrangian (fluid-comoving) domain $\CD$ that evolved
from an initial domain $\initial{\CD}$,
\begin{align}
  V_{\CD}(t) &:= \int_{\CD} \diffd \mu_{\mathbf{g}} \; \equiv
  \int_{\CD} J\, \Gdet\, \diffd^3 X,
  \;\;\; a_{\CD} := \initial{a}\,\left(V_{\CD}(t)/V_{\initial{\CD}}\right)^{1/3},
  \label{e-defn-aD-VD}
\end{align}
where the initial spatial metric $\mathbf{G}(\mathbf{X}) := \mathbf{g}(\initial{t},\mathbf{X})$ is used to define
a local volume deformation $J := \left(\det(g_{ij})/\det(G_{ij})\right)^{1/2}$,
an initial normalisation $\Gdet := \left(\det{G_{ij}}\right)^{1/2}$,
and the spatial-metric volume element $\diffd \mu_{\mathbf{g}}$ is rewritten
as $J \, \Gdet \, \diffd^3 X$.
The Lagrangian average for a scalar field $\CA$ is defined \citep[][Eq.~(7)]{BuchRZA2}
\begin{align}
  \initaverageriem{\CA} &= \frac{\int_{\CD} \CA \, \Gdet\,\diffd^3X}
                  {\int_{\initial{\CD}} J(\initial{\mathbf{X}},\initial{t}) \,\Gdet\,\diffd^3X}
                  \;
                  = \frac{\int_{\CD} \CA \,\Gdet\,\diffd^3X}{\int_{\initial{\CD}} \,\Gdet\,\diffd^3X}
                  \;
                  = \frac{\int_{\CD} \CA \,\Gdet\,\diffd^3X}{V_{\initial{\CD}}} \,,
                  \label{e-defn-calI-average}
\end{align}
with the consequence that the Riemannian average can be written \citep[][Eq~(8)]{BuchRZA2}
\begin{align}
  \average{\CA} &= \frac{\initaverageriem{\CA J}}{\initaverageriem{J}}\,, \;
  \initaverageriem{J} = a_{\CD}^3 \,.
  \label{e-Riem-vs-Lag-average}
\end{align}
The Riemannian average is intended to correspond to the physical intuition of an average (mean),
while the Lagrangian average is normally intended as a convenient tool for both numerical
and analytical purposes, without a simple intuitive interpretation.
An exception is at the initial time, at which
$\initaverageriem{\CA(\initial{t})} \equiv \average{\CA(\initial{t})}$,
whence the subscript $\CI$ and frequent use of this
for describing initial averages.
The peculiar expansion rate of a domain is defined
\begin{align}
  H_{\CD} &:= \frac{\dot{a}_{\CD}}{a_{\CD}} \,.
  \label{e-defn-HD}
\end{align}
A domain that, in the Riemannian (spatial) averaging sense,
expands in the same way as the reference model at a
given time $t'$ would have $H_{\CD}(t') = H(t')$.

We can now relate the peculiar volume deformation $\CJ$
to the RZA local volume deformation
${}^{\mathrm{RZA}}J$ and assume that the latter is a fair approximation,
\begin{align}
  J \approx {}^{\mathrm{RZA}}J := a^3 \CJ \,,
  \label{e-J-vs-CJ}
\end{align}
\citep[][Eq.~(41)]{BuchRZA2}
where we have reversed the direction of
definition in order to avoid multiply defining $\CJ$.
Hereafter, we drop the \enquote*{RZA} superscript.

The special case that we consider is then defined by setting the
initial average values of the
second and third invariants
to zero,
\begin{align}
  \initaverageriem{\initial{\invII}} = 0 \;, \;\;\;
  \initaverageriem{\initial{\invIII}}  = 0\;.
  \label{e-II-III-zero}
\end{align}
This includes the plane-symmetric subcase considered above.
In \SSS\ref{s-results-EdS-II-III-zero} we use these together with Eq.~(48) of \cite{BuchRZA2}
to investigate the behaviour of $\OmRD$ at turnaround.

\tversions

\subsubsection{QZA simulations} \label{s-method-QZA}

For the general case, we adopt a standard cosmic-microwave-background normalised
Gaussian-random-fluctuation power spectrum
for the appropriate FLRW reference model
and the growing mode
of perturbations of either the EdS or {\LCDM} FLRW models.
This lets us study the positive spatial curvature associated
with achieving turnaround. We leave deeper investigation of the post-turnaround stages of gravitational
collapse of galaxy and cluster scale objects to future work.

\postrefereeAAchanges{We generate a realisation of $N$-body simulation initial conditions;
  we estimate the peculiar-expansion tensor;
  we average the principal scalar invariants of this tensor within
  individual domains $\CD$;
  we calculate $\CQ_\CD$ evolution using the relativistic Zel'dovich approximation for each domain;
  we calculate evolution of the effective scale factor $a_{\CD}$ for each domain using the average Raychaudhuri equation;
  and we infer the evolution of the per-domain expansion rate $H_{\CD}$ and of
  the cosmological functionals (\enquote*{Omegas})
  in each domain.}

\postrefereeAAchanges{More specifically,
we} use the same method as stated in
\cite{Roukema17silvir}
in \SSSS{}3.1--3.4, 3.5.1--3.5.3,
3.5.4, 3.5.5,
and the first paragraph of 3.7.
In particular, we again use the domain-normalised Hamiltonian constraint
for $\OmRD$ given at the end of (42) in \cite{Roukema17silvir}, rather than
calculating $\average{\CR}$ directly using Eqs~(13) and (54)
of \cite{BuchRZA2}.
This way, we effectively use Eq.~(60) of \cite{BuchRZA2},
which satisfies the integral constraint
between the Raychaudhuri equation and the Hamiltonian constraint,
instead of Eq.~(54) of \cite{BuchRZA2},
which is expected to be less accurate. There should be no difference between
these in the special case
considered in \SSS\ref{s-results-EdS-II-III-zero},
and in general there should be a disagreement between the methods
of the order of
$\initaverageriem{\initial{\invII}}$ \citep[][Eq.~(60)]{BuchRZA2}.

We use {\sc mpgrafic}
to generate initial conditions
\citep{Bert01grafic,PrunetPichon08mpgrafic}, {\DTFE}
to estimate the peculiar-expansion tensor
\citep{SchaapvdWeygaert00,vdWeygaertSchaap07,CW11,Kennel04kdtreeDTFE},
\inhomog{} to carry out
effective scale factor and
\postrefereeAAchanges{$\CQ_\CD$}
evolution using the relativistic Zel'dovich
approximation
\citep{Roukema17silvir},
and {\RAMSESscalav}
\citep{Roukema17silvir}
for using {\RAMSES}
\citep{Teyssier02,GuilletTeyssier11}
as a front end for
reading in initial conditions and calling {\DTFE} and {\inhomog}.
All these packages are free (\enquote*{as in speech}) software.
The versions and {\sc git} commit hashes of the software used for the results shown here
are listed in Table~\ref{t-versions}.

Here, we also consider the case of a {\LCDM} reference model, in order to find out how strongly
positive the pre-turnaround curvature is required to be in order to allow
structure formation in {\LCDM}.
To obtain initial conditions normalised at the CMB epoch, instead of using {\LCDM} as
a proxy to extrapolate from late times back towards CMB-epoch EdS parameters
\citep[see][]{RMBO16Hbg1}, we use {\LCDM} as
a proxy for itself.
That is,
for the {\LCDM} case
we use the Planck 2015 \citep[][Table 4, final column]{Planck2015cosmparam} estimates
of a current-epoch matter density parameter
of $\Ommzero = 0.3089$ and a normalisation of
$\sigma_8 = 0.8159$.
Using \cite{Kasai2010}'s formulae,
version {\sc inhomog-0.1.9} includes a speed improvement
of about 4--10
for calculations of the flat, non-EdS growth function (and its first and
second derivatives) over those
performed using the \cite{BildBuchKas} incomplete beta function algorithm.
We set the spacetime unit conversion constant to unity except
where otherwise noted.
A free-licensed script to install system-level packages, to
download, compile, and install user-space packages, to run
{\RAMSES} as a front-end, and to plot and table results that are statistically equivalent
to those presented here is provided online with the aim of convenient
reproducibility.\footnote{\url{https://bitbucket.org/broukema/1902.09064}}

\section{Results} \label{s-results}

\subsection{Hamiltonian constraint} \label{s-results-Hamiltonian-constraint}

We now examine Eq.~\eqref{e-Buch00dust-4a} to see if there are
conditions in which flatness prevents gravitational collapse.
At an initial time $\initial{t}$, we adopt the standard assumption
that the Universe is expanding everywhere, so
$\Theta(\initial{t}) > 0$ holds everywhere, with only weak perturbations.
We do not assume any linearisation of the perturbations;
in this subsection and the folllowing, we only consider
\enquote*{exact perturbations}, as mentioned above.

Given that the Universe is initially expanding everywhere, pointwise collapse
requires the expansion scalar
to decrease from $\Theta(\initial{t}) > 0$
to $\Theta=0$, that is, it has to reach its turnaround epoch.
This requires that the right-hand side of
Eq.~\eqref{e-Buch00dust-4a} be zero.
The density $\rho$ will normally be expected to be positive in order for gravitational
collapse to occur, and a strictly zero density is physically unreasonable in the
cosmological context.
The shear scalar term $\sigma^2$ is necessarily non-negative (see the
definitions in \SSS\ref{s-method-GR-fluid} and
\cite[][Eq.~(79)]{BuchRZA2}).
We are assuming flatness, so $\CR = 0$.
If we also have $\Lambda \ge 0$, as in the EdS or {\LCDM} models,
then the right-hand side must be positive.

Since we do not consider $\Lambda < 0$ here, and we consider zero density to be unrealistic,
especially for a gravitationally collapsing perturbation,
the only way that the right-hand side can
reach zero is with
positive curvature: $\CR >0$.
Positive curvature is the physical, geometrical phenomenon that
can permit an overdensity to slow
its expansion and turn around from expansion to contraction in terms of
proper (\enquote*{physical}) spatial separations.
It is in this sense that flatness prevents the gravitational collapse of
exact density perturbations, by blocking the positive spatial curvature
required for achieving turnaround. We summarise \postrefereeAAchanges{what we have shown} as follows.
\begin{proposition}
  Suppose that a cosmological solution to the Einstein equation
  with a zero or positive cosmological constant
  and a fluid-flow--orthogonal foliation satisfies the conditions
  that there is a spatial hypersurface $\Sigma(t)$ at foliation
  time $t$ such that (i) the model is everywhere 3-Ricci flat since an
  initial time $\initial{t}$, that is,
  $\forall X \in \Sigma, \,\forall t' \in [\initial{t},t], \, \CR(X,t) = 0$;
  (ii) the model expands since the initial time,
  $\forall X \in \Sigma, \,\forall t' \in \closedopen{\initial{t}}{t},\,  \Theta(X,t') > 0$;
  and (iii) the fluid is irrotational, pressure-less.
  In these conditions, pointwise gravitational slowdown \postrefereeAAchanges{and
  \enquote*{turn around}} ($\Theta(t) = 0$) cannot occur on $\Sigma(t)$.  Thus, density
  perturbations on $\Sigma$ cannot (isotropically) gravitationally
  collapse at any point $X$ during the interval $t'\in [\initial{t},t]$.
  \label{e-lemma-flat-no-turnaround}
\end{proposition}
\newcommand{\ELemmaFlatNoTurnaround}{1} 
The caveat on isotropy is required because $\Theta$ is the
trace of the expansion tensor. What might be called \enquote*{anisotropic collapse}, with
expansion outweighing contraction, is not forbidden.

\subsection{Plane-symmetric subcase} \label{s-results-P-only-case}

\postrefereeAAchanges{Proposition}~{\ELemmaFlatNoTurnaround} shows why
\cite[][\SSS{}V.A]{BuchRZA2} found a fundamental difference between
Newtonian and relativistic cosmology in a subcase of the
plane-symmetric solution, as given above with the line element in
Eq.~\eqref{e-plane-collapse-RZA}.
The absence of a growing mode of exact density perturbations
was pointed out in that work as a consequence of the Hamiltonian
constraint, which in the relativistic case includes a scalar curvature
term. However, it was argued there that
pancake collapse could nevertheless occur for this exact solution, despite the absence of
a growing mode.

Here, we examine this case more closely, extending it from the EdS
reference model case to include the option of a {\LCDM} reference
model.  The metric as shown in Eq.~\eqref{e-plane-collapse-RZA}
is exactly 3-Ricci flat: all components $\CR^i_j $ are zero.
As shown in \cite[][Eq.~(67), \SSS{}V.A]{BuchRZA2},
by using the
definitions in \SSS\ref{s-method-Hamiltonian-constraint},
the peculiar-expansion tensor can be written
\begin{align}
  \theta^i_{\,j}
  &= \frac{1}{2} g^{ik}\dot{g}_{kj} - H\,\delta^i_j 
  = \mathrm{diag} \left( 0, 0,\frac{\dot{P}}{1+P} \right) \,,
  \label{e-P-theta-ij}
\end{align}
and it follows that
\begin{align}
  \Theta = 3H + \theta^i_i =: 3H + \theta \;\;
  \textrm{and} \;\;
  \sigma^2 = \frac{1}{3}\left(\frac{\dot{P}}{1+P}\right)^2
  = \frac{1}{3} \theta^2 \,,
  \label{e-P-Theta}
\end{align}
where $H$ is the expansion rate of the EdS or {\LCDM} reference
model.
The conditions of \postrefereeAAchanges{Proposition}~{\ELemmaFlatNoTurnaround}
are satisfied, which is sufficient to show
that isotropic gravitational turnaround cannot occur pointwise anywhere,
and thus collapse cannot occur either.
However, this is insufficient to show that pancake turnaround,
which is anisotropic by definition, cannot occur
pointwise, or that turnaround could occur in terms of averaged properties of a spatial
domain. Pointwise, pancake collapse in one spatial direction could,
in principle, be sufficiently balanced by expansion in the other directions
to give $\Theta(t) > 0$, which would not
violate \postrefereeAAchanges{Proposition}~{\ELemmaFlatNoTurnaround}.

The Einstein equation can now be written in the form of
Eq.~\eqref{e-Buch00dust-4a} together with
an equation closely related to the Raychaudhuri equation,
expressed in the form
\citep[][Eq.~(71)]{BuchRZA2}
\begin{align}
  \dot{\Theta}^i_j + \Theta \Theta^i_j =
  (4\pi G \rho + \Lambda) \delta^i_j \,.
  \label{e-RZA2-accel-eqn}
\end{align}
We now generalise the \cite[][Eq.~(74), \SSS{}V.A]{BuchRZA2} solution to
Eqs~\eqref{e-Buch00dust-4a} and \eqref{e-RZA2-accel-eqn}
to an arbitrary flat FLRW reference model by writing it as
\begin{align}
  P(w,t) &= B(w)+C(w)\frac{\dot{a}}{a}
  = B(w)+C(w)\, H(t)\,,
  \label{e-P-general-flat}
\end{align}
where $B(w)$ and $C(w)$ are functions depending only on the spatial coordinate $w$,
with no temporal dependence,
and we assume that $|B(w)| \ll 1,\; |C(w) \initial{H}| \ll 1$.
\postrefereeAAchanges{These low amplitude assumptions are only required for inequalities and limits
  as $t \rightarrow \infty$; high order terms are not set to zero. In other words, this is a non-perturbative
model.}
We ignore the special (fine-tuned) case where
$B(w) \approx -1,\; C(w) \initial{H} \approx 1,
B(w) + C(w) \initial{H} \ll 1$.

Adopting $|P(w,\initial{t})| \ll 1$,
gravitational collapse at a plane $w$ would
require $C(w) < 0$ and $H$ would have to increase with time $t$, in order
that $1+P$ drops to zero, that is,
to obtain $ 1 + P \rightarrow 0^+$. However,
$H(t)$ is a decreasing function for both the EdS and {\LCDM} reference
models, which can be seen (for example) as follows.
As $t \rightarrow \infty$, in the EdS case $H(t) = 2/(3t) \rightarrow 0^+$
and $\dot{H}(t) \rightarrow 0^-$.
In the {\LCDM} case, the exact FLRW expression using the
Hubble constant $H_0$,
the current-epoch cosmological constant density parameter
$\OmLamzero$ and the current age of the Universe $t_0$, is
\begin{align}
  a(t) = \left[\frac{\sinh\left(\frac{3}{2} H_0 \sqrt{\OmLamzero} \;t\right)}
    {\sinh\left(\frac{3}{2} H_0 \sqrt{\OmLamzero} \;t_0\right)} \right]^{2/3}
  \,,
  \label{e-a-LCDM}
\end{align}
(\cite{Peebles84inflation}, Eq.~(3);
\cite{SahniStaro00exactLambda}, Eq.~(12)),
which yields a reference model
expansion rate dropping from its initial
value towards a limiting
constant expansion rate, $H(t) \rightarrow (H_0 \sqrt{\OmLamzero})^+$
as $t\rightarrow \infty$,
with $\dot{H} < 0 \;\forall t$, and
$\dot{H}(t) \rightarrow 0^-$.
It follows that as $t\rightarrow \infty$,
$1+P(w\postrefereeAAchanges{,t}) \rightarrow 1 + B(w)$ (EdS reference model) or
$1+P(w\postrefereeAAchanges{,t}) \rightarrow 1 + B(w) + C(w) H_0 \sqrt{\OmLamzero}$
({\LCDM}).\footnote{These are one-sided limits, depending on the sign of $C(w)$.}
Thus, compression in the $w$ direction,
$g_{ww} \rightarrow 0^+$, is prevented, suggesting that
pointwise pancake collapse cannot occur.

We can check this using
Eqs~\eqref{e-P-theta-ij},
\eqref{e-P-Theta},
\eqref{e-P-general-flat} and the definitions in
\SSS\ref{s-method-GR-fluid}, which give the expansion scalar
\begin{align}
  \Theta &= 3H - \frac{\dot{P}}{1+P}
  = 3H(t) - \frac{C(w)\,\dot{H}(t)}{1 + B(w) + C(w)\,H(t)}\,,
\end{align}
where in the rightmost expression we write coordinate dependences
explicitly. Since $H(t) \rightarrow 0^+, \dot{H} \rightarrow 0^-$ (EdS)
and $H(t) \rightarrow H_0 \sqrt{\OmLamzero}^+, \dot{H} \rightarrow 0^-$ ({\LCDM}),
we have for both the EdS and {\LCDM} reference models,
\begin{equation}
  \lim_{t \rightarrow \infty} \Theta = 3H \,.
  \label{e-P-Theta-3H}
\end{equation}
In other words, if $\Theta > 0$ (and $H > 0$) initially, then
\postrefereeAAchanges{overdensities and underdensities
will be unable to grow in amplitude, and} the initial
conditions force the universe to approach
\postrefereeAAchanges{a homogeneous expansion rate} in the limit as
$t \rightarrow \infty$.
\postrefereeAAchanges{Thus, there is no growing mode in this exact relativistic
  cosmological solution of the Einstein equation.}
This contrasts to the Newtonian case, which, as discussed by
\cite[][\SSS{}V.A.]{BuchRZA2}, has flat spatial sections but allows
gravity to form overdense and underdense structures.

\postrefereeAAchanges{Proposition}~{\ELemmaFlatNoTurnaround} refers to isotropic collapse.
So $\Theta \rightarrow 3H$ on its own could allow, for example,
$\Theta^w_w \rightarrow 0, \Theta^x_x = \Theta^y_y \rightarrow 3H/2$.
This is not the case under consideration here, however.
Equation~\eqref{e-P-theta-ij} shows that the only non-zero
component of peculiar expansion is $\theta^w_w = \tfrac{\dot{P}}{1+P}$,
so expansion in the $x$ and $y$ directions cannot compensate collapse
in the $w$ direction. The more general plane-symmetric case is interesting,
but not considered in this work.

\cite[][\SSS{}V.A, second last paragraph]{BuchRZA2}
suggested that by considering spatial-domain--averaged behaviour
(\cite{BuchRZA2}, Eqs~(10), (11); written here
as Eq.~\eqref{e-Buch00dust-4a-averaged})
and using the relativistic Zel'dovich approximation
\citep[RZA;][]{BKS00,BuchRZA1,BuchRZA2},
anisotropic pancake-like collapse of the plane-symmetric subcase
could be possible for a domain that
is initially expanding more slowly than the reference model
(has a negative initial extrinsic curvature invariant
$\initaverageriem{\invI}$).
The expansion variance term $\average{\left(\Theta - \average{\Theta}\right)^2}$
does not have a pointwise equivalent, and the behaviour of the spatially averaged
parameters cannot be trivially related to the behaviour of the pointwise parameters.

\begin{figure}
  \begin{center}
    \adjustbox{width=0.55\textwidth}{\begingroup
  \makeatletter
  \providecommand\color[2][]{
    \GenericError{(gnuplot) \space\space\space\@spaces}{
      Package color not loaded in conjunction with
      terminal option `colourtext'
    }{See the gnuplot documentation for explanation.
    }{Either use 'blacktext' in gnuplot or load the package
      color.sty in LaTeX.}
    \renewcommand\color[2][]{}
  }
  \providecommand\includegraphics[2][]{
    \GenericError{(gnuplot) \space\space\space\@spaces}{
      Package graphicx or graphics not loaded
    }{See the gnuplot documentation for explanation.
    }{The gnuplot epslatex terminal needs graphicx.sty or graphics.sty.}
    \renewcommand\includegraphics[2][]{}
  }
  \providecommand\rotatebox[2]{#2}
  \@ifundefined{ifGPcolor}{
    \newif\ifGPcolor
    \GPcolorfalse
  }{}
  \@ifundefined{ifGPblacktext}{
    \newif\ifGPblacktext
    \GPblacktexttrue
  }{}
  \let\gplgaddtomacro\g@addto@macro
  \gdef\gplbacktext{}
  \gdef\gplfronttext{}
  \makeatother
  \ifGPblacktext
    \def\colorrgb#1{}
    \def\colorgray#1{}
  \else
    \ifGPcolor
      \def\colorrgb#1{\color[rgb]{#1}}
      \def\colorgray#1{\color[gray]{#1}}
      \expandafter\def\csname LTw\endcsname{\color{white}}
      \expandafter\def\csname LTb\endcsname{\color{black}}
      \expandafter\def\csname LTa\endcsname{\color{black}}
      \expandafter\def\csname LT0\endcsname{\color[rgb]{1,0,0}}
      \expandafter\def\csname LT1\endcsname{\color[rgb]{0,1,0}}
      \expandafter\def\csname LT2\endcsname{\color[rgb]{0,0,1}}
      \expandafter\def\csname LT3\endcsname{\color[rgb]{1,0,1}}
      \expandafter\def\csname LT4\endcsname{\color[rgb]{0,1,1}}
      \expandafter\def\csname LT5\endcsname{\color[rgb]{1,1,0}}
      \expandafter\def\csname LT6\endcsname{\color[rgb]{0,0,0}}
      \expandafter\def\csname LT7\endcsname{\color[rgb]{1,0.3,0}}
      \expandafter\def\csname LT8\endcsname{\color[rgb]{0.5,0.5,0.5}}
    \else
      \def\colorrgb#1{\color{black}}
      \def\colorgray#1{\color[gray]{#1}}
      \expandafter\def\csname LTw\endcsname{\color{white}}
      \expandafter\def\csname LTb\endcsname{\color{black}}
      \expandafter\def\csname LTa\endcsname{\color{black}}
      \expandafter\def\csname LT0\endcsname{\color{black}}
      \expandafter\def\csname LT1\endcsname{\color{black}}
      \expandafter\def\csname LT2\endcsname{\color{black}}
      \expandafter\def\csname LT3\endcsname{\color{black}}
      \expandafter\def\csname LT4\endcsname{\color{black}}
      \expandafter\def\csname LT5\endcsname{\color{black}}
      \expandafter\def\csname LT6\endcsname{\color{black}}
      \expandafter\def\csname LT7\endcsname{\color{black}}
      \expandafter\def\csname LT8\endcsname{\color{black}}
    \fi
  \fi
    \setlength{\unitlength}{0.0200bp}
    \ifx\gptboxheight\undefined
      \newlength{\gptboxheight}
      \newlength{\gptboxwidth}
      \newsavebox{\gptboxtext}
    \fi
    \setlength{\fboxrule}{0.5pt}
    \setlength{\fboxsep}{1pt}
\begin{picture}(11520.00,8640.00)
    \gplgaddtomacro\gplbacktext{
      \colorrgb{0.00,0.00,0.00}
      \put(1480,1280){\makebox(0,0)[r]{\strut{}-10}}
      \colorrgb{0.00,0.00,0.00}
      \put(1480,3000){\makebox(0,0)[r]{\strut{}-5}}
      \colorrgb{0.00,0.00,0.00}
      \put(1480,4720){\makebox(0,0)[r]{\strut{}0}}
      \colorrgb{0.00,0.00,0.00}
      \put(1480,6439){\makebox(0,0)[r]{\strut{}5}}
      \colorrgb{0.00,0.00,0.00}
      \put(1480,8159){\makebox(0,0)[r]{\strut{}10}}
      \colorrgb{0.00,0.00,0.00}
      \put(1720,880){\makebox(0,0){\strut{}$10^{0}$}}
      \colorrgb{0.00,0.00,0.00}
      \put(4746,880){\makebox(0,0){\strut{}$10^{1}$}}
      \colorrgb{0.00,0.00,0.00}
      \put(7773,880){\makebox(0,0){\strut{}$10^{2}$}}
      \colorrgb{0.00,0.00,0.00}
      \put(10799,880){\makebox(0,0){\strut{}$10^{3}$}}
    }
    \gplgaddtomacro\gplfronttext{
      \colorrgb{0.00,0.00,0.00}
      \put(320,4719){\rotatebox{90}{\makebox(0,0){\strut{}$\OmRD$}}}
      \colorrgb{0.00,0.00,0.00}
      \put(6259,280){\makebox(0,0){\strut{}$H_{\CD}$ (km/s/Mpc)}}
      \colorrgb{0.00,0.00,0.00}
      \put(4135,7793){\makebox(0,0){\footnotesize $ 0.9$~Gyr}}
      \colorrgb{0.00,0.00,0.00}
      \put(4135,7188){\makebox(0,0){\footnotesize $ 4.0$~Gyr}}
      \colorrgb{0.00,0.00,0.00}
      \put(4135,6583){\makebox(0,0){\footnotesize $ 7.0$~Gyr}}
      \colorrgb{0.00,0.00,0.00}
      \put(4135,5978){\makebox(0,0){\footnotesize $10.0$~Gyr}}
      \colorrgb{0.00,0.00,0.00}
      \put(4135,5373){\makebox(0,0){\footnotesize $12.9$~Gyr}}
      \colorrgb{0.00,0.00,0.00}
      \put(2174,1624){\makebox(0,0)[l]{\strut{}$\LD =   2.5$~Mpc/$\hzeroeff$}}
    }
    \gplbacktext
    \put(0,0){\includegraphics[scale=0.4]{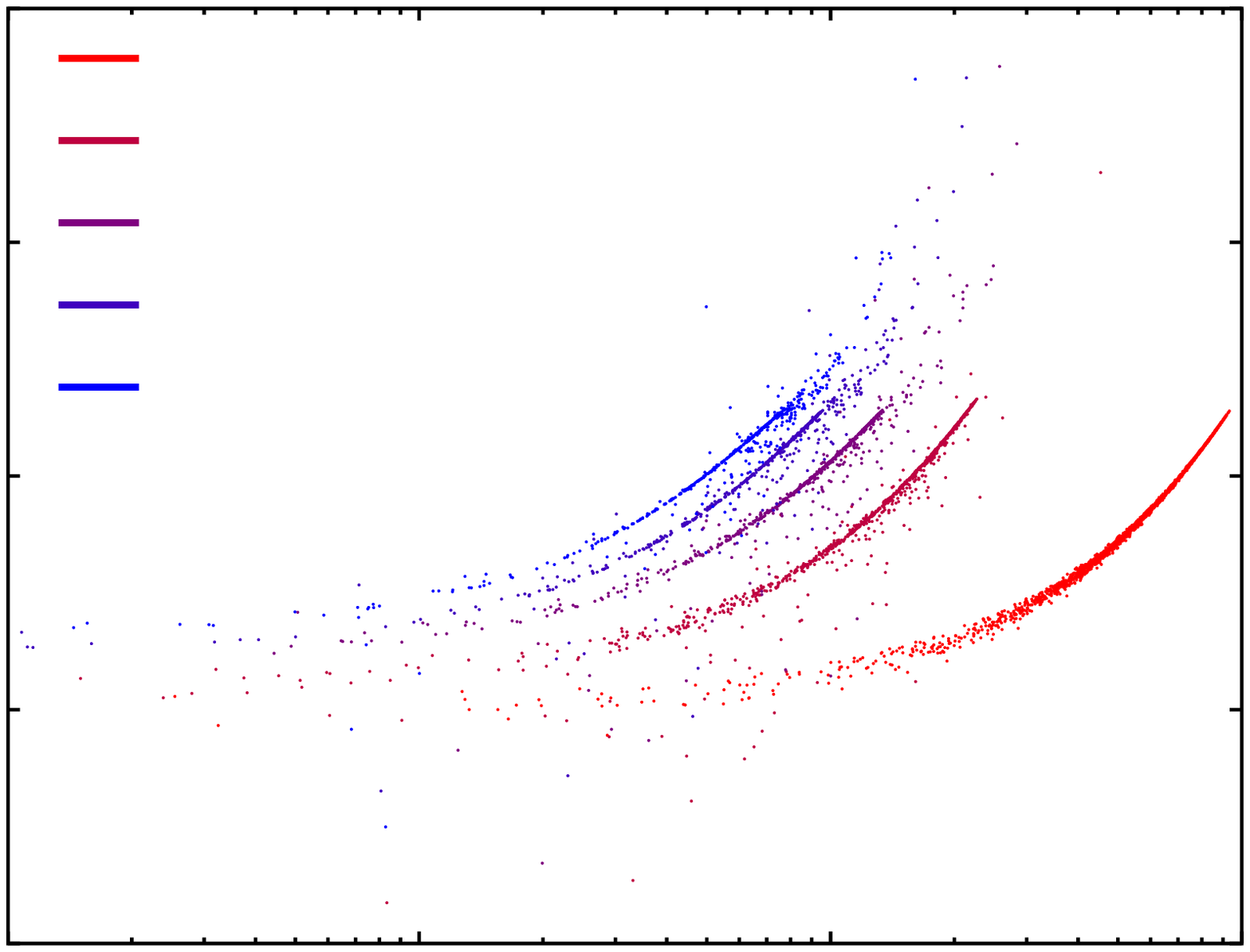}}
    \gplfronttext
  \end{picture}
\endgroup
 }\\
    \adjustbox{width=0.55\textwidth}{\begingroup
  \makeatletter
  \providecommand\color[2][]{
    \GenericError{(gnuplot) \space\space\space\@spaces}{
      Package color not loaded in conjunction with
      terminal option `colourtext'
    }{See the gnuplot documentation for explanation.
    }{Either use 'blacktext' in gnuplot or load the package
      color.sty in LaTeX.}
    \renewcommand\color[2][]{}
  }
  \providecommand\includegraphics[2][]{
    \GenericError{(gnuplot) \space\space\space\@spaces}{
      Package graphicx or graphics not loaded
    }{See the gnuplot documentation for explanation.
    }{The gnuplot epslatex terminal needs graphicx.sty or graphics.sty.}
    \renewcommand\includegraphics[2][]{}
  }
  \providecommand\rotatebox[2]{#2}
  \@ifundefined{ifGPcolor}{
    \newif\ifGPcolor
    \GPcolorfalse
  }{}
  \@ifundefined{ifGPblacktext}{
    \newif\ifGPblacktext
    \GPblacktexttrue
  }{}
  \let\gplgaddtomacro\g@addto@macro
  \gdef\gplbacktext{}
  \gdef\gplfronttext{}
  \makeatother
  \ifGPblacktext
    \def\colorrgb#1{}
    \def\colorgray#1{}
  \else
    \ifGPcolor
      \def\colorrgb#1{\color[rgb]{#1}}
      \def\colorgray#1{\color[gray]{#1}}
      \expandafter\def\csname LTw\endcsname{\color{white}}
      \expandafter\def\csname LTb\endcsname{\color{black}}
      \expandafter\def\csname LTa\endcsname{\color{black}}
      \expandafter\def\csname LT0\endcsname{\color[rgb]{1,0,0}}
      \expandafter\def\csname LT1\endcsname{\color[rgb]{0,1,0}}
      \expandafter\def\csname LT2\endcsname{\color[rgb]{0,0,1}}
      \expandafter\def\csname LT3\endcsname{\color[rgb]{1,0,1}}
      \expandafter\def\csname LT4\endcsname{\color[rgb]{0,1,1}}
      \expandafter\def\csname LT5\endcsname{\color[rgb]{1,1,0}}
      \expandafter\def\csname LT6\endcsname{\color[rgb]{0,0,0}}
      \expandafter\def\csname LT7\endcsname{\color[rgb]{1,0.3,0}}
      \expandafter\def\csname LT8\endcsname{\color[rgb]{0.5,0.5,0.5}}
    \else
      \def\colorrgb#1{\color{black}}
      \def\colorgray#1{\color[gray]{#1}}
      \expandafter\def\csname LTw\endcsname{\color{white}}
      \expandafter\def\csname LTb\endcsname{\color{black}}
      \expandafter\def\csname LTa\endcsname{\color{black}}
      \expandafter\def\csname LT0\endcsname{\color{black}}
      \expandafter\def\csname LT1\endcsname{\color{black}}
      \expandafter\def\csname LT2\endcsname{\color{black}}
      \expandafter\def\csname LT3\endcsname{\color{black}}
      \expandafter\def\csname LT4\endcsname{\color{black}}
      \expandafter\def\csname LT5\endcsname{\color{black}}
      \expandafter\def\csname LT6\endcsname{\color{black}}
      \expandafter\def\csname LT7\endcsname{\color{black}}
      \expandafter\def\csname LT8\endcsname{\color{black}}
    \fi
  \fi
    \setlength{\unitlength}{0.0200bp}
    \ifx\gptboxheight\undefined
      \newlength{\gptboxheight}
      \newlength{\gptboxwidth}
      \newsavebox{\gptboxtext}
    \fi
    \setlength{\fboxrule}{0.5pt}
    \setlength{\fboxsep}{1pt}
\begin{picture}(11520.00,8640.00)
    \gplgaddtomacro\gplbacktext{
      \colorrgb{0.00,0.00,0.00}
      \put(1480,1280){\makebox(0,0)[r]{\strut{}-10}}
      \colorrgb{0.00,0.00,0.00}
      \put(1480,3000){\makebox(0,0)[r]{\strut{}-5}}
      \colorrgb{0.00,0.00,0.00}
      \put(1480,4720){\makebox(0,0)[r]{\strut{}0}}
      \colorrgb{0.00,0.00,0.00}
      \put(1480,6439){\makebox(0,0)[r]{\strut{}5}}
      \colorrgb{0.00,0.00,0.00}
      \put(1480,8159){\makebox(0,0)[r]{\strut{}10}}
      \colorrgb{0.00,0.00,0.00}
      \put(1720,880){\makebox(0,0){\strut{}$10^{0}$}}
      \colorrgb{0.00,0.00,0.00}
      \put(4746,880){\makebox(0,0){\strut{}$10^{1}$}}
      \colorrgb{0.00,0.00,0.00}
      \put(7773,880){\makebox(0,0){\strut{}$10^{2}$}}
      \colorrgb{0.00,0.00,0.00}
      \put(10799,880){\makebox(0,0){\strut{}$10^{3}$}}
    }
    \gplgaddtomacro\gplfronttext{
      \colorrgb{0.00,0.00,0.00}
      \put(320,4719){\rotatebox{90}{\makebox(0,0){\strut{}$\OmRD$}}}
      \colorrgb{0.00,0.00,0.00}
      \put(6259,280){\makebox(0,0){\strut{}$H_{\CD}$ (km/s/Mpc)}}
      \colorrgb{0.00,0.00,0.00}
      \put(4135,7793){\makebox(0,0){\footnotesize $ 1.0$~Gyr}}
      \colorrgb{0.00,0.00,0.00}
      \put(4135,7188){\makebox(0,0){\footnotesize $ 4.1$~Gyr}}
      \colorrgb{0.00,0.00,0.00}
      \put(4135,6583){\makebox(0,0){\footnotesize $ 7.0$~Gyr}}
      \colorrgb{0.00,0.00,0.00}
      \put(4135,5978){\makebox(0,0){\footnotesize $10.0$~Gyr}}
      \colorrgb{0.00,0.00,0.00}
      \put(4135,5373){\makebox(0,0){\footnotesize $12.9$~Gyr}}
      \colorrgb{0.00,0.00,0.00}
      \put(2174,1624){\makebox(0,0)[l]{\strut{}$\LD =  10.0$~Mpc/$\hzeroeff$}}
    }
    \gplbacktext
    \put(0,0){\includegraphics[scale=0.4]{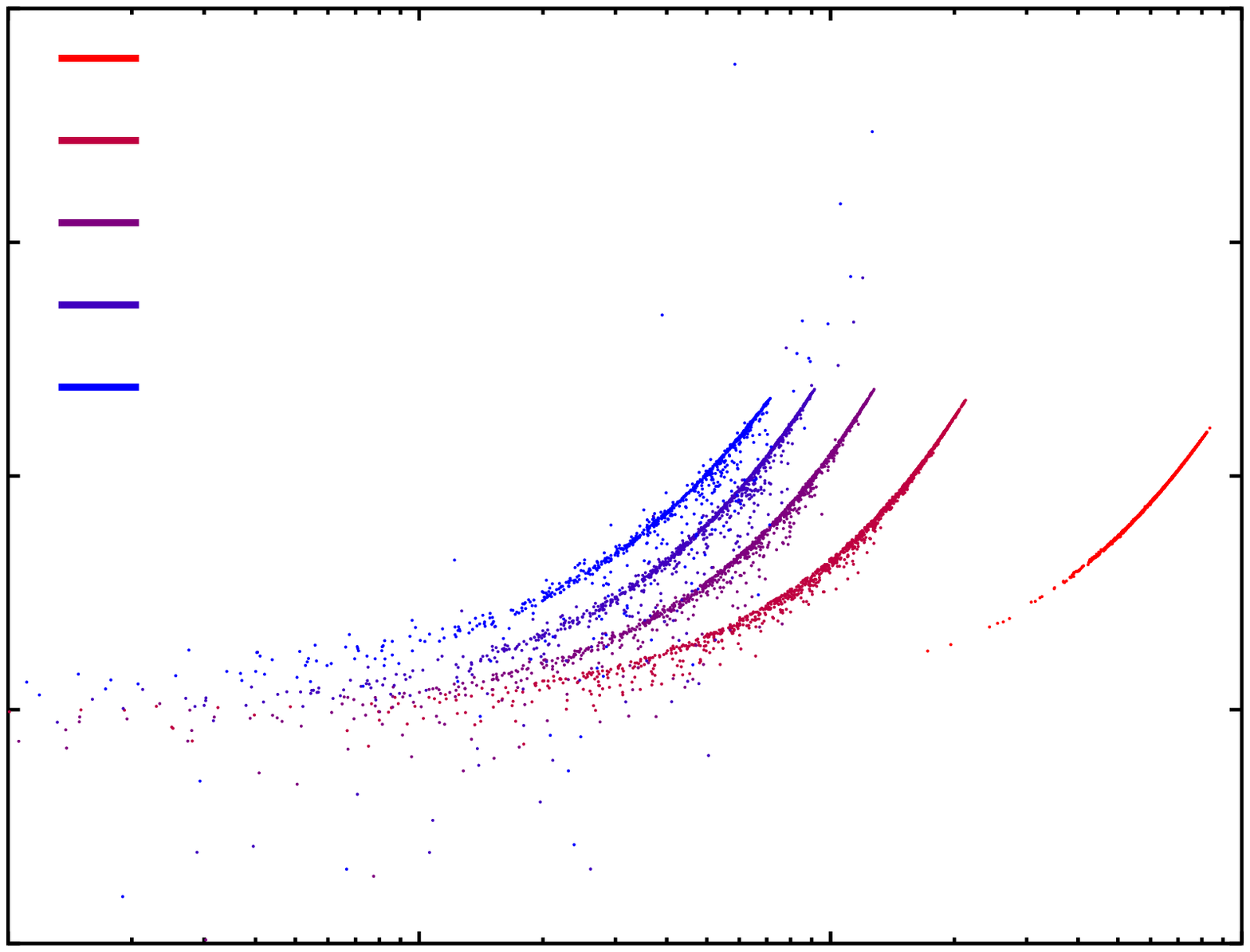}}
    \gplfronttext
  \end{picture}
\endgroup
 }\\
    \adjustbox{width=0.55\textwidth}{\begingroup
  \makeatletter
  \providecommand\color[2][]{
    \GenericError{(gnuplot) \space\space\space\@spaces}{
      Package color not loaded in conjunction with
      terminal option `colourtext'
    }{See the gnuplot documentation for explanation.
    }{Either use 'blacktext' in gnuplot or load the package
      color.sty in LaTeX.}
    \renewcommand\color[2][]{}
  }
  \providecommand\includegraphics[2][]{
    \GenericError{(gnuplot) \space\space\space\@spaces}{
      Package graphicx or graphics not loaded
    }{See the gnuplot documentation for explanation.
    }{The gnuplot epslatex terminal needs graphicx.sty or graphics.sty.}
    \renewcommand\includegraphics[2][]{}
  }
  \providecommand\rotatebox[2]{#2}
  \@ifundefined{ifGPcolor}{
    \newif\ifGPcolor
    \GPcolorfalse
  }{}
  \@ifundefined{ifGPblacktext}{
    \newif\ifGPblacktext
    \GPblacktexttrue
  }{}
  \let\gplgaddtomacro\g@addto@macro
  \gdef\gplbacktext{}
  \gdef\gplfronttext{}
  \makeatother
  \ifGPblacktext
    \def\colorrgb#1{}
    \def\colorgray#1{}
  \else
    \ifGPcolor
      \def\colorrgb#1{\color[rgb]{#1}}
      \def\colorgray#1{\color[gray]{#1}}
      \expandafter\def\csname LTw\endcsname{\color{white}}
      \expandafter\def\csname LTb\endcsname{\color{black}}
      \expandafter\def\csname LTa\endcsname{\color{black}}
      \expandafter\def\csname LT0\endcsname{\color[rgb]{1,0,0}}
      \expandafter\def\csname LT1\endcsname{\color[rgb]{0,1,0}}
      \expandafter\def\csname LT2\endcsname{\color[rgb]{0,0,1}}
      \expandafter\def\csname LT3\endcsname{\color[rgb]{1,0,1}}
      \expandafter\def\csname LT4\endcsname{\color[rgb]{0,1,1}}
      \expandafter\def\csname LT5\endcsname{\color[rgb]{1,1,0}}
      \expandafter\def\csname LT6\endcsname{\color[rgb]{0,0,0}}
      \expandafter\def\csname LT7\endcsname{\color[rgb]{1,0.3,0}}
      \expandafter\def\csname LT8\endcsname{\color[rgb]{0.5,0.5,0.5}}
    \else
      \def\colorrgb#1{\color{black}}
      \def\colorgray#1{\color[gray]{#1}}
      \expandafter\def\csname LTw\endcsname{\color{white}}
      \expandafter\def\csname LTb\endcsname{\color{black}}
      \expandafter\def\csname LTa\endcsname{\color{black}}
      \expandafter\def\csname LT0\endcsname{\color{black}}
      \expandafter\def\csname LT1\endcsname{\color{black}}
      \expandafter\def\csname LT2\endcsname{\color{black}}
      \expandafter\def\csname LT3\endcsname{\color{black}}
      \expandafter\def\csname LT4\endcsname{\color{black}}
      \expandafter\def\csname LT5\endcsname{\color{black}}
      \expandafter\def\csname LT6\endcsname{\color{black}}
      \expandafter\def\csname LT7\endcsname{\color{black}}
      \expandafter\def\csname LT8\endcsname{\color{black}}
    \fi
  \fi
    \setlength{\unitlength}{0.0200bp}
    \ifx\gptboxheight\undefined
      \newlength{\gptboxheight}
      \newlength{\gptboxwidth}
      \newsavebox{\gptboxtext}
    \fi
    \setlength{\fboxrule}{0.5pt}
    \setlength{\fboxsep}{1pt}
\begin{picture}(11520.00,8640.00)
    \gplgaddtomacro\gplbacktext{
      \colorrgb{0.00,0.00,0.00}
      \put(1480,1280){\makebox(0,0)[r]{\strut{}-10}}
      \colorrgb{0.00,0.00,0.00}
      \put(1480,3000){\makebox(0,0)[r]{\strut{}-5}}
      \colorrgb{0.00,0.00,0.00}
      \put(1480,4720){\makebox(0,0)[r]{\strut{}0}}
      \colorrgb{0.00,0.00,0.00}
      \put(1480,6439){\makebox(0,0)[r]{\strut{}5}}
      \colorrgb{0.00,0.00,0.00}
      \put(1480,8159){\makebox(0,0)[r]{\strut{}10}}
      \colorrgb{0.00,0.00,0.00}
      \put(1720,880){\makebox(0,0){\strut{}$10^{0}$}}
      \colorrgb{0.00,0.00,0.00}
      \put(4746,880){\makebox(0,0){\strut{}$10^{1}$}}
      \colorrgb{0.00,0.00,0.00}
      \put(7773,880){\makebox(0,0){\strut{}$10^{2}$}}
      \colorrgb{0.00,0.00,0.00}
      \put(10799,880){\makebox(0,0){\strut{}$10^{3}$}}
    }
    \gplgaddtomacro\gplfronttext{
      \colorrgb{0.00,0.00,0.00}
      \put(320,4719){\rotatebox{90}{\makebox(0,0){\strut{}$\OmRD$}}}
      \colorrgb{0.00,0.00,0.00}
      \put(6259,280){\makebox(0,0){\strut{}$H_{\CD}$ (km/s/Mpc)}}
      \colorrgb{0.00,0.00,0.00}
      \put(4135,7793){\makebox(0,0){\footnotesize $ 1.1$~Gyr}}
      \colorrgb{0.00,0.00,0.00}
      \put(4135,7188){\makebox(0,0){\footnotesize $ 4.0$~Gyr}}
      \colorrgb{0.00,0.00,0.00}
      \put(4135,6583){\makebox(0,0){\footnotesize $ 6.9$~Gyr}}
      \colorrgb{0.00,0.00,0.00}
      \put(4135,5978){\makebox(0,0){\footnotesize $10.0$~Gyr}}
      \colorrgb{0.00,0.00,0.00}
      \put(4135,5373){\makebox(0,0){\footnotesize $13.0$~Gyr}}
      \colorrgb{0.00,0.00,0.00}
      \put(2174,1624){\makebox(0,0)[l]{\strut{}$\LD =  40.0$~Mpc/$\hzeroeff$}}
    }
    \gplbacktext
    \put(0,0){\includegraphics[scale=0.4]{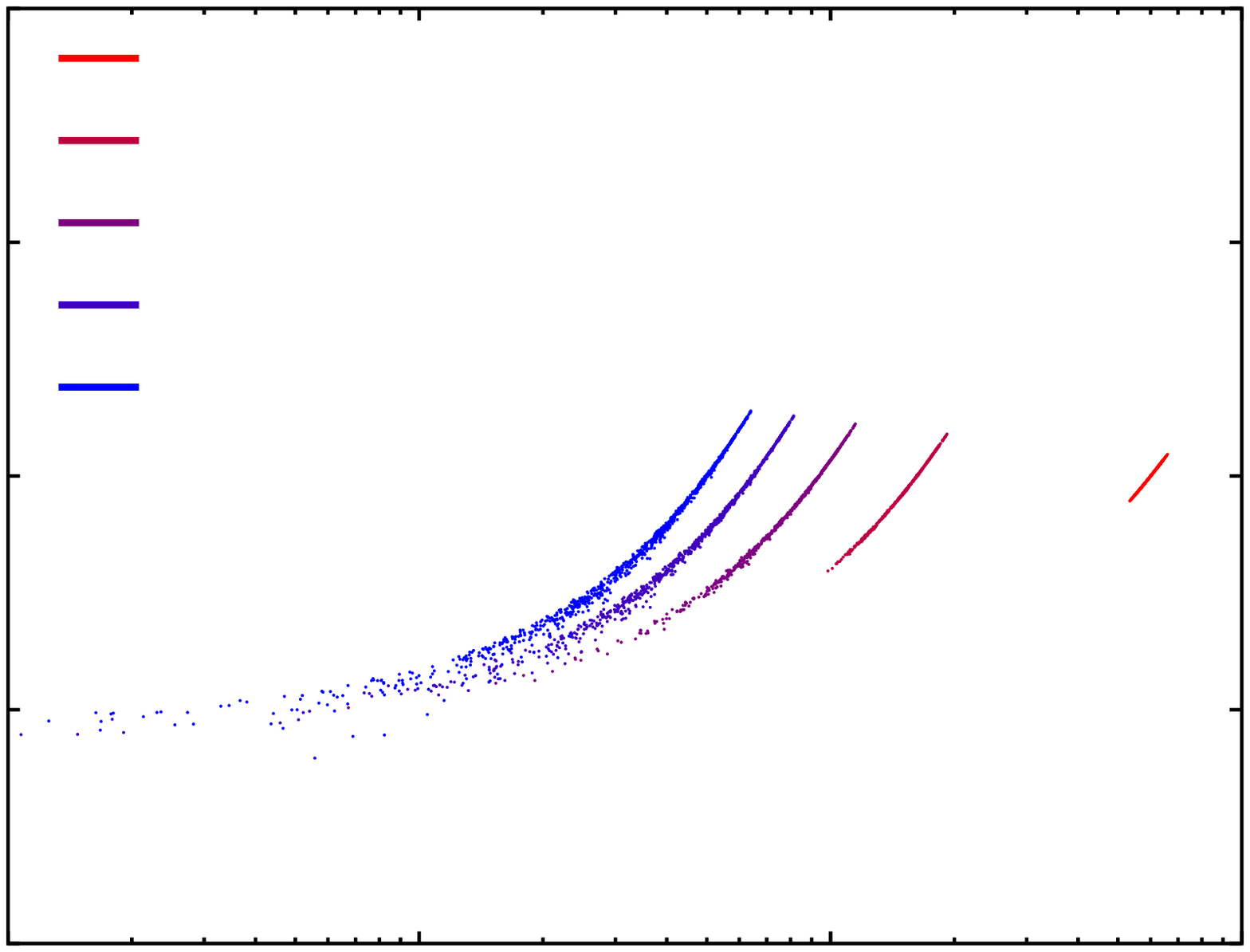}}
    \gplfronttext
  \end{picture}
\endgroup
 }
  \end{center}
  \caption{Domain-averaged scalar curvature functional $\OmRD$ versus
    expansion rate $H_{\CD}$,
    from QZA simulations ($N$-body initial conditions and RZA
    kinematical backreaction $\CQ_{\CD}$ evolution) for
    a CMB-normalised almost-EdS model.
    {\em {\Toptobottom} panels}: averaging scales
    $\LD = 2.5, 10, 40$~Mpc/$\hzeroeff$, respectively, where
    \protect\postrefereeBBchanges{$L_{\mathrm{box}}  = 16 \LD = 64 L_{\mathrm{DTFE}} = 256 L_N$} 
    and
    $N=256^3$ particles.
    Colour hues indicate universe ages from red
    (earliest) to blue (most recent).
    Only pre-turnaround domains are shown. The domains closest to turnaround
    (lowest $H_{\CD}$) show the most negative $\OmRD$ values
    (most positive curvatures).
    The axis ranges are identical in all panels here and in Fig.~\protect\ref{f-OmRDLCDM}.
    \label{f-OmRDEdS}}
\end{figure} 
 
\begin{figure}
  \begin{center}
    \adjustbox{width=0.55\textwidth}{\begingroup
  \makeatletter
  \providecommand\color[2][]{
    \GenericError{(gnuplot) \space\space\space\@spaces}{
      Package color not loaded in conjunction with
      terminal option `colourtext'
    }{See the gnuplot documentation for explanation.
    }{Either use 'blacktext' in gnuplot or load the package
      color.sty in LaTeX.}
    \renewcommand\color[2][]{}
  }
  \providecommand\includegraphics[2][]{
    \GenericError{(gnuplot) \space\space\space\@spaces}{
      Package graphicx or graphics not loaded
    }{See the gnuplot documentation for explanation.
    }{The gnuplot epslatex terminal needs graphicx.sty or graphics.sty.}
    \renewcommand\includegraphics[2][]{}
  }
  \providecommand\rotatebox[2]{#2}
  \@ifundefined{ifGPcolor}{
    \newif\ifGPcolor
    \GPcolorfalse
  }{}
  \@ifundefined{ifGPblacktext}{
    \newif\ifGPblacktext
    \GPblacktexttrue
  }{}
  \let\gplgaddtomacro\g@addto@macro
  \gdef\gplbacktext{}
  \gdef\gplfronttext{}
  \makeatother
  \ifGPblacktext
    \def\colorrgb#1{}
    \def\colorgray#1{}
  \else
    \ifGPcolor
      \def\colorrgb#1{\color[rgb]{#1}}
      \def\colorgray#1{\color[gray]{#1}}
      \expandafter\def\csname LTw\endcsname{\color{white}}
      \expandafter\def\csname LTb\endcsname{\color{black}}
      \expandafter\def\csname LTa\endcsname{\color{black}}
      \expandafter\def\csname LT0\endcsname{\color[rgb]{1,0,0}}
      \expandafter\def\csname LT1\endcsname{\color[rgb]{0,1,0}}
      \expandafter\def\csname LT2\endcsname{\color[rgb]{0,0,1}}
      \expandafter\def\csname LT3\endcsname{\color[rgb]{1,0,1}}
      \expandafter\def\csname LT4\endcsname{\color[rgb]{0,1,1}}
      \expandafter\def\csname LT5\endcsname{\color[rgb]{1,1,0}}
      \expandafter\def\csname LT6\endcsname{\color[rgb]{0,0,0}}
      \expandafter\def\csname LT7\endcsname{\color[rgb]{1,0.3,0}}
      \expandafter\def\csname LT8\endcsname{\color[rgb]{0.5,0.5,0.5}}
    \else
      \def\colorrgb#1{\color{black}}
      \def\colorgray#1{\color[gray]{#1}}
      \expandafter\def\csname LTw\endcsname{\color{white}}
      \expandafter\def\csname LTb\endcsname{\color{black}}
      \expandafter\def\csname LTa\endcsname{\color{black}}
      \expandafter\def\csname LT0\endcsname{\color{black}}
      \expandafter\def\csname LT1\endcsname{\color{black}}
      \expandafter\def\csname LT2\endcsname{\color{black}}
      \expandafter\def\csname LT3\endcsname{\color{black}}
      \expandafter\def\csname LT4\endcsname{\color{black}}
      \expandafter\def\csname LT5\endcsname{\color{black}}
      \expandafter\def\csname LT6\endcsname{\color{black}}
      \expandafter\def\csname LT7\endcsname{\color{black}}
      \expandafter\def\csname LT8\endcsname{\color{black}}
    \fi
  \fi
    \setlength{\unitlength}{0.0200bp}
    \ifx\gptboxheight\undefined
      \newlength{\gptboxheight}
      \newlength{\gptboxwidth}
      \newsavebox{\gptboxtext}
    \fi
    \setlength{\fboxrule}{0.5pt}
    \setlength{\fboxsep}{1pt}
\begin{picture}(11520.00,8640.00)
    \gplgaddtomacro\gplbacktext{
      \colorrgb{0.00,0.00,0.00}
      \put(1480,1280){\makebox(0,0)[r]{\strut{}-10}}
      \colorrgb{0.00,0.00,0.00}
      \put(1480,3000){\makebox(0,0)[r]{\strut{}-5}}
      \colorrgb{0.00,0.00,0.00}
      \put(1480,4720){\makebox(0,0)[r]{\strut{}0}}
      \colorrgb{0.00,0.00,0.00}
      \put(1480,6439){\makebox(0,0)[r]{\strut{}5}}
      \colorrgb{0.00,0.00,0.00}
      \put(1480,8159){\makebox(0,0)[r]{\strut{}10}}
      \colorrgb{0.00,0.00,0.00}
      \put(1720,880){\makebox(0,0){\strut{}$10^{0}$}}
      \colorrgb{0.00,0.00,0.00}
      \put(4746,880){\makebox(0,0){\strut{}$10^{1}$}}
      \colorrgb{0.00,0.00,0.00}
      \put(7773,880){\makebox(0,0){\strut{}$10^{2}$}}
      \colorrgb{0.00,0.00,0.00}
      \put(10799,880){\makebox(0,0){\strut{}$10^{3}$}}
    }
    \gplgaddtomacro\gplfronttext{
      \colorrgb{0.00,0.00,0.00}
      \put(320,4719){\rotatebox{90}{\makebox(0,0){\strut{}$\OmRD$}}}
      \colorrgb{0.00,0.00,0.00}
      \put(6259,280){\makebox(0,0){\strut{}$H_{\CD}$ (km/s/Mpc)}}
      \colorrgb{0.00,0.00,0.00}
      \put(4135,7793){\makebox(0,0){\footnotesize $ 1.1$~Gyr}}
      \colorrgb{0.00,0.00,0.00}
      \put(4135,7188){\makebox(0,0){\footnotesize $ 4.0$~Gyr}}
      \colorrgb{0.00,0.00,0.00}
      \put(4135,6583){\makebox(0,0){\footnotesize $ 7.0$~Gyr}}
      \colorrgb{0.00,0.00,0.00}
      \put(4135,5978){\makebox(0,0){\footnotesize $10.1$~Gyr}}
      \colorrgb{0.00,0.00,0.00}
      \put(4135,5373){\makebox(0,0){\footnotesize $13.0$~Gyr}}
      \colorrgb{0.00,0.00,0.00}
      \put(2174,1624){\makebox(0,0)[l]{\strut{}$\LD =   2.5$~Mpc/$\hzeroeff$}}
    }
    \gplbacktext
    \put(0,0){\includegraphics[scale=0.4]{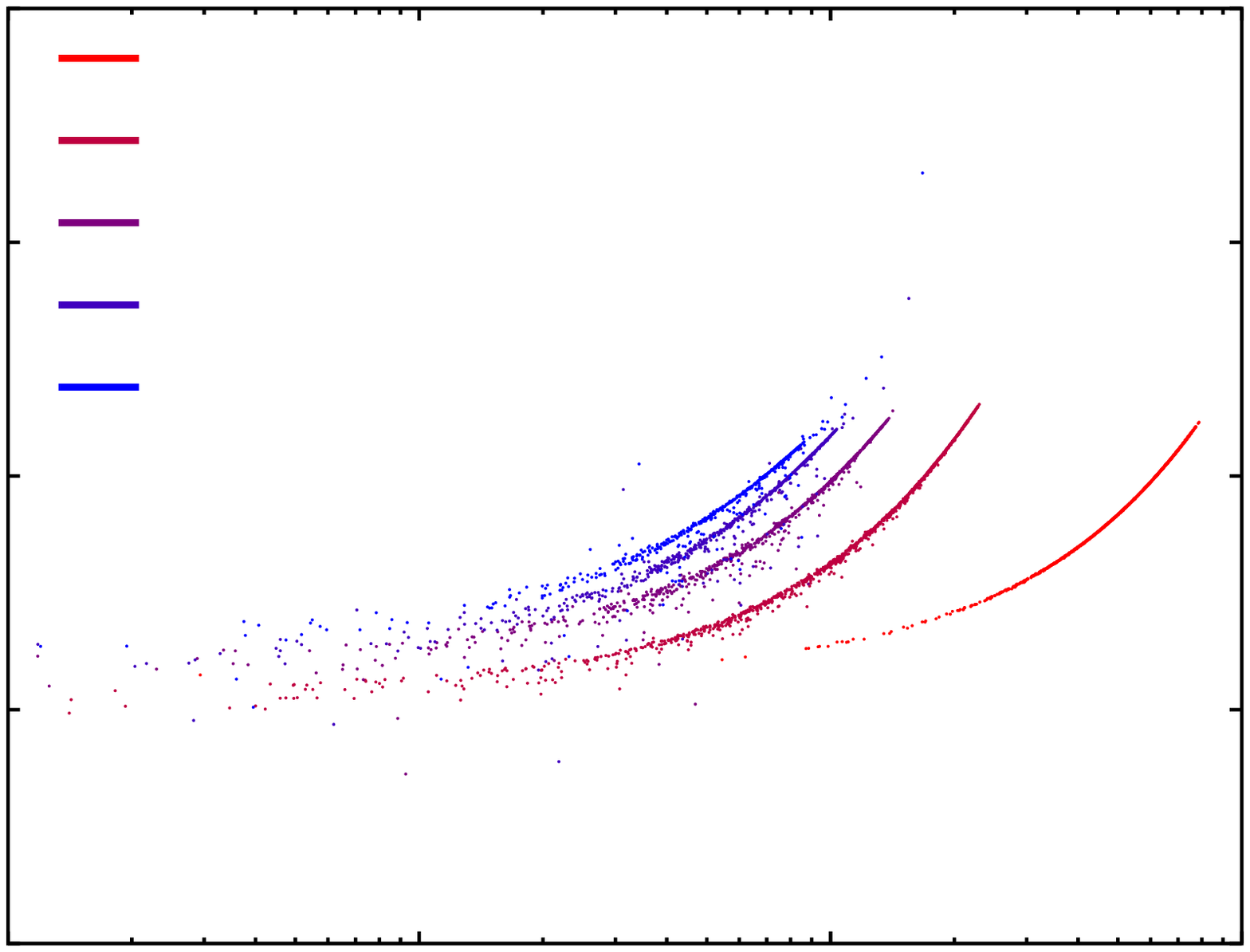}}
    \gplfronttext
  \end{picture}
\endgroup
 }\\
    \adjustbox{width=0.55\textwidth}{\begingroup
  \makeatletter
  \providecommand\color[2][]{
    \GenericError{(gnuplot) \space\space\space\@spaces}{
      Package color not loaded in conjunction with
      terminal option `colourtext'
    }{See the gnuplot documentation for explanation.
    }{Either use 'blacktext' in gnuplot or load the package
      color.sty in LaTeX.}
    \renewcommand\color[2][]{}
  }
  \providecommand\includegraphics[2][]{
    \GenericError{(gnuplot) \space\space\space\@spaces}{
      Package graphicx or graphics not loaded
    }{See the gnuplot documentation for explanation.
    }{The gnuplot epslatex terminal needs graphicx.sty or graphics.sty.}
    \renewcommand\includegraphics[2][]{}
  }
  \providecommand\rotatebox[2]{#2}
  \@ifundefined{ifGPcolor}{
    \newif\ifGPcolor
    \GPcolorfalse
  }{}
  \@ifundefined{ifGPblacktext}{
    \newif\ifGPblacktext
    \GPblacktexttrue
  }{}
  \let\gplgaddtomacro\g@addto@macro
  \gdef\gplbacktext{}
  \gdef\gplfronttext{}
  \makeatother
  \ifGPblacktext
    \def\colorrgb#1{}
    \def\colorgray#1{}
  \else
    \ifGPcolor
      \def\colorrgb#1{\color[rgb]{#1}}
      \def\colorgray#1{\color[gray]{#1}}
      \expandafter\def\csname LTw\endcsname{\color{white}}
      \expandafter\def\csname LTb\endcsname{\color{black}}
      \expandafter\def\csname LTa\endcsname{\color{black}}
      \expandafter\def\csname LT0\endcsname{\color[rgb]{1,0,0}}
      \expandafter\def\csname LT1\endcsname{\color[rgb]{0,1,0}}
      \expandafter\def\csname LT2\endcsname{\color[rgb]{0,0,1}}
      \expandafter\def\csname LT3\endcsname{\color[rgb]{1,0,1}}
      \expandafter\def\csname LT4\endcsname{\color[rgb]{0,1,1}}
      \expandafter\def\csname LT5\endcsname{\color[rgb]{1,1,0}}
      \expandafter\def\csname LT6\endcsname{\color[rgb]{0,0,0}}
      \expandafter\def\csname LT7\endcsname{\color[rgb]{1,0.3,0}}
      \expandafter\def\csname LT8\endcsname{\color[rgb]{0.5,0.5,0.5}}
    \else
      \def\colorrgb#1{\color{black}}
      \def\colorgray#1{\color[gray]{#1}}
      \expandafter\def\csname LTw\endcsname{\color{white}}
      \expandafter\def\csname LTb\endcsname{\color{black}}
      \expandafter\def\csname LTa\endcsname{\color{black}}
      \expandafter\def\csname LT0\endcsname{\color{black}}
      \expandafter\def\csname LT1\endcsname{\color{black}}
      \expandafter\def\csname LT2\endcsname{\color{black}}
      \expandafter\def\csname LT3\endcsname{\color{black}}
      \expandafter\def\csname LT4\endcsname{\color{black}}
      \expandafter\def\csname LT5\endcsname{\color{black}}
      \expandafter\def\csname LT6\endcsname{\color{black}}
      \expandafter\def\csname LT7\endcsname{\color{black}}
      \expandafter\def\csname LT8\endcsname{\color{black}}
    \fi
  \fi
    \setlength{\unitlength}{0.0200bp}
    \ifx\gptboxheight\undefined
      \newlength{\gptboxheight}
      \newlength{\gptboxwidth}
      \newsavebox{\gptboxtext}
    \fi
    \setlength{\fboxrule}{0.5pt}
    \setlength{\fboxsep}{1pt}
\begin{picture}(11520.00,8640.00)
    \gplgaddtomacro\gplbacktext{
      \colorrgb{0.00,0.00,0.00}
      \put(1480,1280){\makebox(0,0)[r]{\strut{}-10}}
      \colorrgb{0.00,0.00,0.00}
      \put(1480,3000){\makebox(0,0)[r]{\strut{}-5}}
      \colorrgb{0.00,0.00,0.00}
      \put(1480,4720){\makebox(0,0)[r]{\strut{}0}}
      \colorrgb{0.00,0.00,0.00}
      \put(1480,6439){\makebox(0,0)[r]{\strut{}5}}
      \colorrgb{0.00,0.00,0.00}
      \put(1480,8159){\makebox(0,0)[r]{\strut{}10}}
      \colorrgb{0.00,0.00,0.00}
      \put(1720,880){\makebox(0,0){\strut{}$10^{0}$}}
      \colorrgb{0.00,0.00,0.00}
      \put(4746,880){\makebox(0,0){\strut{}$10^{1}$}}
      \colorrgb{0.00,0.00,0.00}
      \put(7773,880){\makebox(0,0){\strut{}$10^{2}$}}
      \colorrgb{0.00,0.00,0.00}
      \put(10799,880){\makebox(0,0){\strut{}$10^{3}$}}
    }
    \gplgaddtomacro\gplfronttext{
      \colorrgb{0.00,0.00,0.00}
      \put(320,4719){\rotatebox{90}{\makebox(0,0){\strut{}$\OmRD$}}}
      \colorrgb{0.00,0.00,0.00}
      \put(6259,280){\makebox(0,0){\strut{}$H_{\CD}$ (km/s/Mpc)}}
      \colorrgb{0.00,0.00,0.00}
      \put(4135,7793){\makebox(0,0){\footnotesize $ 1.0$~Gyr}}
      \colorrgb{0.00,0.00,0.00}
      \put(4135,7188){\makebox(0,0){\footnotesize $ 4.0$~Gyr}}
      \colorrgb{0.00,0.00,0.00}
      \put(4135,6583){\makebox(0,0){\footnotesize $ 7.1$~Gyr}}
      \colorrgb{0.00,0.00,0.00}
      \put(4135,5978){\makebox(0,0){\footnotesize $ 9.9$~Gyr}}
      \colorrgb{0.00,0.00,0.00}
      \put(4135,5373){\makebox(0,0){\footnotesize $13.0$~Gyr}}
      \colorrgb{0.00,0.00,0.00}
      \put(2174,1624){\makebox(0,0)[l]{\strut{}$\LD =  10.0$~Mpc/$\hzeroeff$}}
    }
    \gplbacktext
    \put(0,0){\includegraphics[scale=0.4]{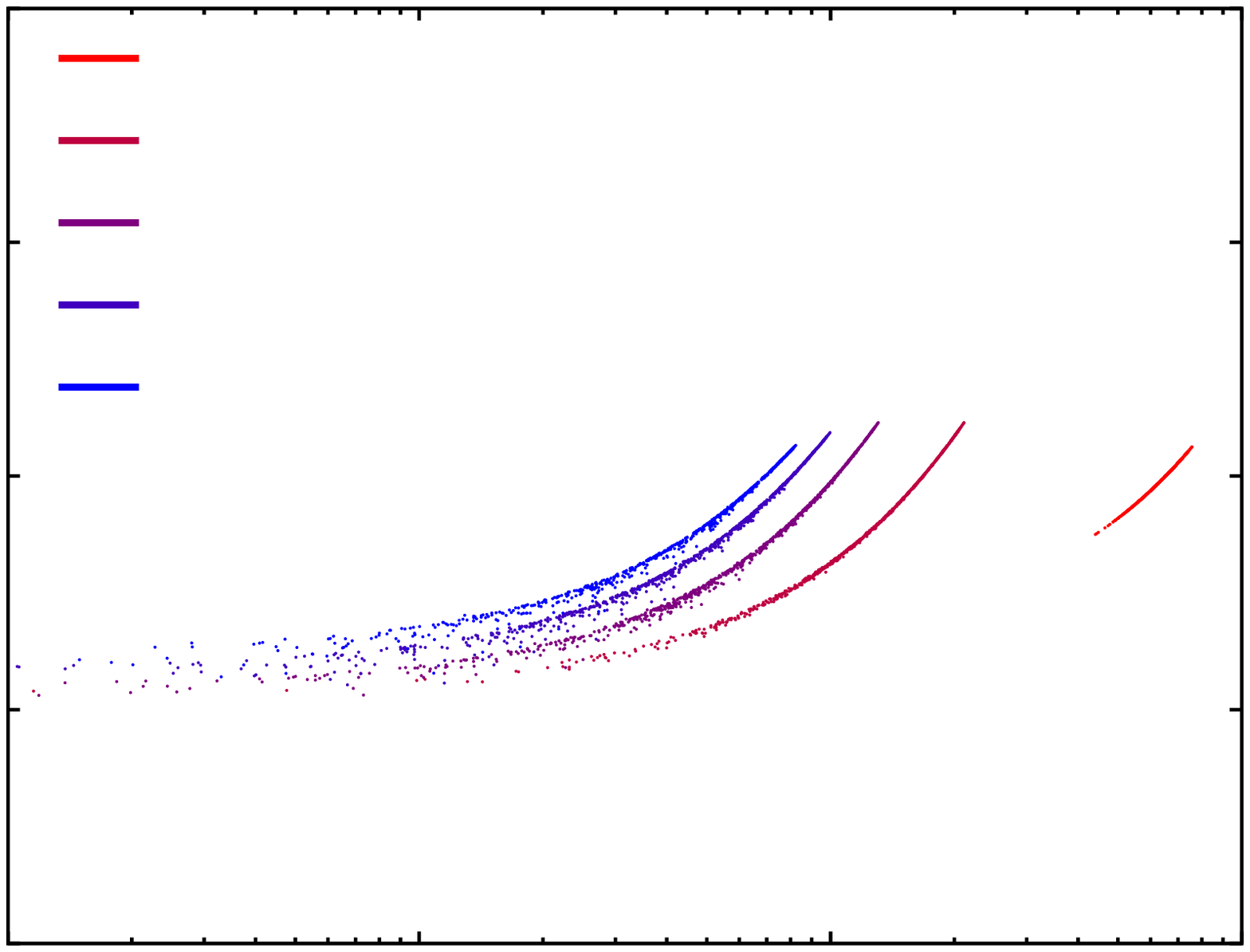}}
    \gplfronttext
  \end{picture}
\endgroup
 }\\
    \adjustbox{width=0.55\textwidth}{\begingroup
  \makeatletter
  \providecommand\color[2][]{
    \GenericError{(gnuplot) \space\space\space\@spaces}{
      Package color not loaded in conjunction with
      terminal option `colourtext'
    }{See the gnuplot documentation for explanation.
    }{Either use 'blacktext' in gnuplot or load the package
      color.sty in LaTeX.}
    \renewcommand\color[2][]{}
  }
  \providecommand\includegraphics[2][]{
    \GenericError{(gnuplot) \space\space\space\@spaces}{
      Package graphicx or graphics not loaded
    }{See the gnuplot documentation for explanation.
    }{The gnuplot epslatex terminal needs graphicx.sty or graphics.sty.}
    \renewcommand\includegraphics[2][]{}
  }
  \providecommand\rotatebox[2]{#2}
  \@ifundefined{ifGPcolor}{
    \newif\ifGPcolor
    \GPcolorfalse
  }{}
  \@ifundefined{ifGPblacktext}{
    \newif\ifGPblacktext
    \GPblacktexttrue
  }{}
  \let\gplgaddtomacro\g@addto@macro
  \gdef\gplbacktext{}
  \gdef\gplfronttext{}
  \makeatother
  \ifGPblacktext
    \def\colorrgb#1{}
    \def\colorgray#1{}
  \else
    \ifGPcolor
      \def\colorrgb#1{\color[rgb]{#1}}
      \def\colorgray#1{\color[gray]{#1}}
      \expandafter\def\csname LTw\endcsname{\color{white}}
      \expandafter\def\csname LTb\endcsname{\color{black}}
      \expandafter\def\csname LTa\endcsname{\color{black}}
      \expandafter\def\csname LT0\endcsname{\color[rgb]{1,0,0}}
      \expandafter\def\csname LT1\endcsname{\color[rgb]{0,1,0}}
      \expandafter\def\csname LT2\endcsname{\color[rgb]{0,0,1}}
      \expandafter\def\csname LT3\endcsname{\color[rgb]{1,0,1}}
      \expandafter\def\csname LT4\endcsname{\color[rgb]{0,1,1}}
      \expandafter\def\csname LT5\endcsname{\color[rgb]{1,1,0}}
      \expandafter\def\csname LT6\endcsname{\color[rgb]{0,0,0}}
      \expandafter\def\csname LT7\endcsname{\color[rgb]{1,0.3,0}}
      \expandafter\def\csname LT8\endcsname{\color[rgb]{0.5,0.5,0.5}}
    \else
      \def\colorrgb#1{\color{black}}
      \def\colorgray#1{\color[gray]{#1}}
      \expandafter\def\csname LTw\endcsname{\color{white}}
      \expandafter\def\csname LTb\endcsname{\color{black}}
      \expandafter\def\csname LTa\endcsname{\color{black}}
      \expandafter\def\csname LT0\endcsname{\color{black}}
      \expandafter\def\csname LT1\endcsname{\color{black}}
      \expandafter\def\csname LT2\endcsname{\color{black}}
      \expandafter\def\csname LT3\endcsname{\color{black}}
      \expandafter\def\csname LT4\endcsname{\color{black}}
      \expandafter\def\csname LT5\endcsname{\color{black}}
      \expandafter\def\csname LT6\endcsname{\color{black}}
      \expandafter\def\csname LT7\endcsname{\color{black}}
      \expandafter\def\csname LT8\endcsname{\color{black}}
    \fi
  \fi
    \setlength{\unitlength}{0.0200bp}
    \ifx\gptboxheight\undefined
      \newlength{\gptboxheight}
      \newlength{\gptboxwidth}
      \newsavebox{\gptboxtext}
    \fi
    \setlength{\fboxrule}{0.5pt}
    \setlength{\fboxsep}{1pt}
\begin{picture}(11520.00,8640.00)
    \gplgaddtomacro\gplbacktext{
      \colorrgb{0.00,0.00,0.00}
      \put(1480,1280){\makebox(0,0)[r]{\strut{}-10}}
      \colorrgb{0.00,0.00,0.00}
      \put(1480,3000){\makebox(0,0)[r]{\strut{}-5}}
      \colorrgb{0.00,0.00,0.00}
      \put(1480,4720){\makebox(0,0)[r]{\strut{}0}}
      \colorrgb{0.00,0.00,0.00}
      \put(1480,6439){\makebox(0,0)[r]{\strut{}5}}
      \colorrgb{0.00,0.00,0.00}
      \put(1480,8159){\makebox(0,0)[r]{\strut{}10}}
      \colorrgb{0.00,0.00,0.00}
      \put(1720,880){\makebox(0,0){\strut{}$10^{0}$}}
      \colorrgb{0.00,0.00,0.00}
      \put(4746,880){\makebox(0,0){\strut{}$10^{1}$}}
      \colorrgb{0.00,0.00,0.00}
      \put(7773,880){\makebox(0,0){\strut{}$10^{2}$}}
      \colorrgb{0.00,0.00,0.00}
      \put(10799,880){\makebox(0,0){\strut{}$10^{3}$}}
    }
    \gplgaddtomacro\gplfronttext{
      \colorrgb{0.00,0.00,0.00}
      \put(320,4719){\rotatebox{90}{\makebox(0,0){\strut{}$\OmRD$}}}
      \colorrgb{0.00,0.00,0.00}
      \put(6259,280){\makebox(0,0){\strut{}$H_{\CD}$ (km/s/Mpc)}}
      \colorrgb{0.00,0.00,0.00}
      \put(4135,7793){\makebox(0,0){\footnotesize $ 1.0$~Gyr}}
      \colorrgb{0.00,0.00,0.00}
      \put(4135,7188){\makebox(0,0){\footnotesize $ 4.0$~Gyr}}
      \colorrgb{0.00,0.00,0.00}
      \put(4135,6583){\makebox(0,0){\footnotesize $ 7.1$~Gyr}}
      \colorrgb{0.00,0.00,0.00}
      \put(4135,5978){\makebox(0,0){\footnotesize $10.0$~Gyr}}
      \colorrgb{0.00,0.00,0.00}
      \put(4135,5373){\makebox(0,0){\footnotesize $13.0$~Gyr}}
      \colorrgb{0.00,0.00,0.00}
      \put(2174,1624){\makebox(0,0)[l]{\strut{}$\LD =  40.0$~Mpc/$\hzeroeff$}}
    }
    \gplbacktext
    \put(0,0){\includegraphics[scale=0.4]{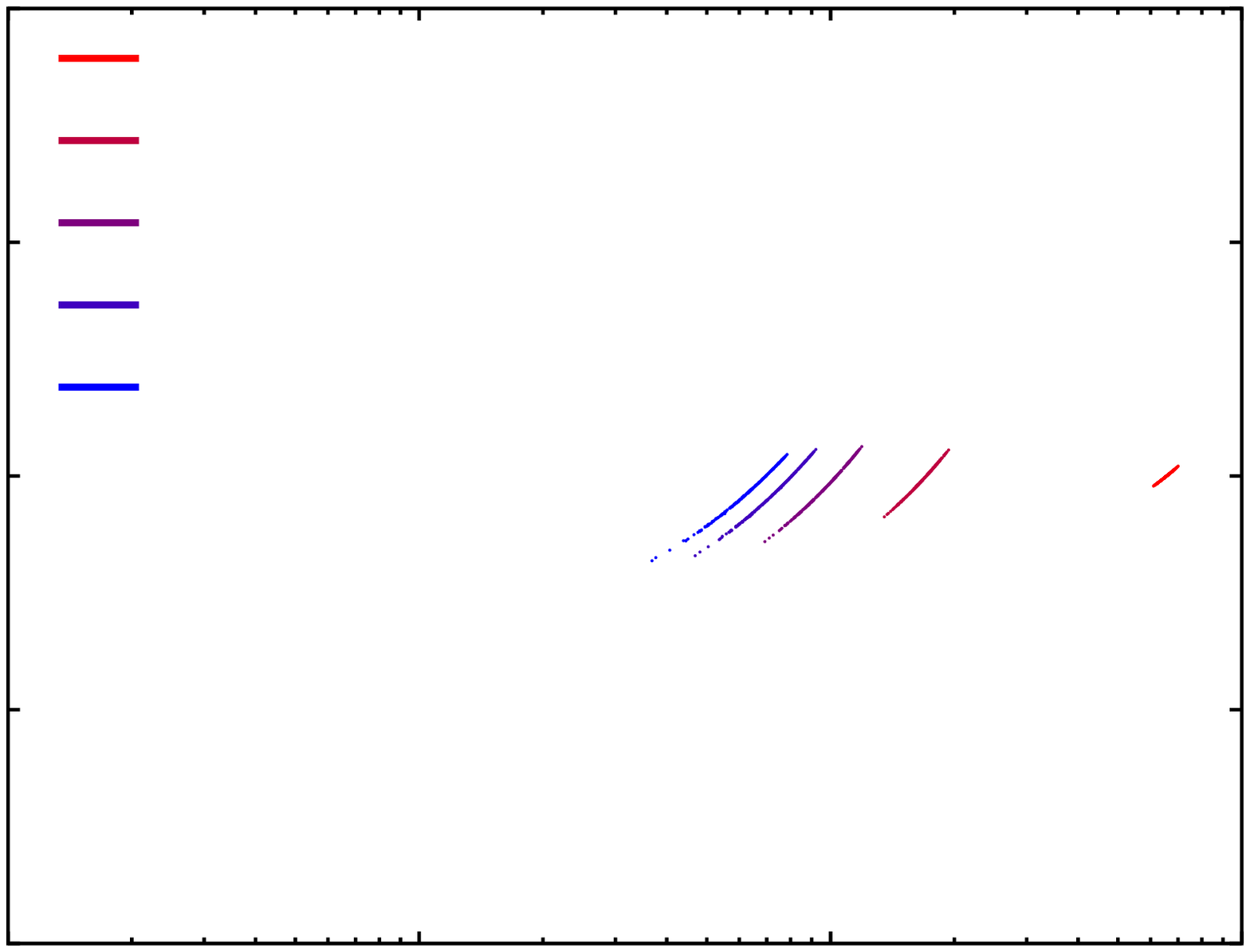}}
    \gplfronttext
  \end{picture}
\endgroup
 }
  \end{center}
  \caption{Averaged scalar curvature functional $\OmRD$ versus
    expansion rate in the same domain $H_{\CD}$,
    for an almost-{\LCDM} model, again from
    {\toptobottom} for
    $\LD = 2.5, 10, 40$~Mpc/$\hzeroeff$, respectively.
    The $\LD = 40$~Mpc/$\hzeroeff$ domains do not reach turnaround.
    \label{f-OmRDLCDM}}
\end{figure} 
 
Nevertheless, Eq.~\eqref{e-P-Theta-3H} implies that the limiting behaviour of the
latter term is
\begin{align}
  \average{\left(\Theta(X,t) - \average{\Theta(X,t)}\right)^2}
  \rightarrow \average{\left(3H(t) - \average{3H(t)}\right)^2} = 0 \,,
\end{align}
given that the pointwise limits are well-controlled by the specific algebraic expressions
above.
Thus, \postrefereeAAchanges{Proposition}~{\ELemmaFlatNoTurnaround} cannot be overridden in this case.
The loophole in the suggestion that pancake collapse could occur in this model
is that it was assumed that the reference model growing mode can be
used as the source of the RZA normalised growing mode function $\xi(t)$.
For consistency with the solution Eq.~\eqref{e-P-general-flat}, only the decaying mode
(the second term, $C(w) H(t)$) should be used. Thus the
expansion variance term approaches zero
(as in the above discussion) rather than diverging, so that no behaviour
beyond that of the pointwise constraint given by \postrefereeAAchanges{Proposition}~{\ELemmaFlatNoTurnaround}
occurs.

\begin{figure}
  \begin{center}
    \adjustbox{width=0.55\textwidth}{\begingroup
  \makeatletter
  \providecommand\color[2][]{
    \GenericError{(gnuplot) \space\space\space\@spaces}{
      Package color not loaded in conjunction with
      terminal option `colourtext'
    }{See the gnuplot documentation for explanation.
    }{Either use 'blacktext' in gnuplot or load the package
      color.sty in LaTeX.}
    \renewcommand\color[2][]{}
  }
  \providecommand\includegraphics[2][]{
    \GenericError{(gnuplot) \space\space\space\@spaces}{
      Package graphicx or graphics not loaded
    }{See the gnuplot documentation for explanation.
    }{The gnuplot epslatex terminal needs graphicx.sty or graphics.sty.}
    \renewcommand\includegraphics[2][]{}
  }
  \providecommand\rotatebox[2]{#2}
  \@ifundefined{ifGPcolor}{
    \newif\ifGPcolor
    \GPcolorfalse
  }{}
  \@ifundefined{ifGPblacktext}{
    \newif\ifGPblacktext
    \GPblacktexttrue
  }{}
  \let\gplgaddtomacro\g@addto@macro
  \gdef\gplbacktext{}
  \gdef\gplfronttext{}
  \makeatother
  \ifGPblacktext
    \def\colorrgb#1{}
    \def\colorgray#1{}
  \else
    \ifGPcolor
      \def\colorrgb#1{\color[rgb]{#1}}
      \def\colorgray#1{\color[gray]{#1}}
      \expandafter\def\csname LTw\endcsname{\color{white}}
      \expandafter\def\csname LTb\endcsname{\color{black}}
      \expandafter\def\csname LTa\endcsname{\color{black}}
      \expandafter\def\csname LT0\endcsname{\color[rgb]{1,0,0}}
      \expandafter\def\csname LT1\endcsname{\color[rgb]{0,1,0}}
      \expandafter\def\csname LT2\endcsname{\color[rgb]{0,0,1}}
      \expandafter\def\csname LT3\endcsname{\color[rgb]{1,0,1}}
      \expandafter\def\csname LT4\endcsname{\color[rgb]{0,1,1}}
      \expandafter\def\csname LT5\endcsname{\color[rgb]{1,1,0}}
      \expandafter\def\csname LT6\endcsname{\color[rgb]{0,0,0}}
      \expandafter\def\csname LT7\endcsname{\color[rgb]{1,0.3,0}}
      \expandafter\def\csname LT8\endcsname{\color[rgb]{0.5,0.5,0.5}}
    \else
      \def\colorrgb#1{\color{black}}
      \def\colorgray#1{\color[gray]{#1}}
      \expandafter\def\csname LTw\endcsname{\color{white}}
      \expandafter\def\csname LTb\endcsname{\color{black}}
      \expandafter\def\csname LTa\endcsname{\color{black}}
      \expandafter\def\csname LT0\endcsname{\color{black}}
      \expandafter\def\csname LT1\endcsname{\color{black}}
      \expandafter\def\csname LT2\endcsname{\color{black}}
      \expandafter\def\csname LT3\endcsname{\color{black}}
      \expandafter\def\csname LT4\endcsname{\color{black}}
      \expandafter\def\csname LT5\endcsname{\color{black}}
      \expandafter\def\csname LT6\endcsname{\color{black}}
      \expandafter\def\csname LT7\endcsname{\color{black}}
      \expandafter\def\csname LT8\endcsname{\color{black}}
    \fi
  \fi
    \setlength{\unitlength}{0.0200bp}
    \ifx\gptboxheight\undefined
      \newlength{\gptboxheight}
      \newlength{\gptboxwidth}
      \newsavebox{\gptboxtext}
    \fi
    \setlength{\fboxrule}{0.5pt}
    \setlength{\fboxsep}{1pt}
\begin{picture}(11520.00,8640.00)
    \gplgaddtomacro\gplbacktext{
      \colorrgb{0.00,0.00,0.00}
      \put(1332,1152){\makebox(0,0)[r]{\strut{}-10}}
      \colorrgb{0.00,0.00,0.00}
      \put(1332,2916){\makebox(0,0)[r]{\strut{}-5}}
      \colorrgb{0.00,0.00,0.00}
      \put(1332,4680){\makebox(0,0)[r]{\strut{}0}}
      \colorrgb{0.00,0.00,0.00}
      \put(1332,6443){\makebox(0,0)[r]{\strut{}5}}
      \colorrgb{0.00,0.00,0.00}
      \put(1332,8207){\makebox(0,0)[r]{\strut{}10}}
      \colorrgb{0.00,0.00,0.00}
      \put(1548,792){\makebox(0,0){\strut{}-0.01}}
      \colorrgb{0.00,0.00,0.00}
      \put(3879,792){\makebox(0,0){\strut{}-0.005}}
      \colorrgb{0.00,0.00,0.00}
      \put(6210,792){\makebox(0,0){\strut{}0}}
      \colorrgb{0.00,0.00,0.00}
      \put(8540,792){\makebox(0,0){\strut{}0.005}}
      \colorrgb{0.00,0.00,0.00}
      \put(10871,792){\makebox(0,0){\strut{}0.01}}
    }
    \gplgaddtomacro\gplfronttext{
      \colorrgb{0.00,0.00,0.00}
      \put(288,4679){\rotatebox{90}{\makebox(0,0){\strut{}$\OmRD\vert_{|H_{\CD}|< 1\mathrm{~km/s/Mpc}}$}}}
      \colorrgb{0.00,0.00,0.00}
      \put(6209,252){\makebox(0,0){\strut{}$\initinvII$}}
      \colorrgb{0.00,0.00,0.00}
      \put(2014,7502){\makebox(0,0)[l]{\strut{}$\LD =   2.5$~Mpc/$\hzeroeff$}}
    }
    \gplbacktext
    \put(0,0){\includegraphics[scale=0.4]{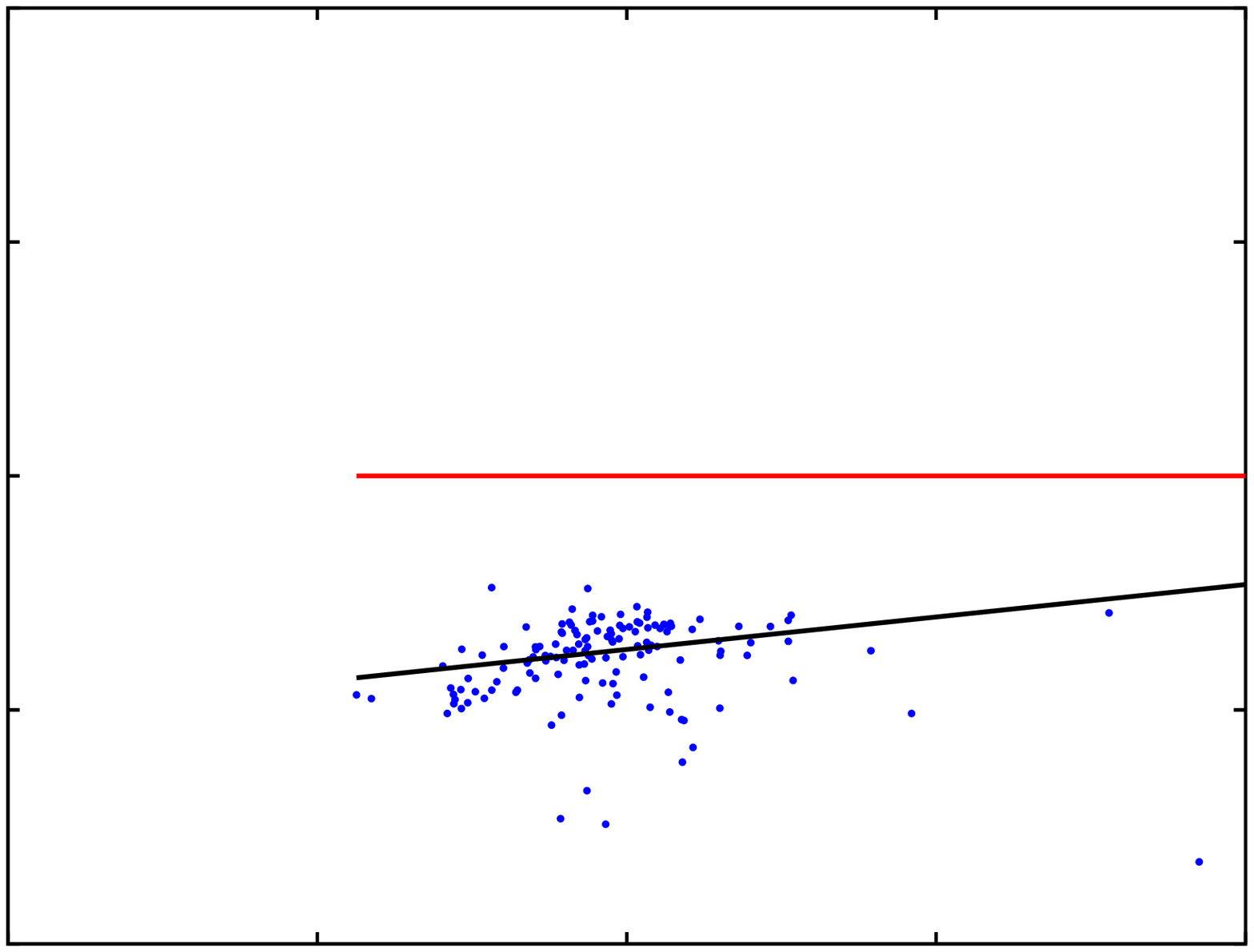}}
    \gplfronttext
  \end{picture}
\endgroup
 }\\
    \adjustbox{width=0.55\textwidth}{\begingroup
  \makeatletter
  \providecommand\color[2][]{
    \GenericError{(gnuplot) \space\space\space\@spaces}{
      Package color not loaded in conjunction with
      terminal option `colourtext'
    }{See the gnuplot documentation for explanation.
    }{Either use 'blacktext' in gnuplot or load the package
      color.sty in LaTeX.}
    \renewcommand\color[2][]{}
  }
  \providecommand\includegraphics[2][]{
    \GenericError{(gnuplot) \space\space\space\@spaces}{
      Package graphicx or graphics not loaded
    }{See the gnuplot documentation for explanation.
    }{The gnuplot epslatex terminal needs graphicx.sty or graphics.sty.}
    \renewcommand\includegraphics[2][]{}
  }
  \providecommand\rotatebox[2]{#2}
  \@ifundefined{ifGPcolor}{
    \newif\ifGPcolor
    \GPcolorfalse
  }{}
  \@ifundefined{ifGPblacktext}{
    \newif\ifGPblacktext
    \GPblacktexttrue
  }{}
  \let\gplgaddtomacro\g@addto@macro
  \gdef\gplbacktext{}
  \gdef\gplfronttext{}
  \makeatother
  \ifGPblacktext
    \def\colorrgb#1{}
    \def\colorgray#1{}
  \else
    \ifGPcolor
      \def\colorrgb#1{\color[rgb]{#1}}
      \def\colorgray#1{\color[gray]{#1}}
      \expandafter\def\csname LTw\endcsname{\color{white}}
      \expandafter\def\csname LTb\endcsname{\color{black}}
      \expandafter\def\csname LTa\endcsname{\color{black}}
      \expandafter\def\csname LT0\endcsname{\color[rgb]{1,0,0}}
      \expandafter\def\csname LT1\endcsname{\color[rgb]{0,1,0}}
      \expandafter\def\csname LT2\endcsname{\color[rgb]{0,0,1}}
      \expandafter\def\csname LT3\endcsname{\color[rgb]{1,0,1}}
      \expandafter\def\csname LT4\endcsname{\color[rgb]{0,1,1}}
      \expandafter\def\csname LT5\endcsname{\color[rgb]{1,1,0}}
      \expandafter\def\csname LT6\endcsname{\color[rgb]{0,0,0}}
      \expandafter\def\csname LT7\endcsname{\color[rgb]{1,0.3,0}}
      \expandafter\def\csname LT8\endcsname{\color[rgb]{0.5,0.5,0.5}}
    \else
      \def\colorrgb#1{\color{black}}
      \def\colorgray#1{\color[gray]{#1}}
      \expandafter\def\csname LTw\endcsname{\color{white}}
      \expandafter\def\csname LTb\endcsname{\color{black}}
      \expandafter\def\csname LTa\endcsname{\color{black}}
      \expandafter\def\csname LT0\endcsname{\color{black}}
      \expandafter\def\csname LT1\endcsname{\color{black}}
      \expandafter\def\csname LT2\endcsname{\color{black}}
      \expandafter\def\csname LT3\endcsname{\color{black}}
      \expandafter\def\csname LT4\endcsname{\color{black}}
      \expandafter\def\csname LT5\endcsname{\color{black}}
      \expandafter\def\csname LT6\endcsname{\color{black}}
      \expandafter\def\csname LT7\endcsname{\color{black}}
      \expandafter\def\csname LT8\endcsname{\color{black}}
    \fi
  \fi
    \setlength{\unitlength}{0.0200bp}
    \ifx\gptboxheight\undefined
      \newlength{\gptboxheight}
      \newlength{\gptboxwidth}
      \newsavebox{\gptboxtext}
    \fi
    \setlength{\fboxrule}{0.5pt}
    \setlength{\fboxsep}{1pt}
\begin{picture}(11520.00,8640.00)
    \gplgaddtomacro\gplbacktext{
      \colorrgb{0.00,0.00,0.00}
      \put(1332,1152){\makebox(0,0)[r]{\strut{}-10}}
      \colorrgb{0.00,0.00,0.00}
      \put(1332,2916){\makebox(0,0)[r]{\strut{}-5}}
      \colorrgb{0.00,0.00,0.00}
      \put(1332,4680){\makebox(0,0)[r]{\strut{}0}}
      \colorrgb{0.00,0.00,0.00}
      \put(1332,6443){\makebox(0,0)[r]{\strut{}5}}
      \colorrgb{0.00,0.00,0.00}
      \put(1332,8207){\makebox(0,0)[r]{\strut{}10}}
      \colorrgb{0.00,0.00,0.00}
      \put(1548,792){\makebox(0,0){\strut{}-0.01}}
      \colorrgb{0.00,0.00,0.00}
      \put(3879,792){\makebox(0,0){\strut{}-0.005}}
      \colorrgb{0.00,0.00,0.00}
      \put(6210,792){\makebox(0,0){\strut{}0}}
      \colorrgb{0.00,0.00,0.00}
      \put(8540,792){\makebox(0,0){\strut{}0.005}}
      \colorrgb{0.00,0.00,0.00}
      \put(10871,792){\makebox(0,0){\strut{}0.01}}
    }
    \gplgaddtomacro\gplfronttext{
      \colorrgb{0.00,0.00,0.00}
      \put(288,4679){\rotatebox{90}{\makebox(0,0){\strut{}$\OmRD\vert_{|H_{\CD}|< 1\mathrm{~km/s/Mpc}}$}}}
      \colorrgb{0.00,0.00,0.00}
      \put(6209,252){\makebox(0,0){\strut{}$\initinvII$}}
      \colorrgb{0.00,0.00,0.00}
      \put(2014,7502){\makebox(0,0)[l]{\strut{}$\LD =  10.0$~Mpc/$\hzeroeff$}}
    }
    \gplbacktext
    \put(0,0){\includegraphics[scale=0.4]{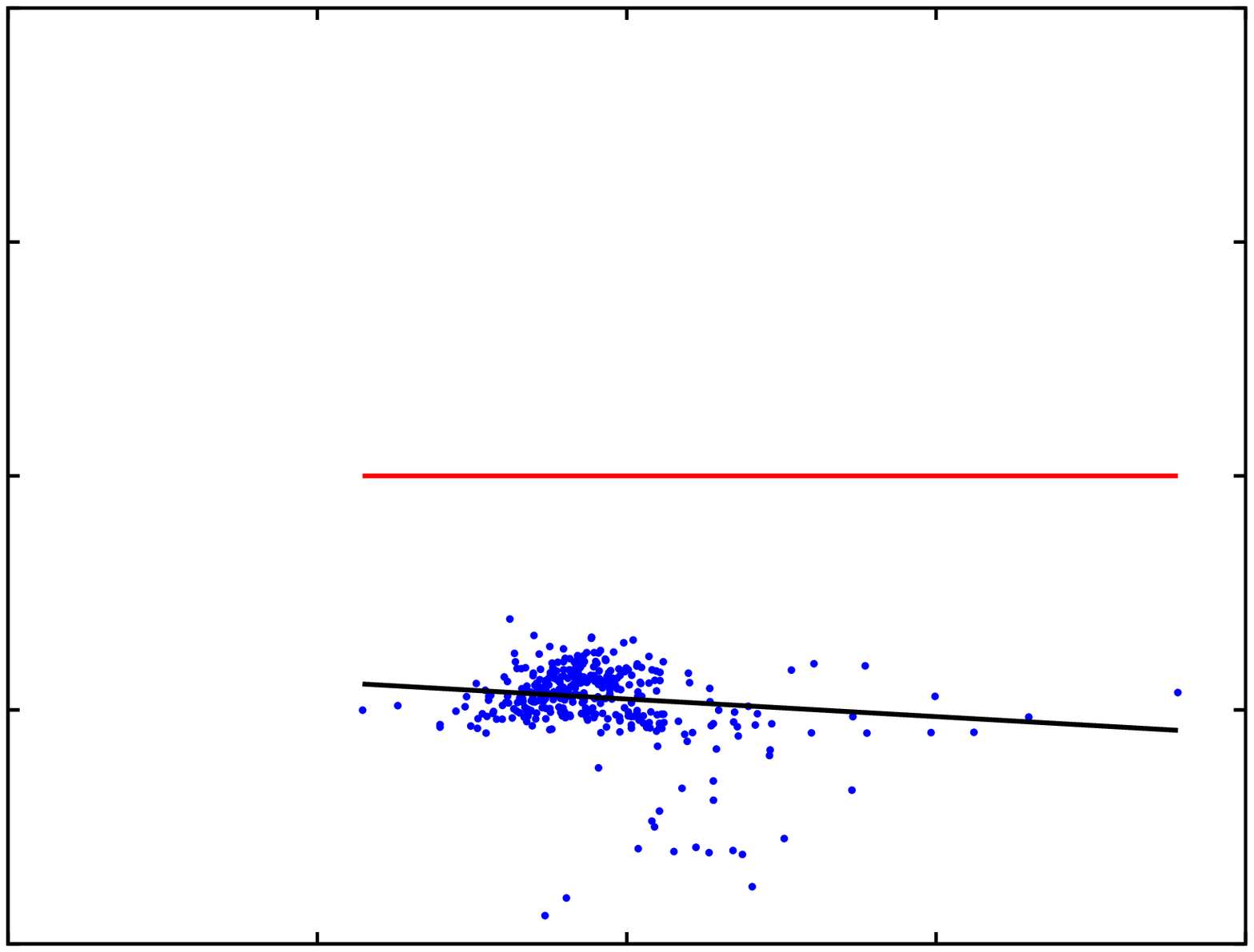}}
    \gplfronttext
  \end{picture}
\endgroup
 }\\
    \adjustbox{width=0.55\textwidth}{\begingroup
  \makeatletter
  \providecommand\color[2][]{
    \GenericError{(gnuplot) \space\space\space\@spaces}{
      Package color not loaded in conjunction with
      terminal option `colourtext'
    }{See the gnuplot documentation for explanation.
    }{Either use 'blacktext' in gnuplot or load the package
      color.sty in LaTeX.}
    \renewcommand\color[2][]{}
  }
  \providecommand\includegraphics[2][]{
    \GenericError{(gnuplot) \space\space\space\@spaces}{
      Package graphicx or graphics not loaded
    }{See the gnuplot documentation for explanation.
    }{The gnuplot epslatex terminal needs graphicx.sty or graphics.sty.}
    \renewcommand\includegraphics[2][]{}
  }
  \providecommand\rotatebox[2]{#2}
  \@ifundefined{ifGPcolor}{
    \newif\ifGPcolor
    \GPcolorfalse
  }{}
  \@ifundefined{ifGPblacktext}{
    \newif\ifGPblacktext
    \GPblacktexttrue
  }{}
  \let\gplgaddtomacro\g@addto@macro
  \gdef\gplbacktext{}
  \gdef\gplfronttext{}
  \makeatother
  \ifGPblacktext
    \def\colorrgb#1{}
    \def\colorgray#1{}
  \else
    \ifGPcolor
      \def\colorrgb#1{\color[rgb]{#1}}
      \def\colorgray#1{\color[gray]{#1}}
      \expandafter\def\csname LTw\endcsname{\color{white}}
      \expandafter\def\csname LTb\endcsname{\color{black}}
      \expandafter\def\csname LTa\endcsname{\color{black}}
      \expandafter\def\csname LT0\endcsname{\color[rgb]{1,0,0}}
      \expandafter\def\csname LT1\endcsname{\color[rgb]{0,1,0}}
      \expandafter\def\csname LT2\endcsname{\color[rgb]{0,0,1}}
      \expandafter\def\csname LT3\endcsname{\color[rgb]{1,0,1}}
      \expandafter\def\csname LT4\endcsname{\color[rgb]{0,1,1}}
      \expandafter\def\csname LT5\endcsname{\color[rgb]{1,1,0}}
      \expandafter\def\csname LT6\endcsname{\color[rgb]{0,0,0}}
      \expandafter\def\csname LT7\endcsname{\color[rgb]{1,0.3,0}}
      \expandafter\def\csname LT8\endcsname{\color[rgb]{0.5,0.5,0.5}}
    \else
      \def\colorrgb#1{\color{black}}
      \def\colorgray#1{\color[gray]{#1}}
      \expandafter\def\csname LTw\endcsname{\color{white}}
      \expandafter\def\csname LTb\endcsname{\color{black}}
      \expandafter\def\csname LTa\endcsname{\color{black}}
      \expandafter\def\csname LT0\endcsname{\color{black}}
      \expandafter\def\csname LT1\endcsname{\color{black}}
      \expandafter\def\csname LT2\endcsname{\color{black}}
      \expandafter\def\csname LT3\endcsname{\color{black}}
      \expandafter\def\csname LT4\endcsname{\color{black}}
      \expandafter\def\csname LT5\endcsname{\color{black}}
      \expandafter\def\csname LT6\endcsname{\color{black}}
      \expandafter\def\csname LT7\endcsname{\color{black}}
      \expandafter\def\csname LT8\endcsname{\color{black}}
    \fi
  \fi
    \setlength{\unitlength}{0.0200bp}
    \ifx\gptboxheight\undefined
      \newlength{\gptboxheight}
      \newlength{\gptboxwidth}
      \newsavebox{\gptboxtext}
    \fi
    \setlength{\fboxrule}{0.5pt}
    \setlength{\fboxsep}{1pt}
\begin{picture}(11520.00,8640.00)
    \gplgaddtomacro\gplbacktext{
      \colorrgb{0.00,0.00,0.00}
      \put(1332,1152){\makebox(0,0)[r]{\strut{}-10}}
      \colorrgb{0.00,0.00,0.00}
      \put(1332,2916){\makebox(0,0)[r]{\strut{}-5}}
      \colorrgb{0.00,0.00,0.00}
      \put(1332,4680){\makebox(0,0)[r]{\strut{}0}}
      \colorrgb{0.00,0.00,0.00}
      \put(1332,6443){\makebox(0,0)[r]{\strut{}5}}
      \colorrgb{0.00,0.00,0.00}
      \put(1332,8207){\makebox(0,0)[r]{\strut{}10}}
      \colorrgb{0.00,0.00,0.00}
      \put(1548,792){\makebox(0,0){\strut{}-0.01}}
      \colorrgb{0.00,0.00,0.00}
      \put(3879,792){\makebox(0,0){\strut{}-0.005}}
      \colorrgb{0.00,0.00,0.00}
      \put(6210,792){\makebox(0,0){\strut{}0}}
      \colorrgb{0.00,0.00,0.00}
      \put(8540,792){\makebox(0,0){\strut{}0.005}}
      \colorrgb{0.00,0.00,0.00}
      \put(10871,792){\makebox(0,0){\strut{}0.01}}
    }
    \gplgaddtomacro\gplfronttext{
      \colorrgb{0.00,0.00,0.00}
      \put(288,4679){\rotatebox{90}{\makebox(0,0){\strut{}$\OmRD\vert_{|H_{\CD}|< 1\mathrm{~km/s/Mpc}}$}}}
      \colorrgb{0.00,0.00,0.00}
      \put(6209,252){\makebox(0,0){\strut{}$\initinvII$}}
      \colorrgb{0.00,0.00,0.00}
      \put(2014,7502){\makebox(0,0)[l]{\strut{}$\LD =  40.0$~Mpc/$\hzeroeff$}}
    }
    \gplbacktext
    \put(0,0){\includegraphics[scale=0.4]{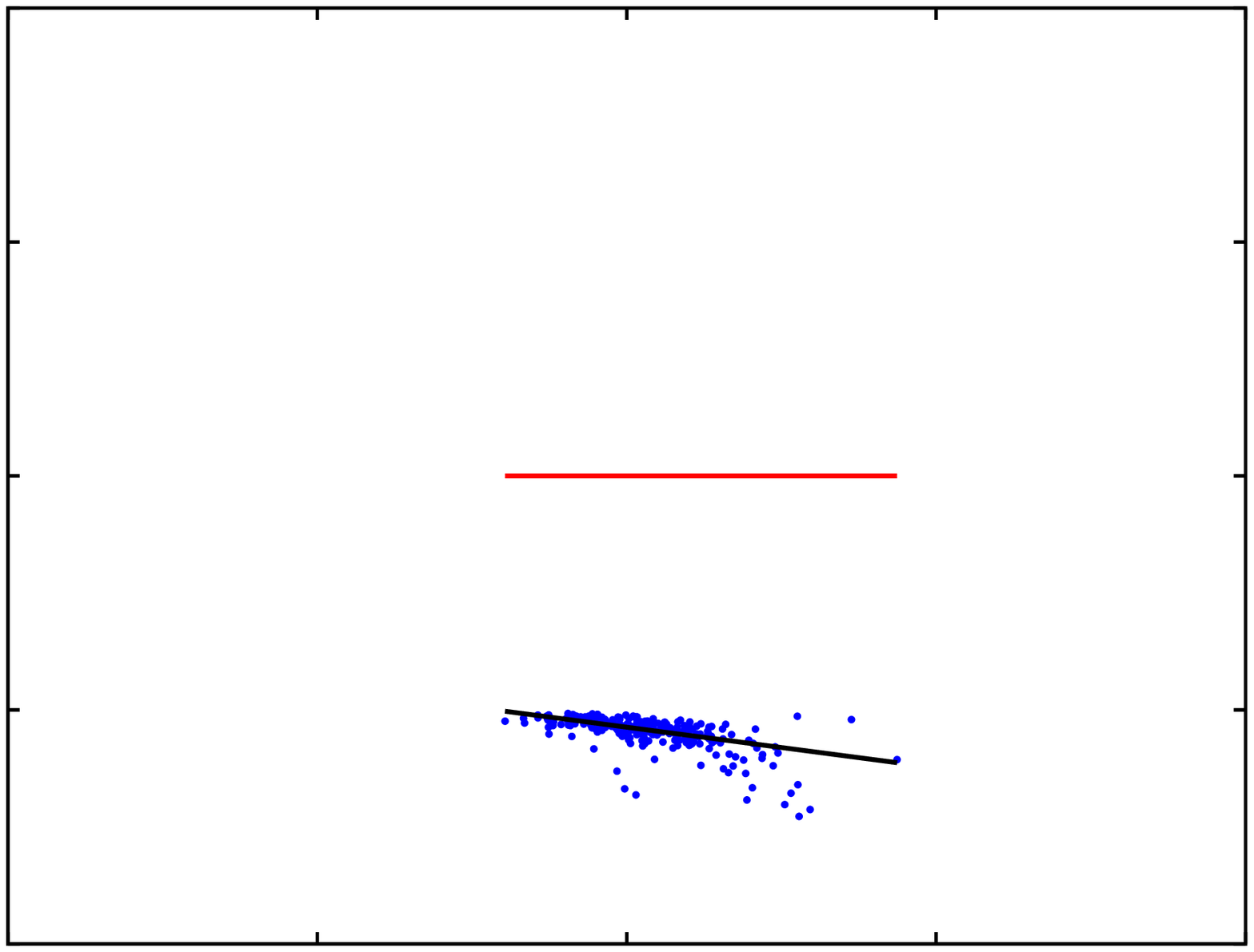}}
    \gplfronttext
  \end{picture}
\endgroup
 }
  \end{center}
  \caption{Curvature functional $\OmRD$ against
    initial
    average peculiar-expansion tensor
    second invariant $\initaverageriem{\invII}$ at
    turnaround, defined as $|H_{\CD}| < 1$~km/s/Mpc, for
    an almost-EdS model. A
    Theil--Sen robust linear fit
    is shown, with the zero points listed in
    Table~\protect\ref{t-OmRDTA}.
    A red line indicates $\OmRD=0$. It is clear that no domains
    were negatively curved ($\OmRD > 0$) at turnaround. The distributions
    of $\OmRD$ are clearly separated from the negative curvature
    region.
    \label{f-II-OmRD-EdS}}
\end{figure} 
 
\begin{figure}
  \begin{center}
    \adjustbox{width=0.55\textwidth}{\begingroup
  \makeatletter
  \providecommand\color[2][]{
    \GenericError{(gnuplot) \space\space\space\@spaces}{
      Package color not loaded in conjunction with
      terminal option `colourtext'
    }{See the gnuplot documentation for explanation.
    }{Either use 'blacktext' in gnuplot or load the package
      color.sty in LaTeX.}
    \renewcommand\color[2][]{}
  }
  \providecommand\includegraphics[2][]{
    \GenericError{(gnuplot) \space\space\space\@spaces}{
      Package graphicx or graphics not loaded
    }{See the gnuplot documentation for explanation.
    }{The gnuplot epslatex terminal needs graphicx.sty or graphics.sty.}
    \renewcommand\includegraphics[2][]{}
  }
  \providecommand\rotatebox[2]{#2}
  \@ifundefined{ifGPcolor}{
    \newif\ifGPcolor
    \GPcolorfalse
  }{}
  \@ifundefined{ifGPblacktext}{
    \newif\ifGPblacktext
    \GPblacktexttrue
  }{}
  \let\gplgaddtomacro\g@addto@macro
  \gdef\gplbacktext{}
  \gdef\gplfronttext{}
  \makeatother
  \ifGPblacktext
    \def\colorrgb#1{}
    \def\colorgray#1{}
  \else
    \ifGPcolor
      \def\colorrgb#1{\color[rgb]{#1}}
      \def\colorgray#1{\color[gray]{#1}}
      \expandafter\def\csname LTw\endcsname{\color{white}}
      \expandafter\def\csname LTb\endcsname{\color{black}}
      \expandafter\def\csname LTa\endcsname{\color{black}}
      \expandafter\def\csname LT0\endcsname{\color[rgb]{1,0,0}}
      \expandafter\def\csname LT1\endcsname{\color[rgb]{0,1,0}}
      \expandafter\def\csname LT2\endcsname{\color[rgb]{0,0,1}}
      \expandafter\def\csname LT3\endcsname{\color[rgb]{1,0,1}}
      \expandafter\def\csname LT4\endcsname{\color[rgb]{0,1,1}}
      \expandafter\def\csname LT5\endcsname{\color[rgb]{1,1,0}}
      \expandafter\def\csname LT6\endcsname{\color[rgb]{0,0,0}}
      \expandafter\def\csname LT7\endcsname{\color[rgb]{1,0.3,0}}
      \expandafter\def\csname LT8\endcsname{\color[rgb]{0.5,0.5,0.5}}
    \else
      \def\colorrgb#1{\color{black}}
      \def\colorgray#1{\color[gray]{#1}}
      \expandafter\def\csname LTw\endcsname{\color{white}}
      \expandafter\def\csname LTb\endcsname{\color{black}}
      \expandafter\def\csname LTa\endcsname{\color{black}}
      \expandafter\def\csname LT0\endcsname{\color{black}}
      \expandafter\def\csname LT1\endcsname{\color{black}}
      \expandafter\def\csname LT2\endcsname{\color{black}}
      \expandafter\def\csname LT3\endcsname{\color{black}}
      \expandafter\def\csname LT4\endcsname{\color{black}}
      \expandafter\def\csname LT5\endcsname{\color{black}}
      \expandafter\def\csname LT6\endcsname{\color{black}}
      \expandafter\def\csname LT7\endcsname{\color{black}}
      \expandafter\def\csname LT8\endcsname{\color{black}}
    \fi
  \fi
    \setlength{\unitlength}{0.0200bp}
    \ifx\gptboxheight\undefined
      \newlength{\gptboxheight}
      \newlength{\gptboxwidth}
      \newsavebox{\gptboxtext}
    \fi
    \setlength{\fboxrule}{0.5pt}
    \setlength{\fboxsep}{1pt}
\begin{picture}(11520.00,8640.00)
    \gplgaddtomacro\gplbacktext{
      \colorrgb{0.00,0.00,0.00}
      \put(1332,1152){\makebox(0,0)[r]{\strut{}-10}}
      \colorrgb{0.00,0.00,0.00}
      \put(1332,2916){\makebox(0,0)[r]{\strut{}-5}}
      \colorrgb{0.00,0.00,0.00}
      \put(1332,4680){\makebox(0,0)[r]{\strut{}0}}
      \colorrgb{0.00,0.00,0.00}
      \put(1332,6443){\makebox(0,0)[r]{\strut{}5}}
      \colorrgb{0.00,0.00,0.00}
      \put(1332,8207){\makebox(0,0)[r]{\strut{}10}}
      \colorrgb{0.00,0.00,0.00}
      \put(1548,792){\makebox(0,0){\strut{}-0.01}}
      \colorrgb{0.00,0.00,0.00}
      \put(3879,792){\makebox(0,0){\strut{}-0.005}}
      \colorrgb{0.00,0.00,0.00}
      \put(6210,792){\makebox(0,0){\strut{}0}}
      \colorrgb{0.00,0.00,0.00}
      \put(8540,792){\makebox(0,0){\strut{}0.005}}
      \colorrgb{0.00,0.00,0.00}
      \put(10871,792){\makebox(0,0){\strut{}0.01}}
    }
    \gplgaddtomacro\gplfronttext{
      \colorrgb{0.00,0.00,0.00}
      \put(288,4679){\rotatebox{90}{\makebox(0,0){\strut{}$\OmRD\vert_{|H_{\CD}|< 1\mathrm{~km/s/Mpc}}$}}}
      \colorrgb{0.00,0.00,0.00}
      \put(6209,252){\makebox(0,0){\strut{}$\initinvII$}}
      \colorrgb{0.00,0.00,0.00}
      \put(2014,7502){\makebox(0,0)[l]{\strut{}$\LD =   2.5$~Mpc/$\hzeroeff$}}
    }
    \gplbacktext
    \put(0,0){\includegraphics[scale=0.4]{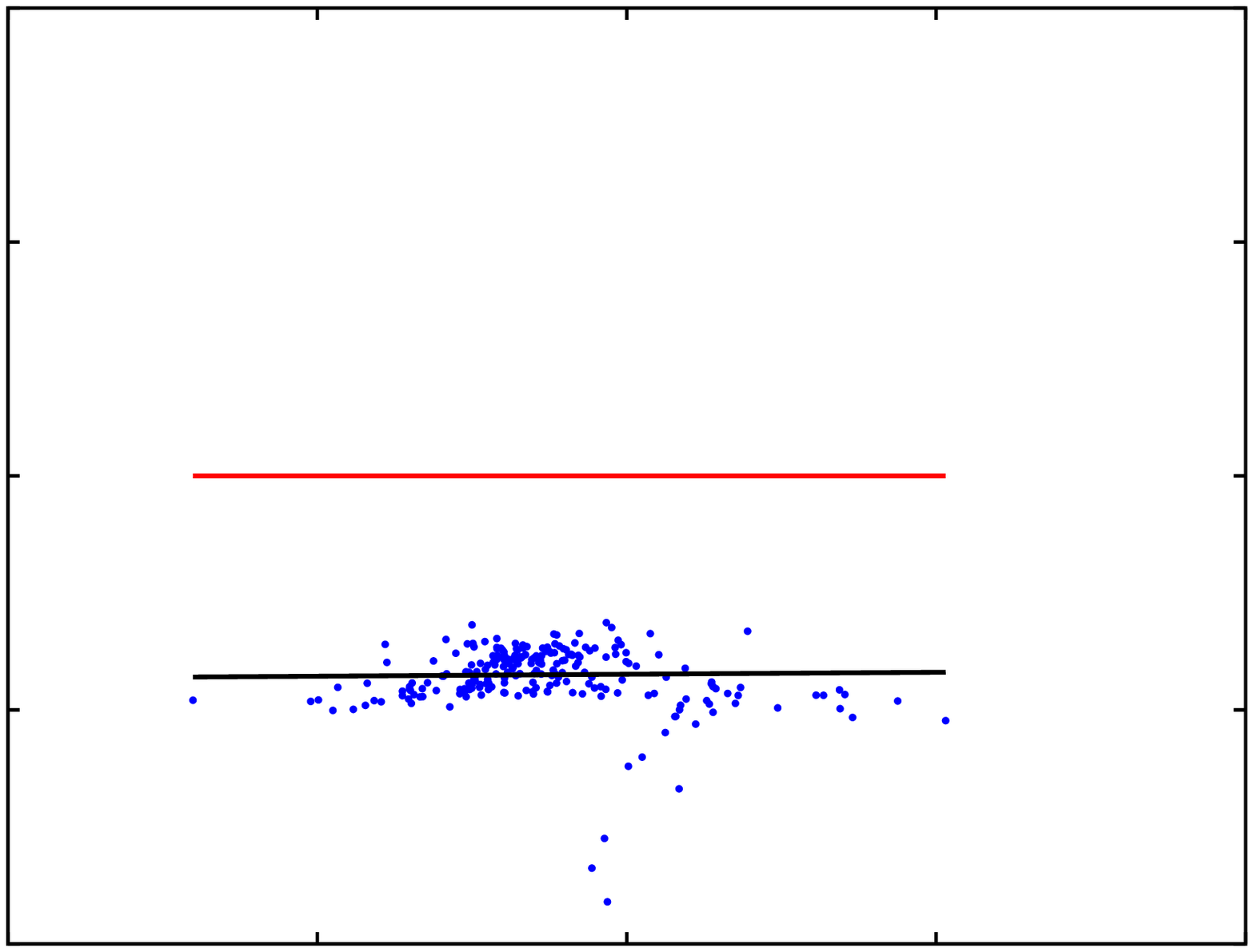}}
    \gplfronttext
  \end{picture}
\endgroup
 }\\
    \adjustbox{width=0.55\textwidth}{\begingroup
  \makeatletter
  \providecommand\color[2][]{
    \GenericError{(gnuplot) \space\space\space\@spaces}{
      Package color not loaded in conjunction with
      terminal option `colourtext'
    }{See the gnuplot documentation for explanation.
    }{Either use 'blacktext' in gnuplot or load the package
      color.sty in LaTeX.}
    \renewcommand\color[2][]{}
  }
  \providecommand\includegraphics[2][]{
    \GenericError{(gnuplot) \space\space\space\@spaces}{
      Package graphicx or graphics not loaded
    }{See the gnuplot documentation for explanation.
    }{The gnuplot epslatex terminal needs graphicx.sty or graphics.sty.}
    \renewcommand\includegraphics[2][]{}
  }
  \providecommand\rotatebox[2]{#2}
  \@ifundefined{ifGPcolor}{
    \newif\ifGPcolor
    \GPcolorfalse
  }{}
  \@ifundefined{ifGPblacktext}{
    \newif\ifGPblacktext
    \GPblacktexttrue
  }{}
  \let\gplgaddtomacro\g@addto@macro
  \gdef\gplbacktext{}
  \gdef\gplfronttext{}
  \makeatother
  \ifGPblacktext
    \def\colorrgb#1{}
    \def\colorgray#1{}
  \else
    \ifGPcolor
      \def\colorrgb#1{\color[rgb]{#1}}
      \def\colorgray#1{\color[gray]{#1}}
      \expandafter\def\csname LTw\endcsname{\color{white}}
      \expandafter\def\csname LTb\endcsname{\color{black}}
      \expandafter\def\csname LTa\endcsname{\color{black}}
      \expandafter\def\csname LT0\endcsname{\color[rgb]{1,0,0}}
      \expandafter\def\csname LT1\endcsname{\color[rgb]{0,1,0}}
      \expandafter\def\csname LT2\endcsname{\color[rgb]{0,0,1}}
      \expandafter\def\csname LT3\endcsname{\color[rgb]{1,0,1}}
      \expandafter\def\csname LT4\endcsname{\color[rgb]{0,1,1}}
      \expandafter\def\csname LT5\endcsname{\color[rgb]{1,1,0}}
      \expandafter\def\csname LT6\endcsname{\color[rgb]{0,0,0}}
      \expandafter\def\csname LT7\endcsname{\color[rgb]{1,0.3,0}}
      \expandafter\def\csname LT8\endcsname{\color[rgb]{0.5,0.5,0.5}}
    \else
      \def\colorrgb#1{\color{black}}
      \def\colorgray#1{\color[gray]{#1}}
      \expandafter\def\csname LTw\endcsname{\color{white}}
      \expandafter\def\csname LTb\endcsname{\color{black}}
      \expandafter\def\csname LTa\endcsname{\color{black}}
      \expandafter\def\csname LT0\endcsname{\color{black}}
      \expandafter\def\csname LT1\endcsname{\color{black}}
      \expandafter\def\csname LT2\endcsname{\color{black}}
      \expandafter\def\csname LT3\endcsname{\color{black}}
      \expandafter\def\csname LT4\endcsname{\color{black}}
      \expandafter\def\csname LT5\endcsname{\color{black}}
      \expandafter\def\csname LT6\endcsname{\color{black}}
      \expandafter\def\csname LT7\endcsname{\color{black}}
      \expandafter\def\csname LT8\endcsname{\color{black}}
    \fi
  \fi
    \setlength{\unitlength}{0.0200bp}
    \ifx\gptboxheight\undefined
      \newlength{\gptboxheight}
      \newlength{\gptboxwidth}
      \newsavebox{\gptboxtext}
    \fi
    \setlength{\fboxrule}{0.5pt}
    \setlength{\fboxsep}{1pt}
\begin{picture}(11520.00,8640.00)
    \gplgaddtomacro\gplbacktext{
      \colorrgb{0.00,0.00,0.00}
      \put(1332,1152){\makebox(0,0)[r]{\strut{}-10}}
      \colorrgb{0.00,0.00,0.00}
      \put(1332,2916){\makebox(0,0)[r]{\strut{}-5}}
      \colorrgb{0.00,0.00,0.00}
      \put(1332,4680){\makebox(0,0)[r]{\strut{}0}}
      \colorrgb{0.00,0.00,0.00}
      \put(1332,6443){\makebox(0,0)[r]{\strut{}5}}
      \colorrgb{0.00,0.00,0.00}
      \put(1332,8207){\makebox(0,0)[r]{\strut{}10}}
      \colorrgb{0.00,0.00,0.00}
      \put(1548,792){\makebox(0,0){\strut{}-0.01}}
      \colorrgb{0.00,0.00,0.00}
      \put(3879,792){\makebox(0,0){\strut{}-0.005}}
      \colorrgb{0.00,0.00,0.00}
      \put(6210,792){\makebox(0,0){\strut{}0}}
      \colorrgb{0.00,0.00,0.00}
      \put(8540,792){\makebox(0,0){\strut{}0.005}}
      \colorrgb{0.00,0.00,0.00}
      \put(10871,792){\makebox(0,0){\strut{}0.01}}
    }
    \gplgaddtomacro\gplfronttext{
      \colorrgb{0.00,0.00,0.00}
      \put(288,4679){\rotatebox{90}{\makebox(0,0){\strut{}$\OmRD\vert_{|H_{\CD}|< 1\mathrm{~km/s/Mpc}}$}}}
      \colorrgb{0.00,0.00,0.00}
      \put(6209,252){\makebox(0,0){\strut{}$\initinvII$}}
      \colorrgb{0.00,0.00,0.00}
      \put(2014,7502){\makebox(0,0)[l]{\strut{}$\LD =  10.0$~Mpc/$\hzeroeff$}}
    }
    \gplbacktext
    \put(0,0){\includegraphics[scale=0.4]{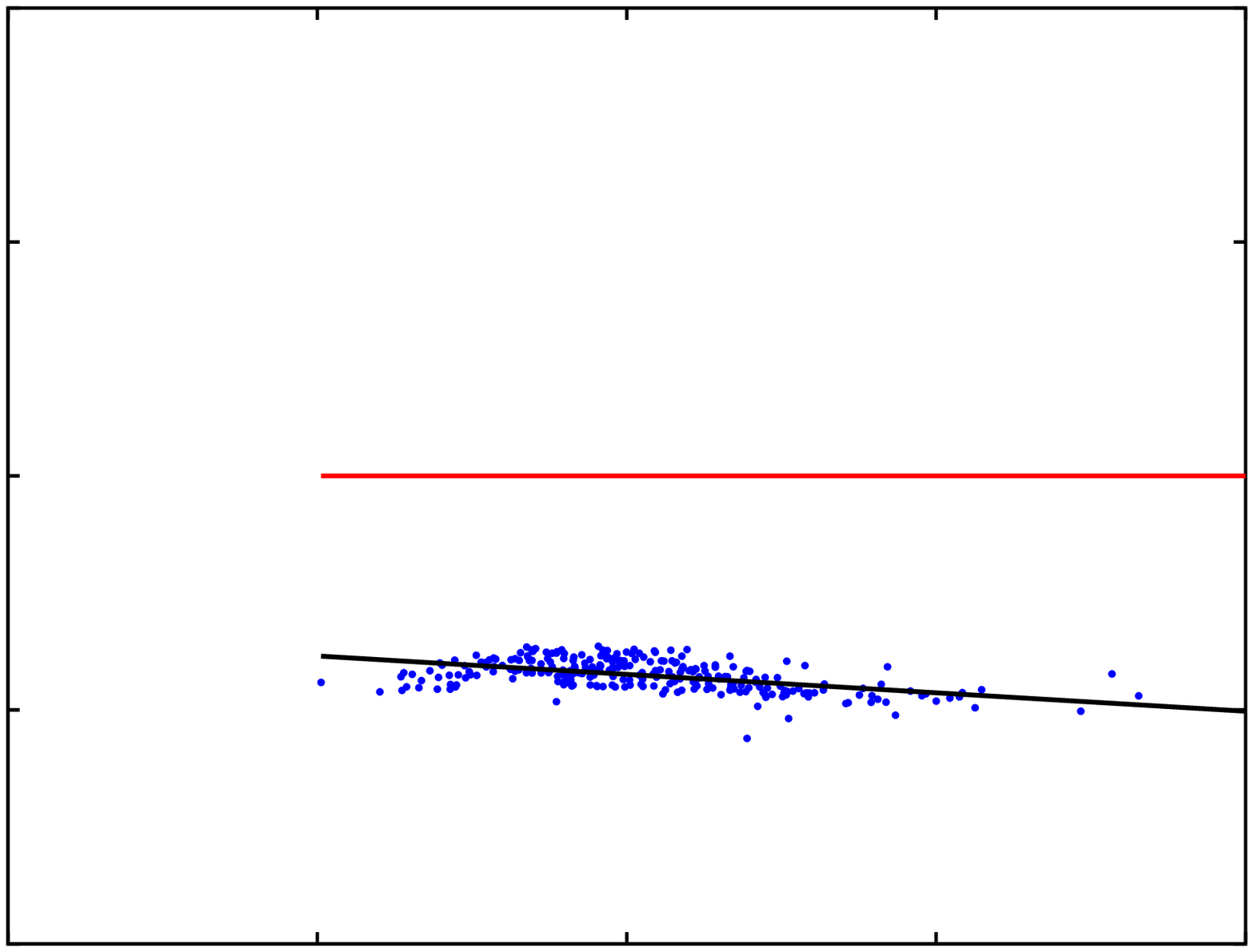}}
    \gplfronttext
  \end{picture}
\endgroup
 }
  \end{center}
  \caption{Curvature functional $\OmRD$ against $\initaverageriem{\invII}$ at
    turnaround, as in Fig.~\ref{f-II-OmRD-EdS},
    for an almost-{\LCDM} model. Turnaround is not reached
    at $\LD=40$~Mpc/$\hzeroeff$.
    \label{f-II-OmRD-LCDM}}
\end{figure}

This resolution of the \cite{BuchRZA2} paradox
implies an important corollary for numerical
implementations of the RZA approach
\postrefereeAAchanges{\citep{BuchRZA1,BuchRZA2}}, which until now have assumed that
the growing modes of the background cosmological model can be
associated with standard choices of Gaussian random field initial
perturbations \citep{BuchRZA2,Roukema17silvir}.  The growing mode
is not guaranteed to be relativistically valid (allowed by the
equations that solve the Einstein equation) for a given perturbation
or domain;
\postrefereeAAchanges{in at least the subcase under discussion here,
  the growing mode is invalid.}
The model
being studied and the perturbation under consideration
must allow positive spatial curvature to accompany the
perturbation as it slows down and reaches its turnaround
epoch, unless a positive expansion variance term,
or to be more specific, a positive kinematical backreaction
(see Eq.~\eqref{e-defn-Q}) is strong enough to allow
reaching turnaround, or unless the behaviour is complex enough (for example,
strongly anisotropic) for average properties to override
\postrefereeAAchanges{Proposition}~{\ELemmaFlatNoTurnaround}.
In \cite{Roukema17silvir}, there is no constraint forcing zero
curvature or blocking kinematical backreaction evolution, but the
existence of the growing mode is effectively an assumption there, rather
than a solution that is guaranteed to be consistent with the Einstein
equation.

In the following subsection, we quantify the positive spatial
curvatures that should be associated with typical scales of
gravitational collapse on galaxy dark matter halo, galaxy cluster and
supercluster scales.

\subsection{Almost-Einstein--de~Sitter and almost-{\LCDM} models} \label{s-results-EdS-LCDM}

\subsubsection{Almost-EdS special case}
\label{s-results-EdS-II-III-zero}
Given the null initial average second and third invariants
(Eq.~\eqref{e-II-III-zero}), the additivity
of the Lagrangian average over its arguments that follows
from its definition in Eq.~\eqref{e-defn-calI-average},
and the spatial independence of $\xi$,
we can rewrite Eq.~\eqref{e-CJ-defn} and its derivatives
as
\begin{align}
  \initaverageriem{\CJ} &= \initaverageriem{1 + \xi \initial{\invI}
  + \xi^2 \initial{\invII}
  + \xi^3 \initial{\invIII}}
  \,= \initaverageriem{1 + \xi \initial{\invI}}
  \,, \nonumber \\
  \initaverageriem{\dot{\CJ}} &= \initaverageriem{\dot{\xi} \initial{\invI}
  + 2\xi \dot{\xi} \initial{\invII}
  + 3 \xi^2 \dot{\xi} \initial{\invIII}}
  \,= \dot{\xi} \initaverageriem{ \initial{\invI}}
  \,, \nonumber \\
  \initaverageriem{\ddot{\CJ}} &= \initaverageriem{\ddot{\xi} \initial{\invI}
  + \left(2\dot{\xi}^2 + 2 \xi \ddot{\xi}\right) \initial{\invII}
  + \left(6 \xi \dot{\xi}^2 + 3 \xi^2 \ddot{\xi}\right) \initial{\invIII}}
  \,= \ddot{\xi} \initaverageriem{ \initial{\invI}}
  \,.
  \label{e-CJ-derivs}
\end{align}
The same relations, together with
the general RZA expressions for the evolution of the invariants,
given in Eq.~(48) of \cite{BuchRZA2}, show that
\begin{align}
  \frac{\initaverageriem{\ddot{\CJ}}}{\initaverageriem{{\CJ}}} &=
  \frac{\ddot{\xi}}{\dot{\xi}}
  \frac{\initaverageriem{\dot{\CJ}}}{\initaverageriem{{\CJ}}}
  \,.
  \label{e-replace-CJ-ddot}
\end{align}
Thus
\begin{align}
  \average{\invII} = 0
\end{align}
at all times.
A similar calculation yields
\begin{align}
  \average{\invIII} = 0 \,,
\end{align}
again, at all times for which the RZA remains valid.

\tOmRDTA

\tOmRDTAplanesym

\tOmRDTAstats

We now consider the turnaround condition, which we set as
\begin{align}
  \average{\invI} &= -3\frac{H}{\initial{H}} \,,
  \label{e-TA-condn}
\end{align}
since the collapse of the domain $\CD$ has to exactly balance
the reference model expansion and the factor of three
follows from $\invI$ being the trace of $\theta^i_j$.
The denominator $\initial{H}$ follows
from using the same convention of dimensionless invariants as in
\cite[][Eq.~(19)]{Roukema17silvir} and much of \cite{BuchRZA2}.
Using the first part of Eq.~(48) of \cite{BuchRZA2},
appropriately normalised and without the \enquote*{RZA} superscript,
\begin{align}
  \invI = \frac{1}{\initial{H}} \frac{\dot{\CJ}}{\CJ}\,,
\end{align}
and
Eqs~\eqref{e-defn-aD-VD}, \eqref{e-defn-calI-average}, \eqref{e-Riem-vs-Lag-average}, and \eqref{e-J-vs-CJ},
we can write
\begin{align}
  \average{\invI}
  &= \frac{\initaverageriem{\invI J}}{\initaverageriem{J}} 
  \;
   = \frac{\initaverageriem{\invI J}}{a_{\CD}^3} 
  \;
  = \frac{\initaverageriem{a^3 \dot{\CJ}}}{\initial{H} a_{\CD}^3 } 
  \;
  = \frac{\initaverageriem{\dot{\CJ}}}{\initial{H} \initaverageriem{\CJ}} \,.
  \label{e-ID-is-Jdot-J}
\end{align}

We are now ready to evaluate the curvature using Eqs~(13) and (53) of \cite{BuchRZA2},
together with the flatness of the EdS reference model, which give
\begin{align}
  \average{\CR} &=
  -\, \left[
    \frac{\initaverageriem{\ddot{\CJ}}}{\initaverageriem{{\CJ}}} +
    3 \left( \frac{\ddot{\xi}}{\dot{\xi}} + 4 \frac{\dot{a}}{a} \right)
    \frac{\initaverageriem{\dot{\CJ}}}{\initaverageriem{{\CJ}}}
    \right] \,.
\end{align}
Using Eq.~\eqref{e-replace-CJ-ddot},
the EdS normalised, zero-pointed growth function
(Eq.~\eqref{e-xi-EdS}),
and the EdS
deceleration parameter $q := -{a\,\ddot{a}}/{\dot{a}^2} = 1/2$,
we rewrite this as
\begin{align}
  \average{\CR} &=
  -\,
  \frac{\initaverageriem{\dot{\CJ}}}{\initaverageriem{{\CJ}}}
  \left( 4 \frac{\ddot{\xi}}{\dot{\xi}} + 12 \frac{\dot{a}}{a} \right)
  \nonumber \\
  &
  =
  -\,
  \frac{\initaverageriem{\dot{\CJ}}}{\initaverageriem{{\CJ}}}
  \left( 4 \frac{\ddot{a}}{\dot{a}} + 12 \frac{\dot{a}}{a} \right)
  \nonumber \\
  &=
  -10 H\,
  \frac{\initaverageriem{\dot{\CJ}}}{\initaverageriem{{\CJ}}}
  \,.
  \label{e-RD-minus10H-Jdot-J}
\end{align}
The turnaround condition Eq.~\eqref{e-TA-condn}, together with
Eq.~\eqref{e-ID-is-Jdot-J}, yield
\begin{align}
  \average{\CR} &= 30 H^2 \,.
\end{align}
The curvature functional (Eq.~\eqref{e-defn-OmRD}) is thus
\begin{align}
  \OmRD &= -5 \alpha^2
  \,,
  \label{e-TA-OmRD-minus5}
\end{align}
where we have defined $\alpha$ as the ratio of
the reference
model expansion rate $H$ to the effective expansion rate $\Heff$,
\begin{align}
  \alpha := \frac{H}{\Heff} \,.
  \label{e-defn-alpha}
\end{align}

An equivalent expression to Eq.~\eqref{e-TA-OmRD-minus5}
can be obtained for the
{\LCDM} case by combining the first line of
Eq.~\eqref{e-RD-minus10H-Jdot-J}, an appropriate expression for the
{\LCDM} growth rate and its first and second time derivatives,
and again equating the turnaround definition
(Eq.~\eqref{e-TA-condn}) to the rightmost expression in
Eq.~\eqref{e-ID-is-Jdot-J}, yielding
\begin{align}
  \OmRD &= -\left(\frac{2}{H}\frac{\ddot{\xi}}{\dot{\xi}} + 6\right) \alpha^2
  \,.
  \label{e-TA-OmRD-LCDM}
\end{align}
We thus define a parameter for testing numerical calculations,
\begin{align}
  \OmRDPS &:= \frac{\OmRD}{\alpha^2} + \frac{2}{H} \frac{\ddot{\xi}}{\dot{\xi}}\,,
  \label{e-OmRDTA-planesym}
\end{align}
which in this special case should have the value $\OmRDPS = -6$ and
for the almost-EdS case can be evaluated as $\OmRDPS = \OmRD/\alpha^2 -1.$

The other main cosmological functionals at turnaround
follow almost immediately. The globally normalised
$\Lambda$-free version
of the Hamiltonian constraint can be written using
\citep[][Eqs~(18), (19)]{WiegBuch10}, the definition
of $H_{\CD}$ in Eq.~\eqref{e-defn-HD}, and by multiplying by
$\left(H_{\CD}/\Heff\right)^2$, as
\begin{equation}
  \OmmD +
  \OmRD +
  \OmQD = \alpha^2\,\frac{H_{\CD}^2}{H^2} \,,
  \label{e-hamilton-Omegas-glob-normalised}
\end{equation}
where $\OmmD := 8\pi G \average{\rho} /(3 \Heff^2) $
is the matter density functional, and
\begin{align}
  \OmQD := -\CQ_{\CD}/(6 \Heff^2)
  \label{e-defn-OmQD}
\end{align}
is the kinematical backreaction
(Eq.~\eqref{e-defn-Q})
functional. Since the condition in Eq.~\eqref{e-TA-condn} is equivalent to $H_{\CD} = 0$,
a more strict derivation of Eq.~\eqref{e-hamilton-Omegas-glob-normalised}
would divide by $\Heff^2$ directly instead of using the intervening step of
division by $H_{\CD}^2$.

For an EdS plus structure formation model to match
observations, $\alpha$ must decrease from about unity at an early epoch
to $\alpha = 47.24/67.74 \approx 0.679$ at the current epoch
for Planck 2015 estimates \citep[][\SSS{}3, Eqs~(12)]{RMBO16Hbg1},
which satisfies
the order of magnitude scalar-averaging key values
listed in Eq.~(16) of \cite{RMBO16Hbg1}.
Thus, observationally realistic bounds from the CMB epoch
to the present are $1 \gtapprox \alpha^2 \gtapprox 0.5$, respectively,
for the almost-EdS case.
{\LCDM} values of $\alpha^2$ are likely to deviate only a few percent
from unity \citep[e.g.][]{Bolejko17styro,Bolejko17styroLCDM}.

\tnonnegOmRDTA

At turnaround for the EdS case, using Eq.~\eqref{e-TA-condn}, or equivalently
$H_{\CD} = 0$, Eq.~\eqref{e-hamilton-Omegas-glob-normalised} becomes
\begin{equation}
  \OmmD +
  \OmRD +
  \OmQD = 0 \,.
  \label{e-hamilton-TA}
\end{equation}

We can now rewrite the kinematical backreaction, defined above
in Eq.~\eqref{e-defn-Q}, as in \cite[][Eq.~(11)]{BuchRZA2},
\begin{align}
  \CQ_{\CD}
  &= \initial{H}^2 \left[ 2 \average{\invII} - \frac{2}{3} \average{\invI}^2 \right]
  \;
  = - \frac{2}{3} \initial{H}^2 \average{\invI}^2
  \;
  = -6 H^2,
  \label{e-evaluate-Q}
\end{align}
giving $\OmQD = \alpha^2$,
and thus, in summary, a triplet of EdS critical
values for the turnaround epoch
of a domain with zero second
and third initial average invariants,
\begin{align}
    \OmRD = -5\alpha^2,\;\; \OmQD = \alpha^2,\;\; \OmmD = 4\alpha^2\,.
    \label{e-TA-critical-values}
\end{align}

By equating the turnaround definition,
Eq.~\eqref{e-TA-condn}, and the rightmost expression
in Eq.~\eqref{e-ID-is-Jdot-J}, and using
the EdS growth function (Eq.~\eqref{e-xi-EdS})
together with Eq.~\eqref{e-CJ-derivs}, it follows
that
\begin{align}
  \dot{a} \, \initaverageriem{\initial{\invI}} &=
  -3H \left[ \initial{a} + \left(a - \initial{a}\right)
    \initaverageriem{\initial{\invI}}\right]\,.
\end{align}
This yields expressions for the turnaround epoch
in terms of the EdS scale factor
and cosmological time as a function of the
averaged initial invariant
(see \cite{VignBuch19} for a similar derivation of $\aturnaround$):
\begin{align}
  \aturnaround &= \frac{3 \initial{a}}{4}
      \left( 1 - \frac{1}{\initaverageriem{\initial{\invI}}} \right) \,,
  \quad
  \tturnaround = \initial{t} \left( \frac{\aturnaround}{\initial{a}} \right)^{3/2} \,.
  \label{e-aTA-tTA-subcase}
\end{align}
Using Eqs~\eqref{e-J-vs-CJ} and \eqref{e-CJ-derivs} to rewrite $a_{\CD}$ and differentiating,
it follows that $\dot{a}_{\CD}\left(\tturnaround\right) = 0$, confirming turnaround in the
averaged sense. The reference model expansion rate at turnaround
can now be written
\begin{align}
  \Hturnaround^2 &= \frac{256 \, \initaverageriem{\initial{\invI}}^3}{243 \,\initial{t}^2
    \left(\initaverageriem{\initial{\invI}}-1\right)^3}
  \,.
\end{align}

\subsubsection{QZA numerical simulations}
\label{s-results-EdS-LCDM-subsub}

As described in \SSS\ref{s-method-EdS-LCDM}, we ran QZA simulations for
the almost-EdS and almost-{\LCDM} cases.
These required defining,
as in \cite[][\SSS{}3.2]{Roukema17silvir},
the FLRW
comoving side length of the 3-torus fundamental domain $\Lbox$
(loosely called the \enquote*{box size}), the initial \enquote*{interparticle} separation
$\LN := \Lbox / N^{1/3}$ for an initial condition set of $N$ particles,
the cell size $\LDTFE$ for estimating the initial extrinsic curvature invariants needed for
the $\CQ_{\CD}$ evolution equation with the {\DTFE} library, and $\LD$, the
current size of an averaging domain, all in effective (global average)
comoving units (see \cite{RMBO16Hbg1} for discussion of 10\%-level effects on $\aeff$ between
an effective model and FLRW models).
We considered
averaging scales covering typical cosmic web scales,
$\LD = 2.5, 10,$ and 40~Mpc$/\hzeroeff$, where $\hzeroeff = 0.6774$ is
the Planck 2015 \citep[][Table 4, final column]{Planck2015cosmparam}
normalised Hubble--Lema\^{i}tre constant.

Figures~\ref{f-OmRDEdS} and \ref{f-OmRDLCDM} show the
curvature functional $\OmRD$.
If an FLRW model is interpreted literally, then $\OmRD$
is zero by assumption: $\OmRD \equiv \Omk = 0$ in an EdS or {\LCDM}
model, and as was shown above, the overdensities cannot, at least pointwise,
slow down to reach their turnaround epoch.
For a more realistic, almost-FLRW model,
Figs~\ref{f-OmRDEdS} and \ref{f-OmRDLCDM} show that both the EdS and {\LCDM}
models require very strong positive curvatures prior to turnaround.
These $\OmRD$--$H_{\CD}$ relations are insensitive to particle resolution,
and the ratios between the fundamental domain size $\Lbox$,
the averaging scale $\LD$,
the DTFE scale $\LDTFE$,
and the particle resolution scale
$\LN$. For example,
smaller QZA simulations with $L_{\mathrm{box}}  = 16 \LD = 2 \LDTFE = 2 \LN,
N=64^3$ look visually almost indistinguishable from Figs~\ref{f-OmRDEdS} and
\ref{f-OmRDLCDM}.

It is clear in these figures that at any reasonable scale at which
overdensities are normally thought to be able to collapse gravitationally,
that is, from 2.5 to 40~Mpc/$\hzeroeff$, most spatial domains pass
through a positive average curvature phase as they approach their turnaround epoch,
that is, for the lowest values of $H_{\CD}$ at the left of the panels.
A striking feature of the plots is that at any given epoch,
the relation between spatial curvature and expansion rate is
mostly quite tight, especially at high expansion rates and early times.
There is increasing scatter in the relation, mostly at lower values of $H_{\CD}$,
$\LD=40$~Mpc/$\hzeroeff$ {\LCDM} panel, in which turnaround is not reached.
In \SSS\ref{s-OmRD-HCD-relation} below,
\postrefereeAAchanges{the ways that the $\OmRD$--$H_{\CD}$ relation could be used numerically or observationally
  are briefly discussed.}

\begin{figure}
  \begin{center}
    \adjustbox{width=0.55\textwidth}{\begingroup
  \makeatletter
  \providecommand\color[2][]{
    \GenericError{(gnuplot) \space\space\space\@spaces}{
      Package color not loaded in conjunction with
      terminal option `colourtext'
    }{See the gnuplot documentation for explanation.
    }{Either use 'blacktext' in gnuplot or load the package
      color.sty in LaTeX.}
    \renewcommand\color[2][]{}
  }
  \providecommand\includegraphics[2][]{
    \GenericError{(gnuplot) \space\space\space\@spaces}{
      Package graphicx or graphics not loaded
    }{See the gnuplot documentation for explanation.
    }{The gnuplot epslatex terminal needs graphicx.sty or graphics.sty.}
    \renewcommand\includegraphics[2][]{}
  }
  \providecommand\rotatebox[2]{#2}
  \@ifundefined{ifGPcolor}{
    \newif\ifGPcolor
    \GPcolorfalse
  }{}
  \@ifundefined{ifGPblacktext}{
    \newif\ifGPblacktext
    \GPblacktexttrue
  }{}
  \let\gplgaddtomacro\g@addto@macro
  \gdef\gplbacktext{}
  \gdef\gplfronttext{}
  \makeatother
  \ifGPblacktext
    \def\colorrgb#1{}
    \def\colorgray#1{}
  \else
    \ifGPcolor
      \def\colorrgb#1{\color[rgb]{#1}}
      \def\colorgray#1{\color[gray]{#1}}
      \expandafter\def\csname LTw\endcsname{\color{white}}
      \expandafter\def\csname LTb\endcsname{\color{black}}
      \expandafter\def\csname LTa\endcsname{\color{black}}
      \expandafter\def\csname LT0\endcsname{\color[rgb]{1,0,0}}
      \expandafter\def\csname LT1\endcsname{\color[rgb]{0,1,0}}
      \expandafter\def\csname LT2\endcsname{\color[rgb]{0,0,1}}
      \expandafter\def\csname LT3\endcsname{\color[rgb]{1,0,1}}
      \expandafter\def\csname LT4\endcsname{\color[rgb]{0,1,1}}
      \expandafter\def\csname LT5\endcsname{\color[rgb]{1,1,0}}
      \expandafter\def\csname LT6\endcsname{\color[rgb]{0,0,0}}
      \expandafter\def\csname LT7\endcsname{\color[rgb]{1,0.3,0}}
      \expandafter\def\csname LT8\endcsname{\color[rgb]{0.5,0.5,0.5}}
    \else
      \def\colorrgb#1{\color{black}}
      \def\colorgray#1{\color[gray]{#1}}
      \expandafter\def\csname LTw\endcsname{\color{white}}
      \expandafter\def\csname LTb\endcsname{\color{black}}
      \expandafter\def\csname LTa\endcsname{\color{black}}
      \expandafter\def\csname LT0\endcsname{\color{black}}
      \expandafter\def\csname LT1\endcsname{\color{black}}
      \expandafter\def\csname LT2\endcsname{\color{black}}
      \expandafter\def\csname LT3\endcsname{\color{black}}
      \expandafter\def\csname LT4\endcsname{\color{black}}
      \expandafter\def\csname LT5\endcsname{\color{black}}
      \expandafter\def\csname LT6\endcsname{\color{black}}
      \expandafter\def\csname LT7\endcsname{\color{black}}
      \expandafter\def\csname LT8\endcsname{\color{black}}
    \fi
  \fi
    \setlength{\unitlength}{0.0200bp}
    \ifx\gptboxheight\undefined
      \newlength{\gptboxheight}
      \newlength{\gptboxwidth}
      \newsavebox{\gptboxtext}
    \fi
    \setlength{\fboxrule}{0.5pt}
    \setlength{\fboxsep}{1pt}
\begin{picture}(11520.00,8640.00)
    \gplgaddtomacro\gplbacktext{
      \colorrgb{0.00,0.00,0.00}
      \put(1116,1858){\makebox(0,0)[r]{\strut{}-4}}
      \colorrgb{0.00,0.00,0.00}
      \put(1116,3269){\makebox(0,0)[r]{\strut{}-2}}
      \colorrgb{0.00,0.00,0.00}
      \put(1116,4680){\makebox(0,0)[r]{\strut{}0}}
      \colorrgb{0.00,0.00,0.00}
      \put(1116,6091){\makebox(0,0)[r]{\strut{}2}}
      \colorrgb{0.00,0.00,0.00}
      \put(1116,7502){\makebox(0,0)[r]{\strut{}4}}
      \colorrgb{0.00,0.00,0.00}
      \put(1332,792){\makebox(0,0){\strut{}0}}
      \colorrgb{0.00,0.00,0.00}
      \put(3240,792){\makebox(0,0){\strut{}2}}
      \colorrgb{0.00,0.00,0.00}
      \put(5148,792){\makebox(0,0){\strut{}4}}
      \colorrgb{0.00,0.00,0.00}
      \put(7055,792){\makebox(0,0){\strut{}6}}
      \colorrgb{0.00,0.00,0.00}
      \put(8963,792){\makebox(0,0){\strut{}8}}
      \colorrgb{0.00,0.00,0.00}
      \put(10871,792){\makebox(0,0){\strut{}10}}
    }
    \gplgaddtomacro\gplfronttext{
      \colorrgb{0.00,0.00,0.00}
      \put(288,4679){\rotatebox{90}{\makebox(0,0){\strut{}$\OmQD$}}}
      \colorrgb{0.00,0.00,0.00}
      \put(6101,252){\makebox(0,0){\strut{}$\OmmD$}}
      \colorrgb{0.00,0.00,0.00}
      \put(6102,7854){\makebox(0,0)[l]{\strut{}$\LD =   2.5$~Mpc/$\hzeroeff$}}
    }
    \gplbacktext
    \put(0,0){\includegraphics[scale=0.4]{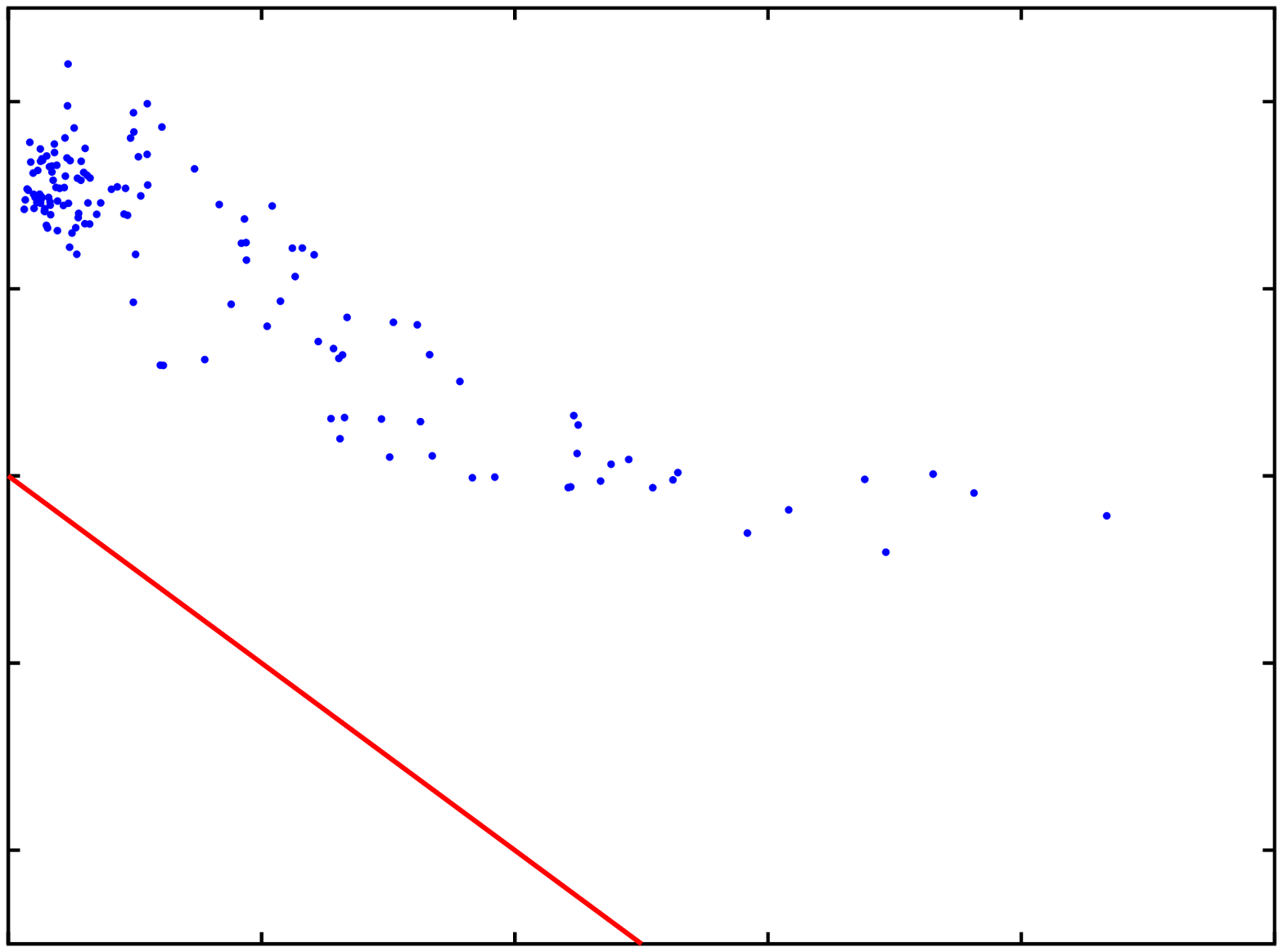}}
    \gplfronttext
  \end{picture}
\endgroup
 }\\
    \adjustbox{width=0.55\textwidth}{\begingroup
  \makeatletter
  \providecommand\color[2][]{
    \GenericError{(gnuplot) \space\space\space\@spaces}{
      Package color not loaded in conjunction with
      terminal option `colourtext'
    }{See the gnuplot documentation for explanation.
    }{Either use 'blacktext' in gnuplot or load the package
      color.sty in LaTeX.}
    \renewcommand\color[2][]{}
  }
  \providecommand\includegraphics[2][]{
    \GenericError{(gnuplot) \space\space\space\@spaces}{
      Package graphicx or graphics not loaded
    }{See the gnuplot documentation for explanation.
    }{The gnuplot epslatex terminal needs graphicx.sty or graphics.sty.}
    \renewcommand\includegraphics[2][]{}
  }
  \providecommand\rotatebox[2]{#2}
  \@ifundefined{ifGPcolor}{
    \newif\ifGPcolor
    \GPcolorfalse
  }{}
  \@ifundefined{ifGPblacktext}{
    \newif\ifGPblacktext
    \GPblacktexttrue
  }{}
  \let\gplgaddtomacro\g@addto@macro
  \gdef\gplbacktext{}
  \gdef\gplfronttext{}
  \makeatother
  \ifGPblacktext
    \def\colorrgb#1{}
    \def\colorgray#1{}
  \else
    \ifGPcolor
      \def\colorrgb#1{\color[rgb]{#1}}
      \def\colorgray#1{\color[gray]{#1}}
      \expandafter\def\csname LTw\endcsname{\color{white}}
      \expandafter\def\csname LTb\endcsname{\color{black}}
      \expandafter\def\csname LTa\endcsname{\color{black}}
      \expandafter\def\csname LT0\endcsname{\color[rgb]{1,0,0}}
      \expandafter\def\csname LT1\endcsname{\color[rgb]{0,1,0}}
      \expandafter\def\csname LT2\endcsname{\color[rgb]{0,0,1}}
      \expandafter\def\csname LT3\endcsname{\color[rgb]{1,0,1}}
      \expandafter\def\csname LT4\endcsname{\color[rgb]{0,1,1}}
      \expandafter\def\csname LT5\endcsname{\color[rgb]{1,1,0}}
      \expandafter\def\csname LT6\endcsname{\color[rgb]{0,0,0}}
      \expandafter\def\csname LT7\endcsname{\color[rgb]{1,0.3,0}}
      \expandafter\def\csname LT8\endcsname{\color[rgb]{0.5,0.5,0.5}}
    \else
      \def\colorrgb#1{\color{black}}
      \def\colorgray#1{\color[gray]{#1}}
      \expandafter\def\csname LTw\endcsname{\color{white}}
      \expandafter\def\csname LTb\endcsname{\color{black}}
      \expandafter\def\csname LTa\endcsname{\color{black}}
      \expandafter\def\csname LT0\endcsname{\color{black}}
      \expandafter\def\csname LT1\endcsname{\color{black}}
      \expandafter\def\csname LT2\endcsname{\color{black}}
      \expandafter\def\csname LT3\endcsname{\color{black}}
      \expandafter\def\csname LT4\endcsname{\color{black}}
      \expandafter\def\csname LT5\endcsname{\color{black}}
      \expandafter\def\csname LT6\endcsname{\color{black}}
      \expandafter\def\csname LT7\endcsname{\color{black}}
      \expandafter\def\csname LT8\endcsname{\color{black}}
    \fi
  \fi
    \setlength{\unitlength}{0.0200bp}
    \ifx\gptboxheight\undefined
      \newlength{\gptboxheight}
      \newlength{\gptboxwidth}
      \newsavebox{\gptboxtext}
    \fi
    \setlength{\fboxrule}{0.5pt}
    \setlength{\fboxsep}{1pt}
\begin{picture}(11520.00,8640.00)
    \gplgaddtomacro\gplbacktext{
      \colorrgb{0.00,0.00,0.00}
      \put(1116,1858){\makebox(0,0)[r]{\strut{}-4}}
      \colorrgb{0.00,0.00,0.00}
      \put(1116,3269){\makebox(0,0)[r]{\strut{}-2}}
      \colorrgb{0.00,0.00,0.00}
      \put(1116,4680){\makebox(0,0)[r]{\strut{}0}}
      \colorrgb{0.00,0.00,0.00}
      \put(1116,6091){\makebox(0,0)[r]{\strut{}2}}
      \colorrgb{0.00,0.00,0.00}
      \put(1116,7502){\makebox(0,0)[r]{\strut{}4}}
      \colorrgb{0.00,0.00,0.00}
      \put(1332,792){\makebox(0,0){\strut{}0}}
      \colorrgb{0.00,0.00,0.00}
      \put(3240,792){\makebox(0,0){\strut{}2}}
      \colorrgb{0.00,0.00,0.00}
      \put(5148,792){\makebox(0,0){\strut{}4}}
      \colorrgb{0.00,0.00,0.00}
      \put(7055,792){\makebox(0,0){\strut{}6}}
      \colorrgb{0.00,0.00,0.00}
      \put(8963,792){\makebox(0,0){\strut{}8}}
      \colorrgb{0.00,0.00,0.00}
      \put(10871,792){\makebox(0,0){\strut{}10}}
    }
    \gplgaddtomacro\gplfronttext{
      \colorrgb{0.00,0.00,0.00}
      \put(288,4679){\rotatebox{90}{\makebox(0,0){\strut{}$\OmQD$}}}
      \colorrgb{0.00,0.00,0.00}
      \put(6101,252){\makebox(0,0){\strut{}$\OmmD$}}
      \colorrgb{0.00,0.00,0.00}
      \put(6102,7854){\makebox(0,0)[l]{\strut{}$\LD =  10.0$~Mpc/$\hzeroeff$}}
    }
    \gplbacktext
    \put(0,0){\includegraphics[scale=0.4]{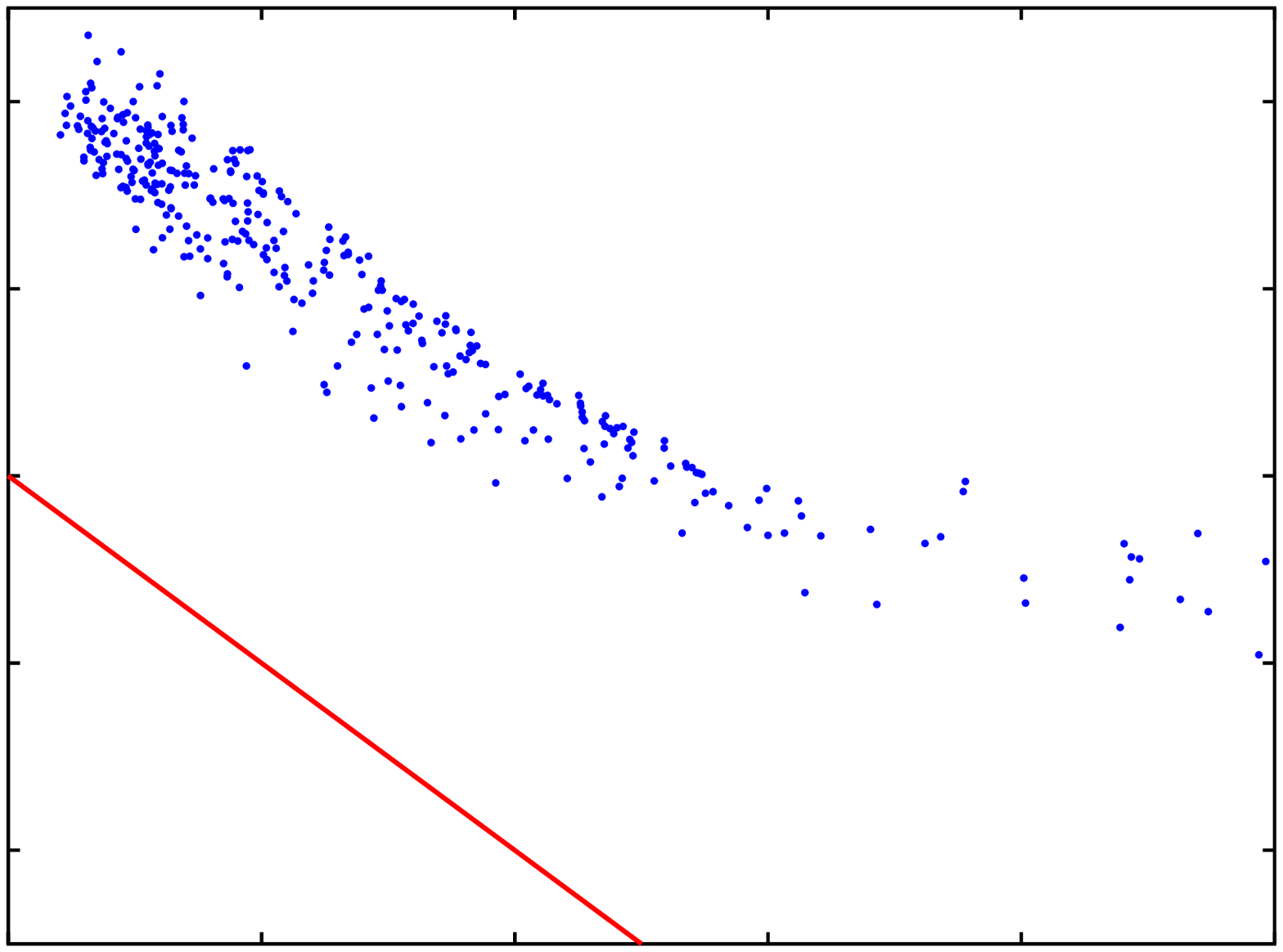}}
    \gplfronttext
  \end{picture}
\endgroup
 }\\
    \adjustbox{width=0.55\textwidth}{\begingroup
  \makeatletter
  \providecommand\color[2][]{
    \GenericError{(gnuplot) \space\space\space\@spaces}{
      Package color not loaded in conjunction with
      terminal option `colourtext'
    }{See the gnuplot documentation for explanation.
    }{Either use 'blacktext' in gnuplot or load the package
      color.sty in LaTeX.}
    \renewcommand\color[2][]{}
  }
  \providecommand\includegraphics[2][]{
    \GenericError{(gnuplot) \space\space\space\@spaces}{
      Package graphicx or graphics not loaded
    }{See the gnuplot documentation for explanation.
    }{The gnuplot epslatex terminal needs graphicx.sty or graphics.sty.}
    \renewcommand\includegraphics[2][]{}
  }
  \providecommand\rotatebox[2]{#2}
  \@ifundefined{ifGPcolor}{
    \newif\ifGPcolor
    \GPcolorfalse
  }{}
  \@ifundefined{ifGPblacktext}{
    \newif\ifGPblacktext
    \GPblacktexttrue
  }{}
  \let\gplgaddtomacro\g@addto@macro
  \gdef\gplbacktext{}
  \gdef\gplfronttext{}
  \makeatother
  \ifGPblacktext
    \def\colorrgb#1{}
    \def\colorgray#1{}
  \else
    \ifGPcolor
      \def\colorrgb#1{\color[rgb]{#1}}
      \def\colorgray#1{\color[gray]{#1}}
      \expandafter\def\csname LTw\endcsname{\color{white}}
      \expandafter\def\csname LTb\endcsname{\color{black}}
      \expandafter\def\csname LTa\endcsname{\color{black}}
      \expandafter\def\csname LT0\endcsname{\color[rgb]{1,0,0}}
      \expandafter\def\csname LT1\endcsname{\color[rgb]{0,1,0}}
      \expandafter\def\csname LT2\endcsname{\color[rgb]{0,0,1}}
      \expandafter\def\csname LT3\endcsname{\color[rgb]{1,0,1}}
      \expandafter\def\csname LT4\endcsname{\color[rgb]{0,1,1}}
      \expandafter\def\csname LT5\endcsname{\color[rgb]{1,1,0}}
      \expandafter\def\csname LT6\endcsname{\color[rgb]{0,0,0}}
      \expandafter\def\csname LT7\endcsname{\color[rgb]{1,0.3,0}}
      \expandafter\def\csname LT8\endcsname{\color[rgb]{0.5,0.5,0.5}}
    \else
      \def\colorrgb#1{\color{black}}
      \def\colorgray#1{\color[gray]{#1}}
      \expandafter\def\csname LTw\endcsname{\color{white}}
      \expandafter\def\csname LTb\endcsname{\color{black}}
      \expandafter\def\csname LTa\endcsname{\color{black}}
      \expandafter\def\csname LT0\endcsname{\color{black}}
      \expandafter\def\csname LT1\endcsname{\color{black}}
      \expandafter\def\csname LT2\endcsname{\color{black}}
      \expandafter\def\csname LT3\endcsname{\color{black}}
      \expandafter\def\csname LT4\endcsname{\color{black}}
      \expandafter\def\csname LT5\endcsname{\color{black}}
      \expandafter\def\csname LT6\endcsname{\color{black}}
      \expandafter\def\csname LT7\endcsname{\color{black}}
      \expandafter\def\csname LT8\endcsname{\color{black}}
    \fi
  \fi
    \setlength{\unitlength}{0.0200bp}
    \ifx\gptboxheight\undefined
      \newlength{\gptboxheight}
      \newlength{\gptboxwidth}
      \newsavebox{\gptboxtext}
    \fi
    \setlength{\fboxrule}{0.5pt}
    \setlength{\fboxsep}{1pt}
\begin{picture}(11520.00,8640.00)
    \gplgaddtomacro\gplbacktext{
      \colorrgb{0.00,0.00,0.00}
      \put(1116,1858){\makebox(0,0)[r]{\strut{}-4}}
      \colorrgb{0.00,0.00,0.00}
      \put(1116,3269){\makebox(0,0)[r]{\strut{}-2}}
      \colorrgb{0.00,0.00,0.00}
      \put(1116,4680){\makebox(0,0)[r]{\strut{}0}}
      \colorrgb{0.00,0.00,0.00}
      \put(1116,6091){\makebox(0,0)[r]{\strut{}2}}
      \colorrgb{0.00,0.00,0.00}
      \put(1116,7502){\makebox(0,0)[r]{\strut{}4}}
      \colorrgb{0.00,0.00,0.00}
      \put(1332,792){\makebox(0,0){\strut{}0}}
      \colorrgb{0.00,0.00,0.00}
      \put(3240,792){\makebox(0,0){\strut{}2}}
      \colorrgb{0.00,0.00,0.00}
      \put(5148,792){\makebox(0,0){\strut{}4}}
      \colorrgb{0.00,0.00,0.00}
      \put(7055,792){\makebox(0,0){\strut{}6}}
      \colorrgb{0.00,0.00,0.00}
      \put(8963,792){\makebox(0,0){\strut{}8}}
      \colorrgb{0.00,0.00,0.00}
      \put(10871,792){\makebox(0,0){\strut{}10}}
    }
    \gplgaddtomacro\gplfronttext{
      \colorrgb{0.00,0.00,0.00}
      \put(288,4679){\rotatebox{90}{\makebox(0,0){\strut{}$\OmQD$}}}
      \colorrgb{0.00,0.00,0.00}
      \put(6101,252){\makebox(0,0){\strut{}$\OmmD$}}
      \colorrgb{0.00,0.00,0.00}
      \put(6102,7854){\makebox(0,0)[l]{\strut{}$\LD =  40.0$~Mpc/$\hzeroeff$}}
    }
    \gplbacktext
    \put(0,0){\includegraphics[scale=0.4]{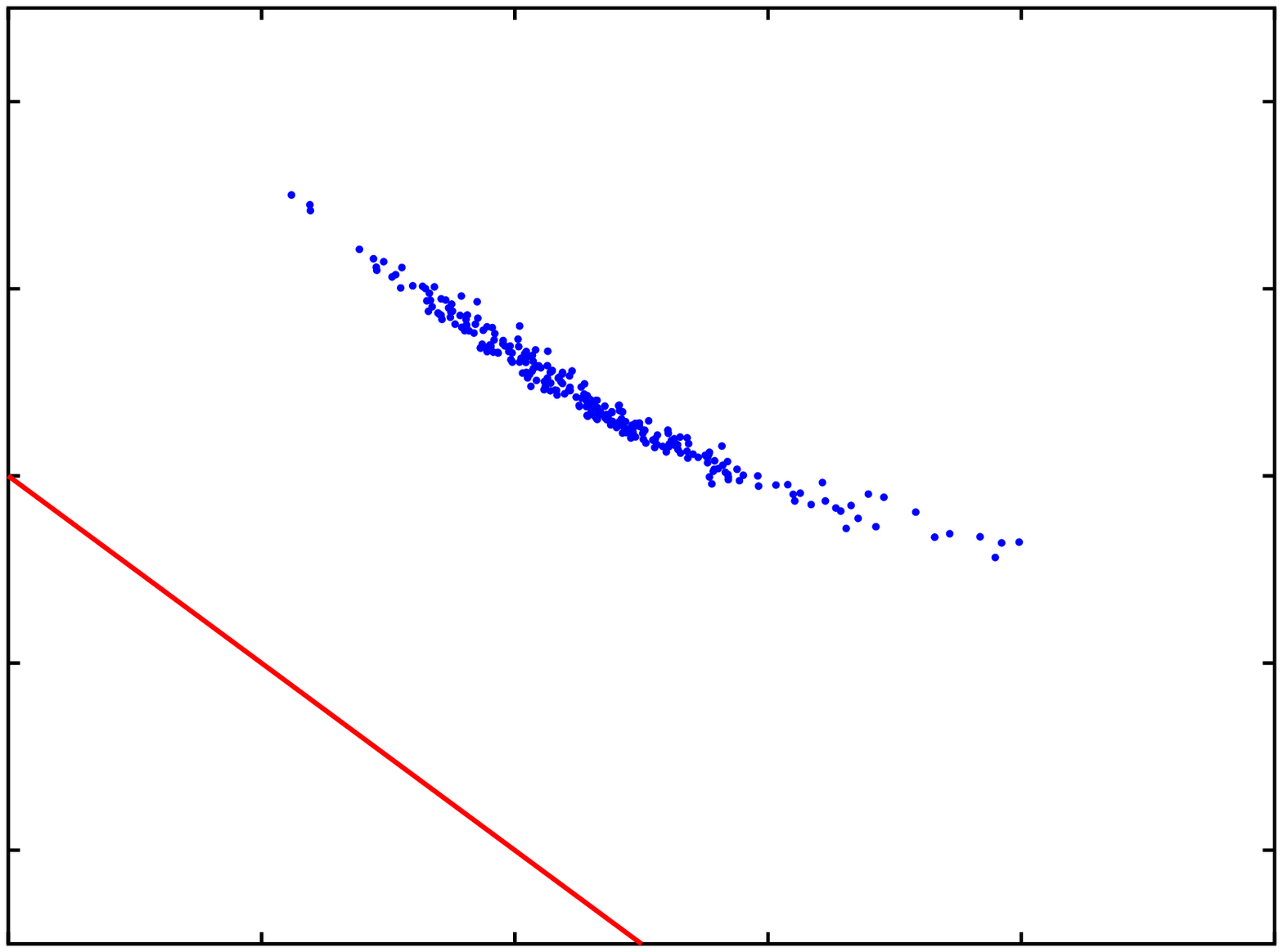}}
    \gplfronttext
  \end{picture}
\endgroup
 }
  \end{center}
  \caption{Averaged kinematical backreaction functional $\OmQD$ versus
    matter density functional $\OmmD$, for domains in which the expansion
    parameter $H_D$ at the final time pre-turnaround time step satisfies
    $|H_D| < 1$~km/s/Mpc,
    for an almost-EdS model,
    {\toptobottom} for
    $\LD = 2.5, 10, 40$~Mpc/$\hzeroeff$, respectively.
    A red line indicates $\OmmD+\OmQD=0$. The region below/left of this
    line correspond to negative curvature ($\OmRD > 0$). The distributions
    of $(\OmmD,\OmQD)$ pairs are clearly separated from the negative curvature
    region.
    \label{f-OmmOmQ-EdS}}
\end{figure} 
 
\begin{figure}
  \begin{center}
    \adjustbox{width=0.55\textwidth}{\begingroup
  \makeatletter
  \providecommand\color[2][]{
    \GenericError{(gnuplot) \space\space\space\@spaces}{
      Package color not loaded in conjunction with
      terminal option `colourtext'
    }{See the gnuplot documentation for explanation.
    }{Either use 'blacktext' in gnuplot or load the package
      color.sty in LaTeX.}
    \renewcommand\color[2][]{}
  }
  \providecommand\includegraphics[2][]{
    \GenericError{(gnuplot) \space\space\space\@spaces}{
      Package graphicx or graphics not loaded
    }{See the gnuplot documentation for explanation.
    }{The gnuplot epslatex terminal needs graphicx.sty or graphics.sty.}
    \renewcommand\includegraphics[2][]{}
  }
  \providecommand\rotatebox[2]{#2}
  \@ifundefined{ifGPcolor}{
    \newif\ifGPcolor
    \GPcolorfalse
  }{}
  \@ifundefined{ifGPblacktext}{
    \newif\ifGPblacktext
    \GPblacktexttrue
  }{}
  \let\gplgaddtomacro\g@addto@macro
  \gdef\gplbacktext{}
  \gdef\gplfronttext{}
  \makeatother
  \ifGPblacktext
    \def\colorrgb#1{}
    \def\colorgray#1{}
  \else
    \ifGPcolor
      \def\colorrgb#1{\color[rgb]{#1}}
      \def\colorgray#1{\color[gray]{#1}}
      \expandafter\def\csname LTw\endcsname{\color{white}}
      \expandafter\def\csname LTb\endcsname{\color{black}}
      \expandafter\def\csname LTa\endcsname{\color{black}}
      \expandafter\def\csname LT0\endcsname{\color[rgb]{1,0,0}}
      \expandafter\def\csname LT1\endcsname{\color[rgb]{0,1,0}}
      \expandafter\def\csname LT2\endcsname{\color[rgb]{0,0,1}}
      \expandafter\def\csname LT3\endcsname{\color[rgb]{1,0,1}}
      \expandafter\def\csname LT4\endcsname{\color[rgb]{0,1,1}}
      \expandafter\def\csname LT5\endcsname{\color[rgb]{1,1,0}}
      \expandafter\def\csname LT6\endcsname{\color[rgb]{0,0,0}}
      \expandafter\def\csname LT7\endcsname{\color[rgb]{1,0.3,0}}
      \expandafter\def\csname LT8\endcsname{\color[rgb]{0.5,0.5,0.5}}
    \else
      \def\colorrgb#1{\color{black}}
      \def\colorgray#1{\color[gray]{#1}}
      \expandafter\def\csname LTw\endcsname{\color{white}}
      \expandafter\def\csname LTb\endcsname{\color{black}}
      \expandafter\def\csname LTa\endcsname{\color{black}}
      \expandafter\def\csname LT0\endcsname{\color{black}}
      \expandafter\def\csname LT1\endcsname{\color{black}}
      \expandafter\def\csname LT2\endcsname{\color{black}}
      \expandafter\def\csname LT3\endcsname{\color{black}}
      \expandafter\def\csname LT4\endcsname{\color{black}}
      \expandafter\def\csname LT5\endcsname{\color{black}}
      \expandafter\def\csname LT6\endcsname{\color{black}}
      \expandafter\def\csname LT7\endcsname{\color{black}}
      \expandafter\def\csname LT8\endcsname{\color{black}}
    \fi
  \fi
    \setlength{\unitlength}{0.0200bp}
    \ifx\gptboxheight\undefined
      \newlength{\gptboxheight}
      \newlength{\gptboxwidth}
      \newsavebox{\gptboxtext}
    \fi
    \setlength{\fboxrule}{0.5pt}
    \setlength{\fboxsep}{1pt}
\begin{picture}(11520.00,8640.00)
    \gplgaddtomacro\gplbacktext{
      \colorrgb{0.00,0.00,0.00}
      \put(1116,1858){\makebox(0,0)[r]{\strut{}-4}}
      \colorrgb{0.00,0.00,0.00}
      \put(1116,3269){\makebox(0,0)[r]{\strut{}-2}}
      \colorrgb{0.00,0.00,0.00}
      \put(1116,4680){\makebox(0,0)[r]{\strut{}0}}
      \colorrgb{0.00,0.00,0.00}
      \put(1116,6091){\makebox(0,0)[r]{\strut{}2}}
      \colorrgb{0.00,0.00,0.00}
      \put(1116,7502){\makebox(0,0)[r]{\strut{}4}}
      \colorrgb{0.00,0.00,0.00}
      \put(1332,792){\makebox(0,0){\strut{}0}}
      \colorrgb{0.00,0.00,0.00}
      \put(3240,792){\makebox(0,0){\strut{}2}}
      \colorrgb{0.00,0.00,0.00}
      \put(5148,792){\makebox(0,0){\strut{}4}}
      \colorrgb{0.00,0.00,0.00}
      \put(7055,792){\makebox(0,0){\strut{}6}}
      \colorrgb{0.00,0.00,0.00}
      \put(8963,792){\makebox(0,0){\strut{}8}}
      \colorrgb{0.00,0.00,0.00}
      \put(10871,792){\makebox(0,0){\strut{}10}}
    }
    \gplgaddtomacro\gplfronttext{
      \colorrgb{0.00,0.00,0.00}
      \put(288,4679){\rotatebox{90}{\makebox(0,0){\strut{}$\OmQD$}}}
      \colorrgb{0.00,0.00,0.00}
      \put(6101,252){\makebox(0,0){\strut{}$\OmmD$}}
      \colorrgb{0.00,0.00,0.00}
      \put(6102,7854){\makebox(0,0)[l]{\strut{}$\LD =   2.5$~Mpc/$\hzeroeff$}}
    }
    \gplbacktext
    \put(0,0){\includegraphics[scale=0.4]{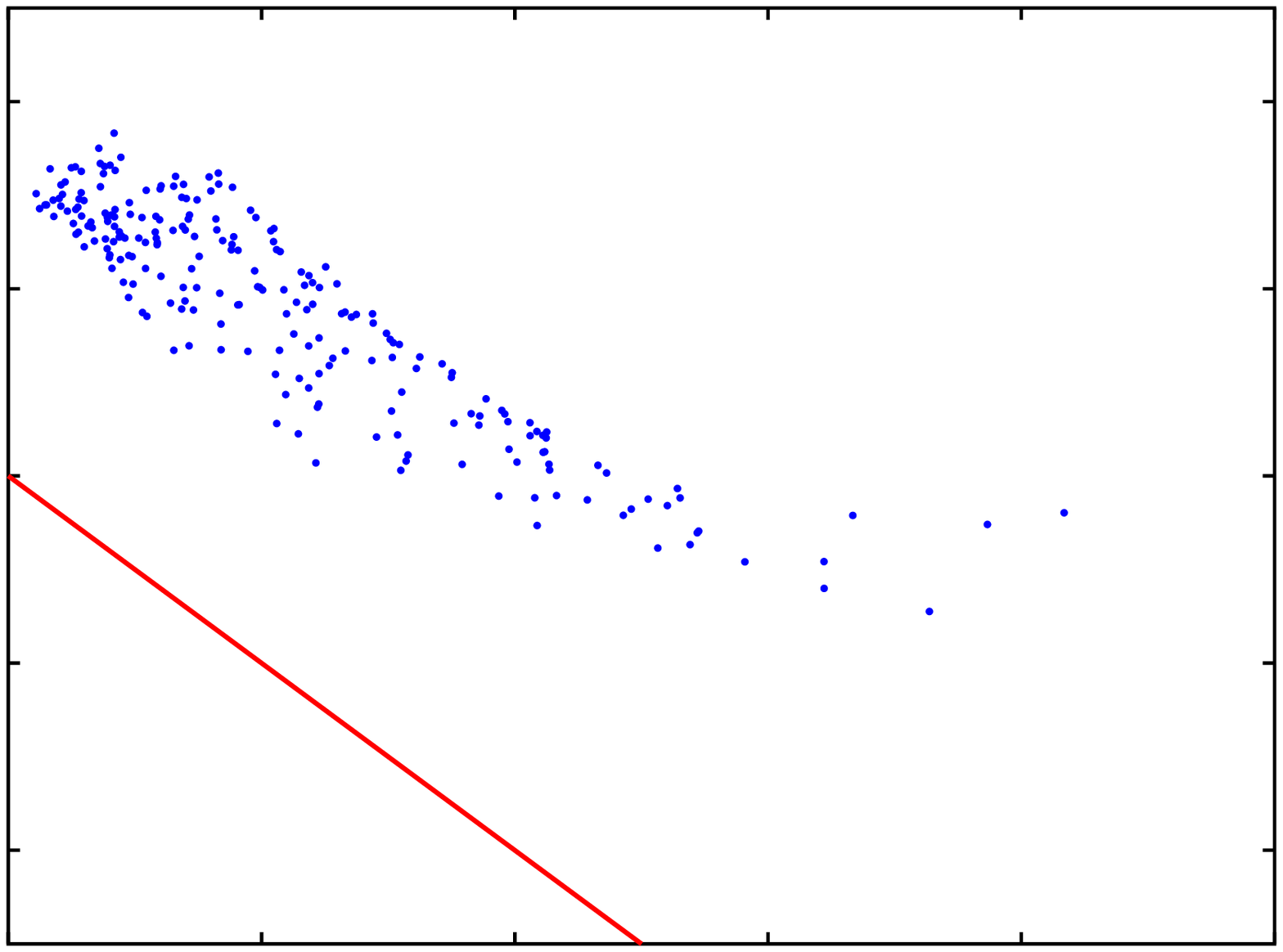}}
    \gplfronttext
  \end{picture}
\endgroup
 }\\
    \adjustbox{width=0.55\textwidth}{\begingroup
  \makeatletter
  \providecommand\color[2][]{
    \GenericError{(gnuplot) \space\space\space\@spaces}{
      Package color not loaded in conjunction with
      terminal option `colourtext'
    }{See the gnuplot documentation for explanation.
    }{Either use 'blacktext' in gnuplot or load the package
      color.sty in LaTeX.}
    \renewcommand\color[2][]{}
  }
  \providecommand\includegraphics[2][]{
    \GenericError{(gnuplot) \space\space\space\@spaces}{
      Package graphicx or graphics not loaded
    }{See the gnuplot documentation for explanation.
    }{The gnuplot epslatex terminal needs graphicx.sty or graphics.sty.}
    \renewcommand\includegraphics[2][]{}
  }
  \providecommand\rotatebox[2]{#2}
  \@ifundefined{ifGPcolor}{
    \newif\ifGPcolor
    \GPcolorfalse
  }{}
  \@ifundefined{ifGPblacktext}{
    \newif\ifGPblacktext
    \GPblacktexttrue
  }{}
  \let\gplgaddtomacro\g@addto@macro
  \gdef\gplbacktext{}
  \gdef\gplfronttext{}
  \makeatother
  \ifGPblacktext
    \def\colorrgb#1{}
    \def\colorgray#1{}
  \else
    \ifGPcolor
      \def\colorrgb#1{\color[rgb]{#1}}
      \def\colorgray#1{\color[gray]{#1}}
      \expandafter\def\csname LTw\endcsname{\color{white}}
      \expandafter\def\csname LTb\endcsname{\color{black}}
      \expandafter\def\csname LTa\endcsname{\color{black}}
      \expandafter\def\csname LT0\endcsname{\color[rgb]{1,0,0}}
      \expandafter\def\csname LT1\endcsname{\color[rgb]{0,1,0}}
      \expandafter\def\csname LT2\endcsname{\color[rgb]{0,0,1}}
      \expandafter\def\csname LT3\endcsname{\color[rgb]{1,0,1}}
      \expandafter\def\csname LT4\endcsname{\color[rgb]{0,1,1}}
      \expandafter\def\csname LT5\endcsname{\color[rgb]{1,1,0}}
      \expandafter\def\csname LT6\endcsname{\color[rgb]{0,0,0}}
      \expandafter\def\csname LT7\endcsname{\color[rgb]{1,0.3,0}}
      \expandafter\def\csname LT8\endcsname{\color[rgb]{0.5,0.5,0.5}}
    \else
      \def\colorrgb#1{\color{black}}
      \def\colorgray#1{\color[gray]{#1}}
      \expandafter\def\csname LTw\endcsname{\color{white}}
      \expandafter\def\csname LTb\endcsname{\color{black}}
      \expandafter\def\csname LTa\endcsname{\color{black}}
      \expandafter\def\csname LT0\endcsname{\color{black}}
      \expandafter\def\csname LT1\endcsname{\color{black}}
      \expandafter\def\csname LT2\endcsname{\color{black}}
      \expandafter\def\csname LT3\endcsname{\color{black}}
      \expandafter\def\csname LT4\endcsname{\color{black}}
      \expandafter\def\csname LT5\endcsname{\color{black}}
      \expandafter\def\csname LT6\endcsname{\color{black}}
      \expandafter\def\csname LT7\endcsname{\color{black}}
      \expandafter\def\csname LT8\endcsname{\color{black}}
    \fi
  \fi
    \setlength{\unitlength}{0.0200bp}
    \ifx\gptboxheight\undefined
      \newlength{\gptboxheight}
      \newlength{\gptboxwidth}
      \newsavebox{\gptboxtext}
    \fi
    \setlength{\fboxrule}{0.5pt}
    \setlength{\fboxsep}{1pt}
\begin{picture}(11520.00,8640.00)
    \gplgaddtomacro\gplbacktext{
      \colorrgb{0.00,0.00,0.00}
      \put(1116,1858){\makebox(0,0)[r]{\strut{}-4}}
      \colorrgb{0.00,0.00,0.00}
      \put(1116,3269){\makebox(0,0)[r]{\strut{}-2}}
      \colorrgb{0.00,0.00,0.00}
      \put(1116,4680){\makebox(0,0)[r]{\strut{}0}}
      \colorrgb{0.00,0.00,0.00}
      \put(1116,6091){\makebox(0,0)[r]{\strut{}2}}
      \colorrgb{0.00,0.00,0.00}
      \put(1116,7502){\makebox(0,0)[r]{\strut{}4}}
      \colorrgb{0.00,0.00,0.00}
      \put(1332,792){\makebox(0,0){\strut{}0}}
      \colorrgb{0.00,0.00,0.00}
      \put(3240,792){\makebox(0,0){\strut{}2}}
      \colorrgb{0.00,0.00,0.00}
      \put(5148,792){\makebox(0,0){\strut{}4}}
      \colorrgb{0.00,0.00,0.00}
      \put(7055,792){\makebox(0,0){\strut{}6}}
      \colorrgb{0.00,0.00,0.00}
      \put(8963,792){\makebox(0,0){\strut{}8}}
      \colorrgb{0.00,0.00,0.00}
      \put(10871,792){\makebox(0,0){\strut{}10}}
    }
    \gplgaddtomacro\gplfronttext{
      \colorrgb{0.00,0.00,0.00}
      \put(288,4679){\rotatebox{90}{\makebox(0,0){\strut{}$\OmQD$}}}
      \colorrgb{0.00,0.00,0.00}
      \put(6101,252){\makebox(0,0){\strut{}$\OmmD$}}
      \colorrgb{0.00,0.00,0.00}
      \put(6102,7854){\makebox(0,0)[l]{\strut{}$\LD =  10.0$~Mpc/$\hzeroeff$}}
    }
    \gplbacktext
    \put(0,0){\includegraphics[scale=0.4]{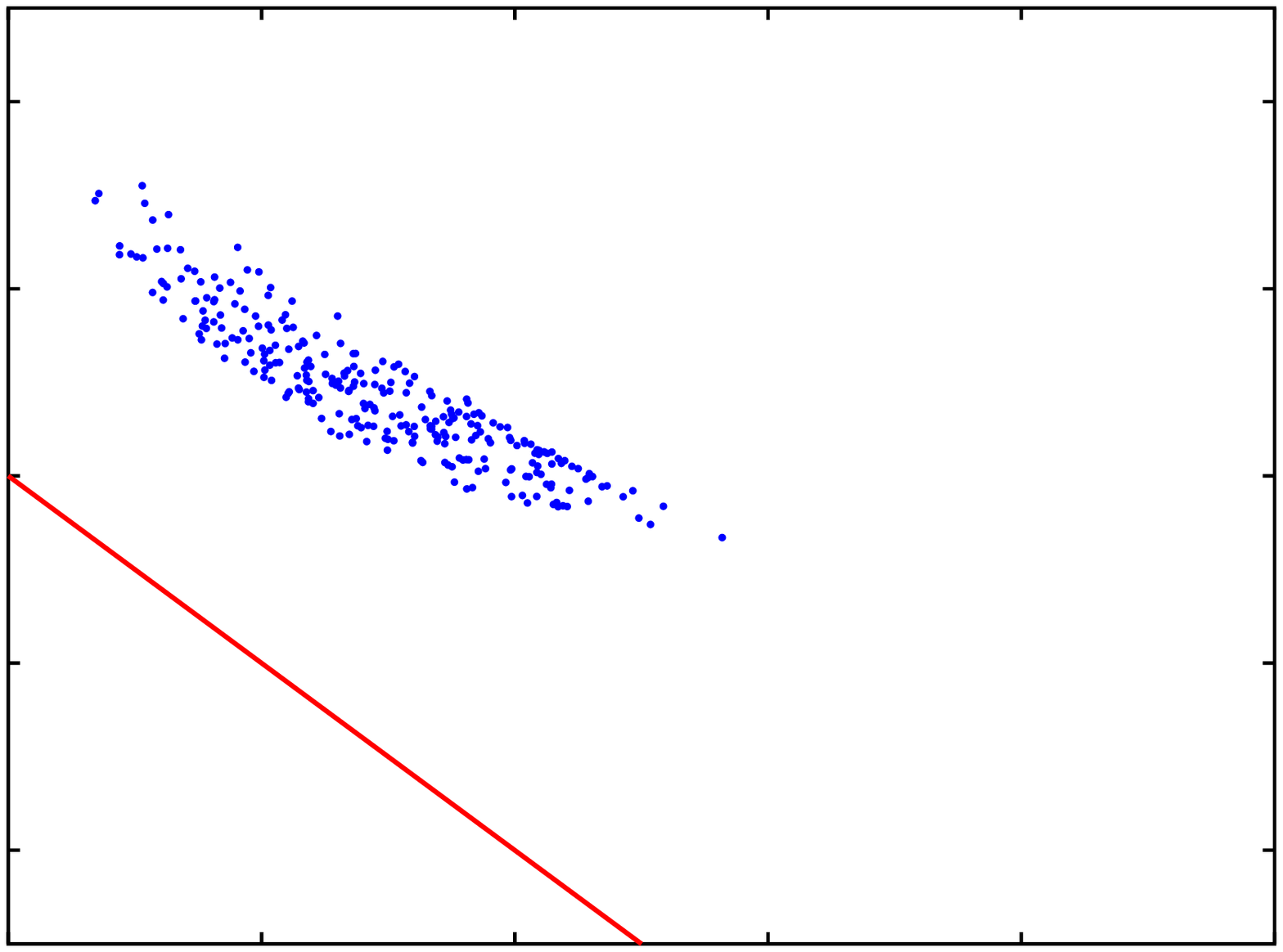}}
    \gplfronttext
  \end{picture}
\endgroup
 }
  \end{center}
  \caption{Averaged kinematical backreaction functional $\OmQD$ versus
    matter density functional $\OmmD$, as in Fig.~\protect\ref{f-OmmOmQ-EdS},
    for an almost-{\LCDM} model,
    $\LD = 2.5$~Mpc/$\hzeroeff$ (top) and $10$~Mpc/$\hzeroeff$ (bottom);
    no turnaround occurs for $\LD = 40$~Mpc/$\hzeroeff$.
    The distributions of $(\OmmD,\OmQD)$ pairs are clearly separated from
    the negative curvature region, which is bounded conservatively
    to lie to the left/below
    the diagonal red line, since $\OmLam^{\CD} \ltapprox 0.7$
    (see Eq.~\protect\eqref{e-hamilton-TA-Omm-OmQ}).
    \label{f-OmmOmQ-LCDM}}
\end{figure} 
 
The curvatures for the domains at low $H_{\CD}$
tend generally to approach the range $\OmRD \sim -5$, which includes
the EdS special case critical value of $\OmRD = -5\alpha^2$,
derived above in Eq.~\eqref{e-TA-OmRD-minus5}.
An exception is the lowest panel of Fig.~\ref{f-OmRDLCDM},
for $\LD=40$~Mpc/$\hzeroeff$, in which strong positive curvatures are not
seen, because turnaround is not reached.
In the simulation shown, the lowest expansion parameter $H_{\CD}$
is $H_{\CD} \sim 30$~km/s/Mpc at $t=13.8$~Gyr.

How close do the turnaround curvatures approach the EdS special case
of $\OmRD = -5\alpha^2$ given in Eq.~\eqref{e-TA-OmRD-minus5}?
Figures~\ref{f-II-OmRD-EdS} and \ref{f-II-OmRD-LCDM} are restricted to
domains shown at the times when they are closest to turnaround, that
is, with absolute values of $H_{\CD}$ selected to satisfy $|H_{\CD}| <
1$~km/s/Mpc.
The EdS special case occurs when the second and third
initial average
invariants of the peculiar-expansion tensor,
$\initavinvII$ and $\initavinvIII$, are
zero. The latter is typically weaker than the former, so here
we only show the behaviour with respect to the former, $\initavinvII$.
Robust linear
fits \citep{Theil50,Sen68}\protect\footnote{See
  also\\\url{https://en.wikipedia.org/w/index.php?oldid=865140889}.}
to the $\OmRD(\average{\invII})$ relation are shown in the figures,
and the zero points $\OmRD(\invII=0)$ and their uncertainties are given
in Table~\ref{t-OmRDTA}.

These $\invII$ zero-point estimates of $\OmRD$ are much closer to the
$\OmRD = -5\alpha^2$ special case than to the flatness assumed in an
exact EdS or {\LCDM} model, keeping in mind that $\alpha^2$ is close
to unity at early epochs and can drop to about $0.5$ at late times in
the EdS case, and close to unity at all times in the {\LCDM} case.
There is some scatter in $\OmRD$ in Figs~\ref{f-II-OmRD-EdS} and
\ref{f-II-OmRD-LCDM}, even when $\average{\invII}$ is close to zero,
in particular due to the role of the third invariant
$\average{\invIII}$, which we have not restricted in this analysis.
The average values and scatter of $\OmRD$, independent of the values
of the initial invariants, are given in Table~\ref{t-OmRDTA-stats}
using both non-robust and robust statistics.  The value $\OmRD = -5$
is not a statistical outlier for any of the three EdS
distributions. Thus, for $\alpha \approx 1$, the EdS $\initial{\invII}
=0, \initial{\invIII} =0$ case gives a turnaround curvature that
characterises the curvature distributions to first order.  The {\LCDM}
distributions should not be expected to be identical to those for the
EdS case, but they are not very different, apart from turnaround not
being achieved for $\LD = 40$~Mpc/$\hzeroeff$.

A more direct comparison with the special analytical case
is provided in Table~\ref{t-OmRDTA-planesym}, showing
zero-point estimates of $\OmRDPS$,
in which the part of $\OmRD$ due to growth function derivatives
is removed (Eq.~\eqref{e-OmRDTA-planesym}). In the largest scale almost-EdS
case and the smallest scale almost-{\LCDM} case, the numerical estimates
of the $\OmRDPS$ zero point are statistically indistinguishable from
the expected value of $-6$, but this is most likely a coincidence,
as indicated by the other three estimates, which are inconsistent with $-6$
by up to about $0.5$ in \enquote*{$\Omega$} dimensionless units.
Given the wide vertical scatter in
Figs~\ref{f-II-OmRD-EdS} and \ref{f-II-OmRD-LCDM},
summarised numerically in Table~\ref{t-OmRDTA-stats}, and
the simple fitting procedure adopted here, it is
unsurprising that the agreement with the expected value is only approximate.
The initial power spectrum is generic, and not intended to directly test
the special $\average{\initial{\invII}} = 0, \average{\initial{\invIII}} = 0$
case.

\begin{figure}
  \begin{center}
    \adjustbox{width=0.55\textwidth}{\begingroup
  \makeatletter
  \providecommand\color[2][]{
    \GenericError{(gnuplot) \space\space\space\@spaces}{
      Package color not loaded in conjunction with
      terminal option `colourtext'
    }{See the gnuplot documentation for explanation.
    }{Either use 'blacktext' in gnuplot or load the package
      color.sty in LaTeX.}
    \renewcommand\color[2][]{}
  }
  \providecommand\includegraphics[2][]{
    \GenericError{(gnuplot) \space\space\space\@spaces}{
      Package graphicx or graphics not loaded
    }{See the gnuplot documentation for explanation.
    }{The gnuplot epslatex terminal needs graphicx.sty or graphics.sty.}
    \renewcommand\includegraphics[2][]{}
  }
  \providecommand\rotatebox[2]{#2}
  \@ifundefined{ifGPcolor}{
    \newif\ifGPcolor
    \GPcolorfalse
  }{}
  \@ifundefined{ifGPblacktext}{
    \newif\ifGPblacktext
    \GPblacktexttrue
  }{}
  \let\gplgaddtomacro\g@addto@macro
  \gdef\gplbacktext{}
  \gdef\gplfronttext{}
  \makeatother
  \ifGPblacktext
    \def\colorrgb#1{}
    \def\colorgray#1{}
  \else
    \ifGPcolor
      \def\colorrgb#1{\color[rgb]{#1}}
      \def\colorgray#1{\color[gray]{#1}}
      \expandafter\def\csname LTw\endcsname{\color{white}}
      \expandafter\def\csname LTb\endcsname{\color{black}}
      \expandafter\def\csname LTa\endcsname{\color{black}}
      \expandafter\def\csname LT0\endcsname{\color[rgb]{1,0,0}}
      \expandafter\def\csname LT1\endcsname{\color[rgb]{0,1,0}}
      \expandafter\def\csname LT2\endcsname{\color[rgb]{0,0,1}}
      \expandafter\def\csname LT3\endcsname{\color[rgb]{1,0,1}}
      \expandafter\def\csname LT4\endcsname{\color[rgb]{0,1,1}}
      \expandafter\def\csname LT5\endcsname{\color[rgb]{1,1,0}}
      \expandafter\def\csname LT6\endcsname{\color[rgb]{0,0,0}}
      \expandafter\def\csname LT7\endcsname{\color[rgb]{1,0.3,0}}
      \expandafter\def\csname LT8\endcsname{\color[rgb]{0.5,0.5,0.5}}
    \else
      \def\colorrgb#1{\color{black}}
      \def\colorgray#1{\color[gray]{#1}}
      \expandafter\def\csname LTw\endcsname{\color{white}}
      \expandafter\def\csname LTb\endcsname{\color{black}}
      \expandafter\def\csname LTa\endcsname{\color{black}}
      \expandafter\def\csname LT0\endcsname{\color{black}}
      \expandafter\def\csname LT1\endcsname{\color{black}}
      \expandafter\def\csname LT2\endcsname{\color{black}}
      \expandafter\def\csname LT3\endcsname{\color{black}}
      \expandafter\def\csname LT4\endcsname{\color{black}}
      \expandafter\def\csname LT5\endcsname{\color{black}}
      \expandafter\def\csname LT6\endcsname{\color{black}}
      \expandafter\def\csname LT7\endcsname{\color{black}}
      \expandafter\def\csname LT8\endcsname{\color{black}}
    \fi
  \fi
    \setlength{\unitlength}{0.0200bp}
    \ifx\gptboxheight\undefined
      \newlength{\gptboxheight}
      \newlength{\gptboxwidth}
      \newsavebox{\gptboxtext}
    \fi
    \setlength{\fboxrule}{0.5pt}
    \setlength{\fboxsep}{1pt}
\begin{picture}(11520.00,8640.00)
    \gplgaddtomacro\gplbacktext{
      \colorrgb{0.00,0.00,0.00}
      \put(2200,1280){\makebox(0,0)[r]{\strut{}-3e-07}}
      \colorrgb{0.00,0.00,0.00}
      \put(2200,2427){\makebox(0,0)[r]{\strut{}-2e-07}}
      \colorrgb{0.00,0.00,0.00}
      \put(2200,3573){\makebox(0,0)[r]{\strut{}-1e-07}}
      \colorrgb{0.00,0.00,0.00}
      \put(2200,4720){\makebox(0,0)[r]{\strut{}0}}
      \colorrgb{0.00,0.00,0.00}
      \put(2200,5866){\makebox(0,0)[r]{\strut{}1e-07}}
      \colorrgb{0.00,0.00,0.00}
      \put(2200,7013){\makebox(0,0)[r]{\strut{}2e-07}}
      \colorrgb{0.00,0.00,0.00}
      \put(2200,8159){\makebox(0,0)[r]{\strut{}3e-07}}
      \colorrgb{0.00,0.00,0.00}
      \put(2440,880){\makebox(0,0){\strut{}$10^{0}$}}
      \colorrgb{0.00,0.00,0.00}
      \put(5226,880){\makebox(0,0){\strut{}$10^{1}$}}
      \colorrgb{0.00,0.00,0.00}
      \put(8013,880){\makebox(0,0){\strut{}$10^{2}$}}
      \colorrgb{0.00,0.00,0.00}
      \put(10799,880){\makebox(0,0){\strut{}$10^{3}$}}
    }
    \gplgaddtomacro\gplfronttext{
      \colorrgb{0.00,0.00,0.00}
      \put(320,4719){\rotatebox{90}{\makebox(0,0){\strut{}$\varepsilon$}}}
      \colorrgb{0.00,0.00,0.00}
      \put(6619,280){\makebox(0,0){\strut{}$H_{\CD}$ (km/s/Mpc)}}
      \colorrgb{0.00,0.00,0.00}
      \put(4855,7793){\makebox(0,0){\footnotesize $ 0.9$~Gyr}}
      \colorrgb{0.00,0.00,0.00}
      \put(4855,7188){\makebox(0,0){\footnotesize $ 4.0$~Gyr}}
      \colorrgb{0.00,0.00,0.00}
      \put(4855,6583){\makebox(0,0){\footnotesize $ 7.0$~Gyr}}
      \colorrgb{0.00,0.00,0.00}
      \put(4855,5978){\makebox(0,0){\footnotesize $10.0$~Gyr}}
      \colorrgb{0.00,0.00,0.00}
      \put(4855,5373){\makebox(0,0){\footnotesize $12.9$~Gyr}}
      \colorrgb{0.00,0.00,0.00}
      \put(2858,1624){\makebox(0,0)[l]{\strut{}$\LD =   2.5$~Mpc/$\hzeroeff$}}
    }
    \gplbacktext
    \put(0,0){\includegraphics[scale=0.4]{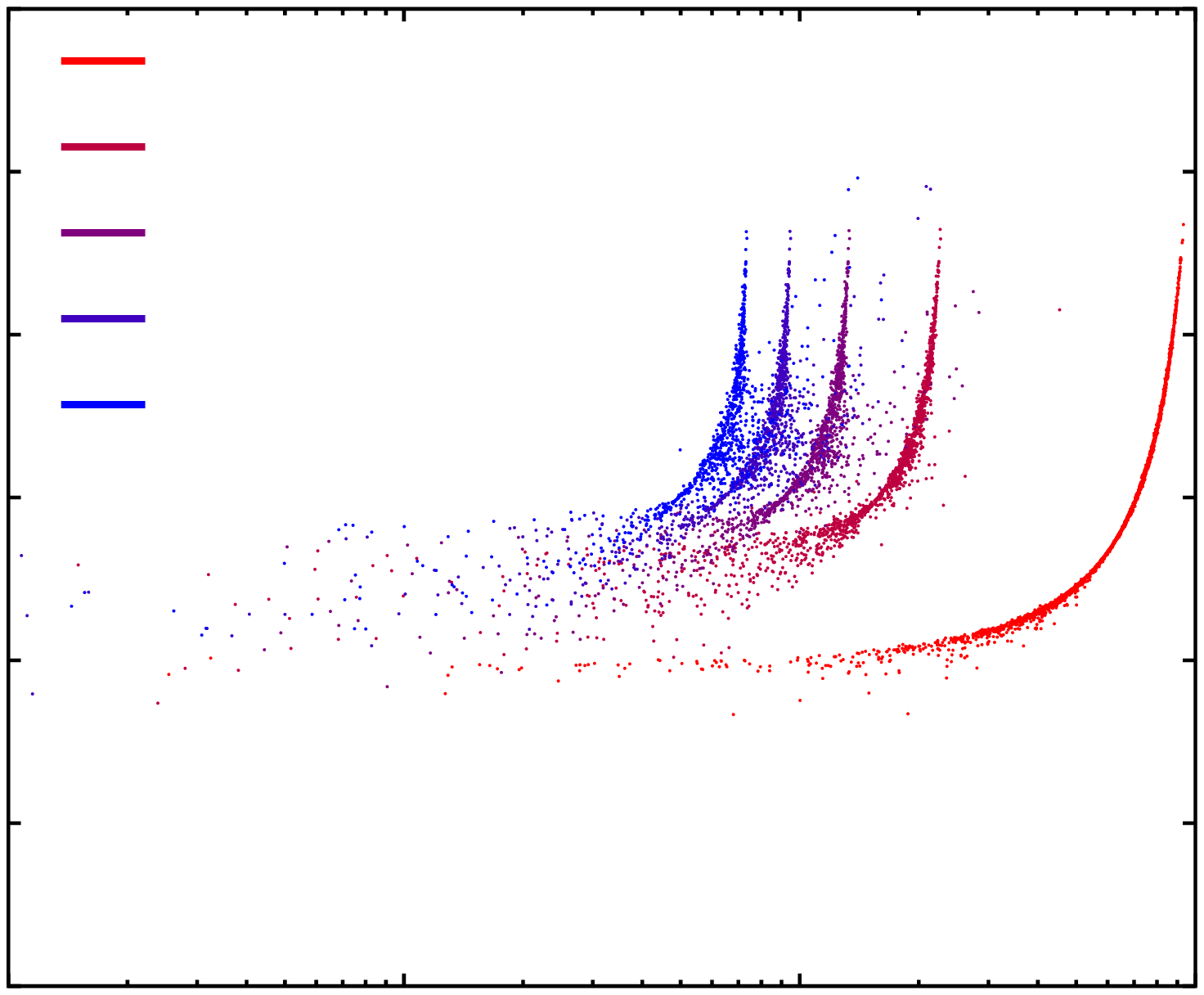}}
    \gplfronttext
  \end{picture}
\endgroup
 }\\
    \adjustbox{width=0.55\textwidth}{\begingroup
  \makeatletter
  \providecommand\color[2][]{
    \GenericError{(gnuplot) \space\space\space\@spaces}{
      Package color not loaded in conjunction with
      terminal option `colourtext'
    }{See the gnuplot documentation for explanation.
    }{Either use 'blacktext' in gnuplot or load the package
      color.sty in LaTeX.}
    \renewcommand\color[2][]{}
  }
  \providecommand\includegraphics[2][]{
    \GenericError{(gnuplot) \space\space\space\@spaces}{
      Package graphicx or graphics not loaded
    }{See the gnuplot documentation for explanation.
    }{The gnuplot epslatex terminal needs graphicx.sty or graphics.sty.}
    \renewcommand\includegraphics[2][]{}
  }
  \providecommand\rotatebox[2]{#2}
  \@ifundefined{ifGPcolor}{
    \newif\ifGPcolor
    \GPcolorfalse
  }{}
  \@ifundefined{ifGPblacktext}{
    \newif\ifGPblacktext
    \GPblacktexttrue
  }{}
  \let\gplgaddtomacro\g@addto@macro
  \gdef\gplbacktext{}
  \gdef\gplfronttext{}
  \makeatother
  \ifGPblacktext
    \def\colorrgb#1{}
    \def\colorgray#1{}
  \else
    \ifGPcolor
      \def\colorrgb#1{\color[rgb]{#1}}
      \def\colorgray#1{\color[gray]{#1}}
      \expandafter\def\csname LTw\endcsname{\color{white}}
      \expandafter\def\csname LTb\endcsname{\color{black}}
      \expandafter\def\csname LTa\endcsname{\color{black}}
      \expandafter\def\csname LT0\endcsname{\color[rgb]{1,0,0}}
      \expandafter\def\csname LT1\endcsname{\color[rgb]{0,1,0}}
      \expandafter\def\csname LT2\endcsname{\color[rgb]{0,0,1}}
      \expandafter\def\csname LT3\endcsname{\color[rgb]{1,0,1}}
      \expandafter\def\csname LT4\endcsname{\color[rgb]{0,1,1}}
      \expandafter\def\csname LT5\endcsname{\color[rgb]{1,1,0}}
      \expandafter\def\csname LT6\endcsname{\color[rgb]{0,0,0}}
      \expandafter\def\csname LT7\endcsname{\color[rgb]{1,0.3,0}}
      \expandafter\def\csname LT8\endcsname{\color[rgb]{0.5,0.5,0.5}}
    \else
      \def\colorrgb#1{\color{black}}
      \def\colorgray#1{\color[gray]{#1}}
      \expandafter\def\csname LTw\endcsname{\color{white}}
      \expandafter\def\csname LTb\endcsname{\color{black}}
      \expandafter\def\csname LTa\endcsname{\color{black}}
      \expandafter\def\csname LT0\endcsname{\color{black}}
      \expandafter\def\csname LT1\endcsname{\color{black}}
      \expandafter\def\csname LT2\endcsname{\color{black}}
      \expandafter\def\csname LT3\endcsname{\color{black}}
      \expandafter\def\csname LT4\endcsname{\color{black}}
      \expandafter\def\csname LT5\endcsname{\color{black}}
      \expandafter\def\csname LT6\endcsname{\color{black}}
      \expandafter\def\csname LT7\endcsname{\color{black}}
      \expandafter\def\csname LT8\endcsname{\color{black}}
    \fi
  \fi
    \setlength{\unitlength}{0.0200bp}
    \ifx\gptboxheight\undefined
      \newlength{\gptboxheight}
      \newlength{\gptboxwidth}
      \newsavebox{\gptboxtext}
    \fi
    \setlength{\fboxrule}{0.5pt}
    \setlength{\fboxsep}{1pt}
\begin{picture}(11520.00,8640.00)
    \gplgaddtomacro\gplbacktext{
      \colorrgb{0.00,0.00,0.00}
      \put(2200,1280){\makebox(0,0)[r]{\strut{}-3e-06}}
      \colorrgb{0.00,0.00,0.00}
      \put(2200,2427){\makebox(0,0)[r]{\strut{}-2e-06}}
      \colorrgb{0.00,0.00,0.00}
      \put(2200,3573){\makebox(0,0)[r]{\strut{}-1e-06}}
      \colorrgb{0.00,0.00,0.00}
      \put(2200,4720){\makebox(0,0)[r]{\strut{}0}}
      \colorrgb{0.00,0.00,0.00}
      \put(2200,5866){\makebox(0,0)[r]{\strut{}1e-06}}
      \colorrgb{0.00,0.00,0.00}
      \put(2200,7012){\makebox(0,0)[r]{\strut{}2e-06}}
      \colorrgb{0.00,0.00,0.00}
      \put(2200,8159){\makebox(0,0)[r]{\strut{}3e-06}}
      \colorrgb{0.00,0.00,0.00}
      \put(2440,880){\makebox(0,0){\strut{}$10^{0}$}}
      \colorrgb{0.00,0.00,0.00}
      \put(5226,880){\makebox(0,0){\strut{}$10^{1}$}}
      \colorrgb{0.00,0.00,0.00}
      \put(8013,880){\makebox(0,0){\strut{}$10^{2}$}}
      \colorrgb{0.00,0.00,0.00}
      \put(10799,880){\makebox(0,0){\strut{}$10^{3}$}}
    }
    \gplgaddtomacro\gplfronttext{
      \colorrgb{0.00,0.00,0.00}
      \put(320,4719){\rotatebox{90}{\makebox(0,0){\strut{}$\varepsilon$}}}
      \colorrgb{0.00,0.00,0.00}
      \put(6619,280){\makebox(0,0){\strut{}$H_{\CD}$ (km/s/Mpc)}}
      \colorrgb{0.00,0.00,0.00}
      \put(4855,7793){\makebox(0,0){\footnotesize $ 1.0$~Gyr}}
      \colorrgb{0.00,0.00,0.00}
      \put(4855,7188){\makebox(0,0){\footnotesize $ 4.1$~Gyr}}
      \colorrgb{0.00,0.00,0.00}
      \put(4855,6583){\makebox(0,0){\footnotesize $ 7.0$~Gyr}}
      \colorrgb{0.00,0.00,0.00}
      \put(4855,5978){\makebox(0,0){\footnotesize $10.0$~Gyr}}
      \colorrgb{0.00,0.00,0.00}
      \put(4855,5373){\makebox(0,0){\footnotesize $12.9$~Gyr}}
      \colorrgb{0.00,0.00,0.00}
      \put(2858,1624){\makebox(0,0)[l]{\strut{}$\LD =  10.0$~Mpc/$\hzeroeff$}}
    }
    \gplbacktext
    \put(0,0){\includegraphics[scale=0.4]{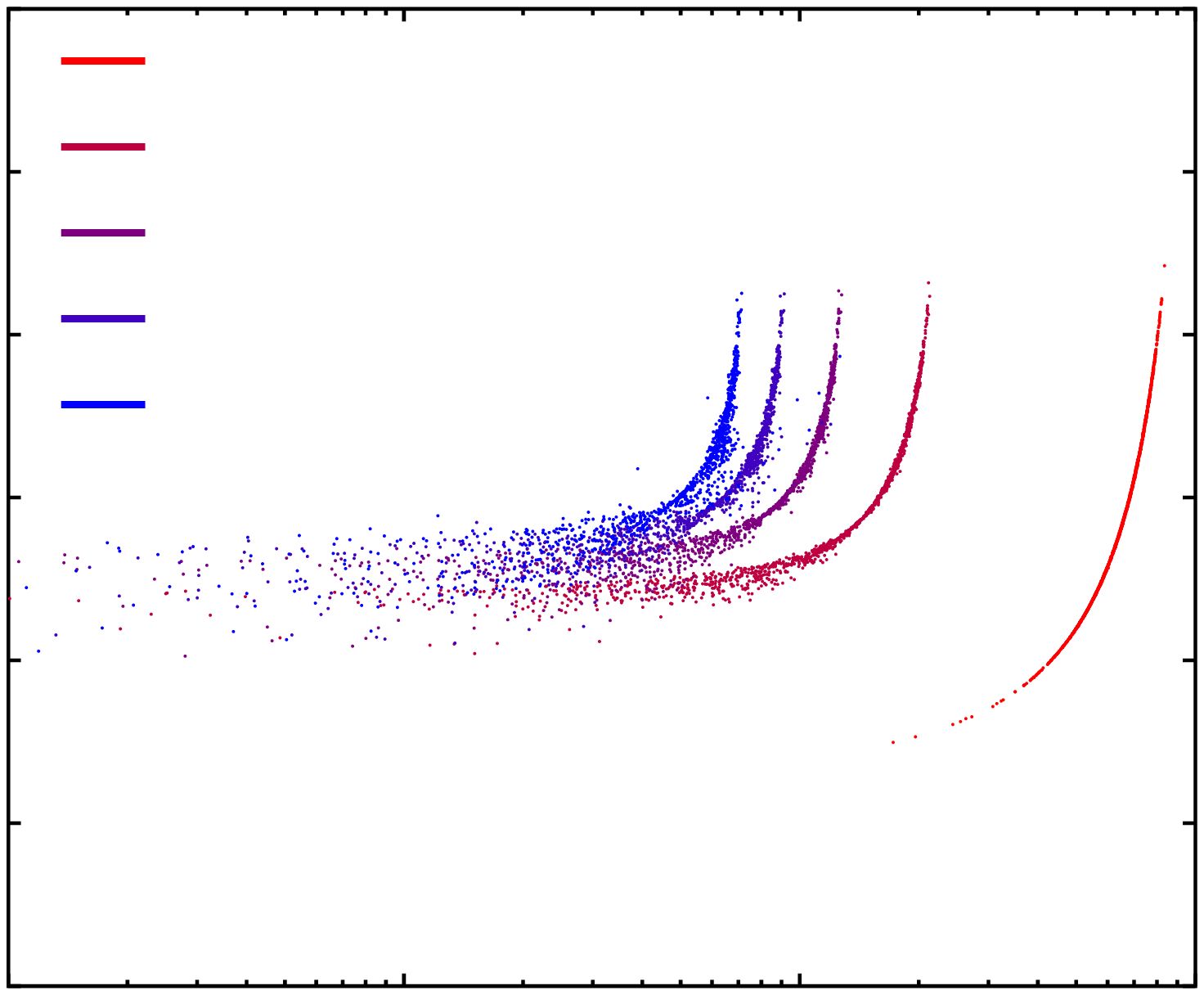}}
    \gplfronttext
  \end{picture}
\endgroup
 }\\
    \adjustbox{width=0.55\textwidth}{\begingroup
  \makeatletter
  \providecommand\color[2][]{
    \GenericError{(gnuplot) \space\space\space\@spaces}{
      Package color not loaded in conjunction with
      terminal option `colourtext'
    }{See the gnuplot documentation for explanation.
    }{Either use 'blacktext' in gnuplot or load the package
      color.sty in LaTeX.}
    \renewcommand\color[2][]{}
  }
  \providecommand\includegraphics[2][]{
    \GenericError{(gnuplot) \space\space\space\@spaces}{
      Package graphicx or graphics not loaded
    }{See the gnuplot documentation for explanation.
    }{The gnuplot epslatex terminal needs graphicx.sty or graphics.sty.}
    \renewcommand\includegraphics[2][]{}
  }
  \providecommand\rotatebox[2]{#2}
  \@ifundefined{ifGPcolor}{
    \newif\ifGPcolor
    \GPcolorfalse
  }{}
  \@ifundefined{ifGPblacktext}{
    \newif\ifGPblacktext
    \GPblacktexttrue
  }{}
  \let\gplgaddtomacro\g@addto@macro
  \gdef\gplbacktext{}
  \gdef\gplfronttext{}
  \makeatother
  \ifGPblacktext
    \def\colorrgb#1{}
    \def\colorgray#1{}
  \else
    \ifGPcolor
      \def\colorrgb#1{\color[rgb]{#1}}
      \def\colorgray#1{\color[gray]{#1}}
      \expandafter\def\csname LTw\endcsname{\color{white}}
      \expandafter\def\csname LTb\endcsname{\color{black}}
      \expandafter\def\csname LTa\endcsname{\color{black}}
      \expandafter\def\csname LT0\endcsname{\color[rgb]{1,0,0}}
      \expandafter\def\csname LT1\endcsname{\color[rgb]{0,1,0}}
      \expandafter\def\csname LT2\endcsname{\color[rgb]{0,0,1}}
      \expandafter\def\csname LT3\endcsname{\color[rgb]{1,0,1}}
      \expandafter\def\csname LT4\endcsname{\color[rgb]{0,1,1}}
      \expandafter\def\csname LT5\endcsname{\color[rgb]{1,1,0}}
      \expandafter\def\csname LT6\endcsname{\color[rgb]{0,0,0}}
      \expandafter\def\csname LT7\endcsname{\color[rgb]{1,0.3,0}}
      \expandafter\def\csname LT8\endcsname{\color[rgb]{0.5,0.5,0.5}}
    \else
      \def\colorrgb#1{\color{black}}
      \def\colorgray#1{\color[gray]{#1}}
      \expandafter\def\csname LTw\endcsname{\color{white}}
      \expandafter\def\csname LTb\endcsname{\color{black}}
      \expandafter\def\csname LTa\endcsname{\color{black}}
      \expandafter\def\csname LT0\endcsname{\color{black}}
      \expandafter\def\csname LT1\endcsname{\color{black}}
      \expandafter\def\csname LT2\endcsname{\color{black}}
      \expandafter\def\csname LT3\endcsname{\color{black}}
      \expandafter\def\csname LT4\endcsname{\color{black}}
      \expandafter\def\csname LT5\endcsname{\color{black}}
      \expandafter\def\csname LT6\endcsname{\color{black}}
      \expandafter\def\csname LT7\endcsname{\color{black}}
      \expandafter\def\csname LT8\endcsname{\color{black}}
    \fi
  \fi
    \setlength{\unitlength}{0.0200bp}
    \ifx\gptboxheight\undefined
      \newlength{\gptboxheight}
      \newlength{\gptboxwidth}
      \newsavebox{\gptboxtext}
    \fi
    \setlength{\fboxrule}{0.5pt}
    \setlength{\fboxsep}{1pt}
\begin{picture}(11520.00,8640.00)
    \gplgaddtomacro\gplbacktext{
      \colorrgb{0.00,0.00,0.00}
      \put(2200,1280){\makebox(0,0)[r]{\strut{}-1e-05}}
      \colorrgb{0.00,0.00,0.00}
      \put(2200,3000){\makebox(0,0)[r]{\strut{}-5e-06}}
      \colorrgb{0.00,0.00,0.00}
      \put(2200,4720){\makebox(0,0)[r]{\strut{}0}}
      \colorrgb{0.00,0.00,0.00}
      \put(2200,6439){\makebox(0,0)[r]{\strut{}5e-06}}
      \colorrgb{0.00,0.00,0.00}
      \put(2200,8159){\makebox(0,0)[r]{\strut{}1e-05}}
      \colorrgb{0.00,0.00,0.00}
      \put(2440,880){\makebox(0,0){\strut{}$10^{0}$}}
      \colorrgb{0.00,0.00,0.00}
      \put(5226,880){\makebox(0,0){\strut{}$10^{1}$}}
      \colorrgb{0.00,0.00,0.00}
      \put(8013,880){\makebox(0,0){\strut{}$10^{2}$}}
      \colorrgb{0.00,0.00,0.00}
      \put(10799,880){\makebox(0,0){\strut{}$10^{3}$}}
    }
    \gplgaddtomacro\gplfronttext{
      \colorrgb{0.00,0.00,0.00}
      \put(320,4719){\rotatebox{90}{\makebox(0,0){\strut{}$\varepsilon$}}}
      \colorrgb{0.00,0.00,0.00}
      \put(6619,280){\makebox(0,0){\strut{}$H_{\CD}$ (km/s/Mpc)}}
      \colorrgb{0.00,0.00,0.00}
      \put(4855,7793){\makebox(0,0){\footnotesize $ 1.1$~Gyr}}
      \colorrgb{0.00,0.00,0.00}
      \put(4855,7188){\makebox(0,0){\footnotesize $ 4.0$~Gyr}}
      \colorrgb{0.00,0.00,0.00}
      \put(4855,6583){\makebox(0,0){\footnotesize $ 6.9$~Gyr}}
      \colorrgb{0.00,0.00,0.00}
      \put(4855,5978){\makebox(0,0){\footnotesize $10.0$~Gyr}}
      \colorrgb{0.00,0.00,0.00}
      \put(4855,5373){\makebox(0,0){\footnotesize $13.0$~Gyr}}
      \colorrgb{0.00,0.00,0.00}
      \put(2858,1624){\makebox(0,0)[l]{\strut{}$\LD =  40.0$~Mpc/$\hzeroeff$}}
    }
    \gplbacktext
    \put(0,0){\includegraphics[scale=0.4]{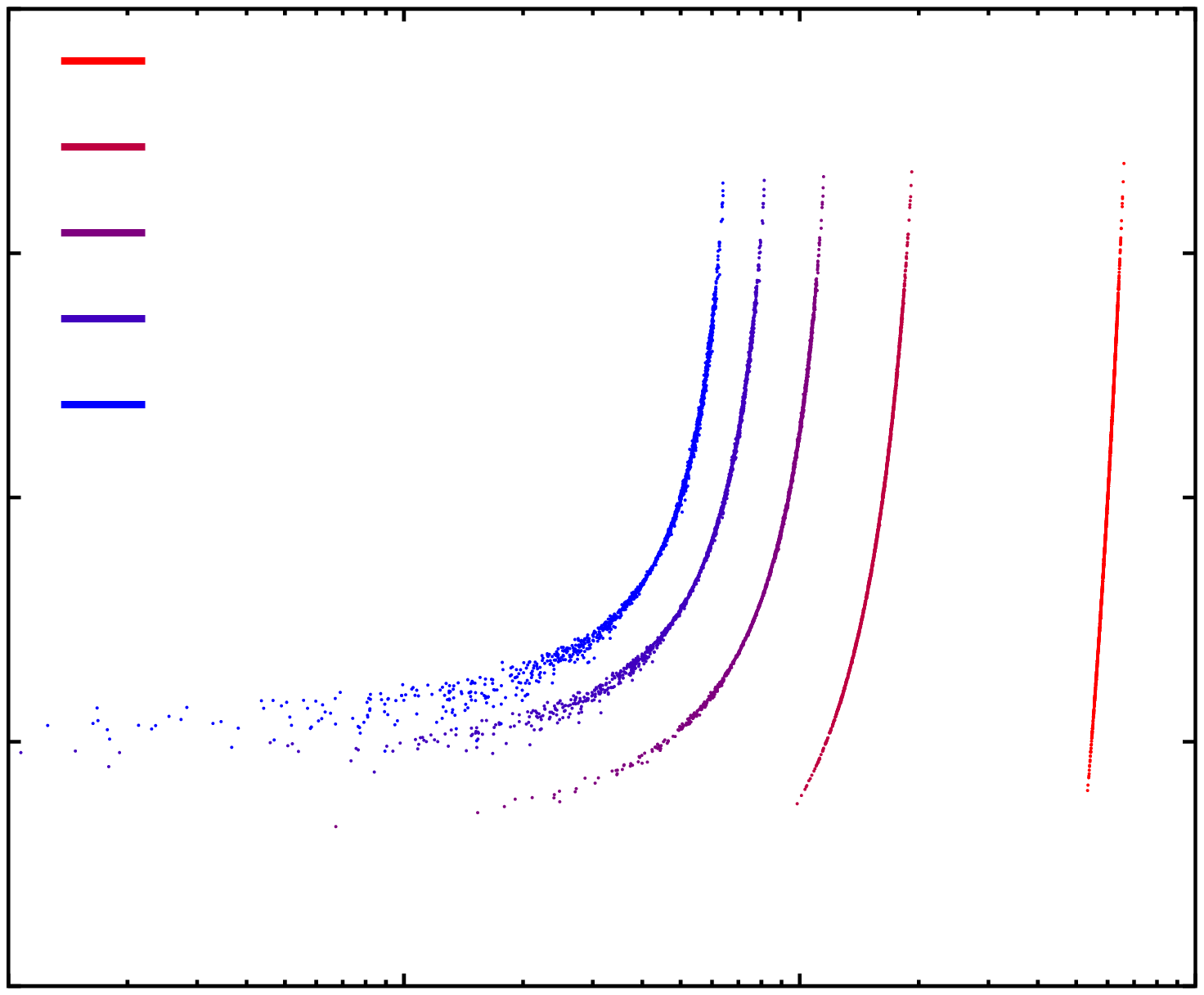}}
    \gplfronttext
  \end{picture}
\endgroup
 }
  \end{center}
  \caption{Curvature deviation parameter $\varepsilon$ as a function of
    expansion rate $H_{\CD}$,
    as in Fig.~\protect\ref{f-OmRDEdS}, for the almost-EdS model.
    The vertical axis ranges vary {\toptobottom} panels
    (different $\LD$ scales); the same ranges are used
    in Fig.~\protect\ref{f-epsilonLCDM}.
    \label{f-epsilonEdS}}
\end{figure} 
 
\begin{figure}
  \begin{center}
    \adjustbox{width=0.55\textwidth}{\begingroup
  \makeatletter
  \providecommand\color[2][]{
    \GenericError{(gnuplot) \space\space\space\@spaces}{
      Package color not loaded in conjunction with
      terminal option `colourtext'
    }{See the gnuplot documentation for explanation.
    }{Either use 'blacktext' in gnuplot or load the package
      color.sty in LaTeX.}
    \renewcommand\color[2][]{}
  }
  \providecommand\includegraphics[2][]{
    \GenericError{(gnuplot) \space\space\space\@spaces}{
      Package graphicx or graphics not loaded
    }{See the gnuplot documentation for explanation.
    }{The gnuplot epslatex terminal needs graphicx.sty or graphics.sty.}
    \renewcommand\includegraphics[2][]{}
  }
  \providecommand\rotatebox[2]{#2}
  \@ifundefined{ifGPcolor}{
    \newif\ifGPcolor
    \GPcolorfalse
  }{}
  \@ifundefined{ifGPblacktext}{
    \newif\ifGPblacktext
    \GPblacktexttrue
  }{}
  \let\gplgaddtomacro\g@addto@macro
  \gdef\gplbacktext{}
  \gdef\gplfronttext{}
  \makeatother
  \ifGPblacktext
    \def\colorrgb#1{}
    \def\colorgray#1{}
  \else
    \ifGPcolor
      \def\colorrgb#1{\color[rgb]{#1}}
      \def\colorgray#1{\color[gray]{#1}}
      \expandafter\def\csname LTw\endcsname{\color{white}}
      \expandafter\def\csname LTb\endcsname{\color{black}}
      \expandafter\def\csname LTa\endcsname{\color{black}}
      \expandafter\def\csname LT0\endcsname{\color[rgb]{1,0,0}}
      \expandafter\def\csname LT1\endcsname{\color[rgb]{0,1,0}}
      \expandafter\def\csname LT2\endcsname{\color[rgb]{0,0,1}}
      \expandafter\def\csname LT3\endcsname{\color[rgb]{1,0,1}}
      \expandafter\def\csname LT4\endcsname{\color[rgb]{0,1,1}}
      \expandafter\def\csname LT5\endcsname{\color[rgb]{1,1,0}}
      \expandafter\def\csname LT6\endcsname{\color[rgb]{0,0,0}}
      \expandafter\def\csname LT7\endcsname{\color[rgb]{1,0.3,0}}
      \expandafter\def\csname LT8\endcsname{\color[rgb]{0.5,0.5,0.5}}
    \else
      \def\colorrgb#1{\color{black}}
      \def\colorgray#1{\color[gray]{#1}}
      \expandafter\def\csname LTw\endcsname{\color{white}}
      \expandafter\def\csname LTb\endcsname{\color{black}}
      \expandafter\def\csname LTa\endcsname{\color{black}}
      \expandafter\def\csname LT0\endcsname{\color{black}}
      \expandafter\def\csname LT1\endcsname{\color{black}}
      \expandafter\def\csname LT2\endcsname{\color{black}}
      \expandafter\def\csname LT3\endcsname{\color{black}}
      \expandafter\def\csname LT4\endcsname{\color{black}}
      \expandafter\def\csname LT5\endcsname{\color{black}}
      \expandafter\def\csname LT6\endcsname{\color{black}}
      \expandafter\def\csname LT7\endcsname{\color{black}}
      \expandafter\def\csname LT8\endcsname{\color{black}}
    \fi
  \fi
    \setlength{\unitlength}{0.0200bp}
    \ifx\gptboxheight\undefined
      \newlength{\gptboxheight}
      \newlength{\gptboxwidth}
      \newsavebox{\gptboxtext}
    \fi
    \setlength{\fboxrule}{0.5pt}
    \setlength{\fboxsep}{1pt}
\begin{picture}(11520.00,8640.00)
    \gplgaddtomacro\gplbacktext{
      \colorrgb{0.00,0.00,0.00}
      \put(2200,1280){\makebox(0,0)[r]{\strut{}-3e-07}}
      \colorrgb{0.00,0.00,0.00}
      \put(2200,2427){\makebox(0,0)[r]{\strut{}-2e-07}}
      \colorrgb{0.00,0.00,0.00}
      \put(2200,3573){\makebox(0,0)[r]{\strut{}-1e-07}}
      \colorrgb{0.00,0.00,0.00}
      \put(2200,4720){\makebox(0,0)[r]{\strut{}0}}
      \colorrgb{0.00,0.00,0.00}
      \put(2200,5866){\makebox(0,0)[r]{\strut{}1e-07}}
      \colorrgb{0.00,0.00,0.00}
      \put(2200,7013){\makebox(0,0)[r]{\strut{}2e-07}}
      \colorrgb{0.00,0.00,0.00}
      \put(2200,8159){\makebox(0,0)[r]{\strut{}3e-07}}
      \colorrgb{0.00,0.00,0.00}
      \put(2440,880){\makebox(0,0){\strut{}$10^{0}$}}
      \colorrgb{0.00,0.00,0.00}
      \put(5226,880){\makebox(0,0){\strut{}$10^{1}$}}
      \colorrgb{0.00,0.00,0.00}
      \put(8013,880){\makebox(0,0){\strut{}$10^{2}$}}
      \colorrgb{0.00,0.00,0.00}
      \put(10799,880){\makebox(0,0){\strut{}$10^{3}$}}
    }
    \gplgaddtomacro\gplfronttext{
      \colorrgb{0.00,0.00,0.00}
      \put(320,4719){\rotatebox{90}{\makebox(0,0){\strut{}$\varepsilon$}}}
      \colorrgb{0.00,0.00,0.00}
      \put(6619,280){\makebox(0,0){\strut{}$H_{\CD}$ (km/s/Mpc)}}
      \colorrgb{0.00,0.00,0.00}
      \put(4855,7793){\makebox(0,0){\footnotesize $ 1.1$~Gyr}}
      \colorrgb{0.00,0.00,0.00}
      \put(4855,7188){\makebox(0,0){\footnotesize $ 4.0$~Gyr}}
      \colorrgb{0.00,0.00,0.00}
      \put(4855,6583){\makebox(0,0){\footnotesize $ 7.0$~Gyr}}
      \colorrgb{0.00,0.00,0.00}
      \put(4855,5978){\makebox(0,0){\footnotesize $10.1$~Gyr}}
      \colorrgb{0.00,0.00,0.00}
      \put(4855,5373){\makebox(0,0){\footnotesize $13.0$~Gyr}}
      \colorrgb{0.00,0.00,0.00}
      \put(2858,1624){\makebox(0,0)[l]{\strut{}$\LD =   2.5$~Mpc/$\hzeroeff$}}
    }
    \gplbacktext
    \put(0,0){\includegraphics[scale=0.4]{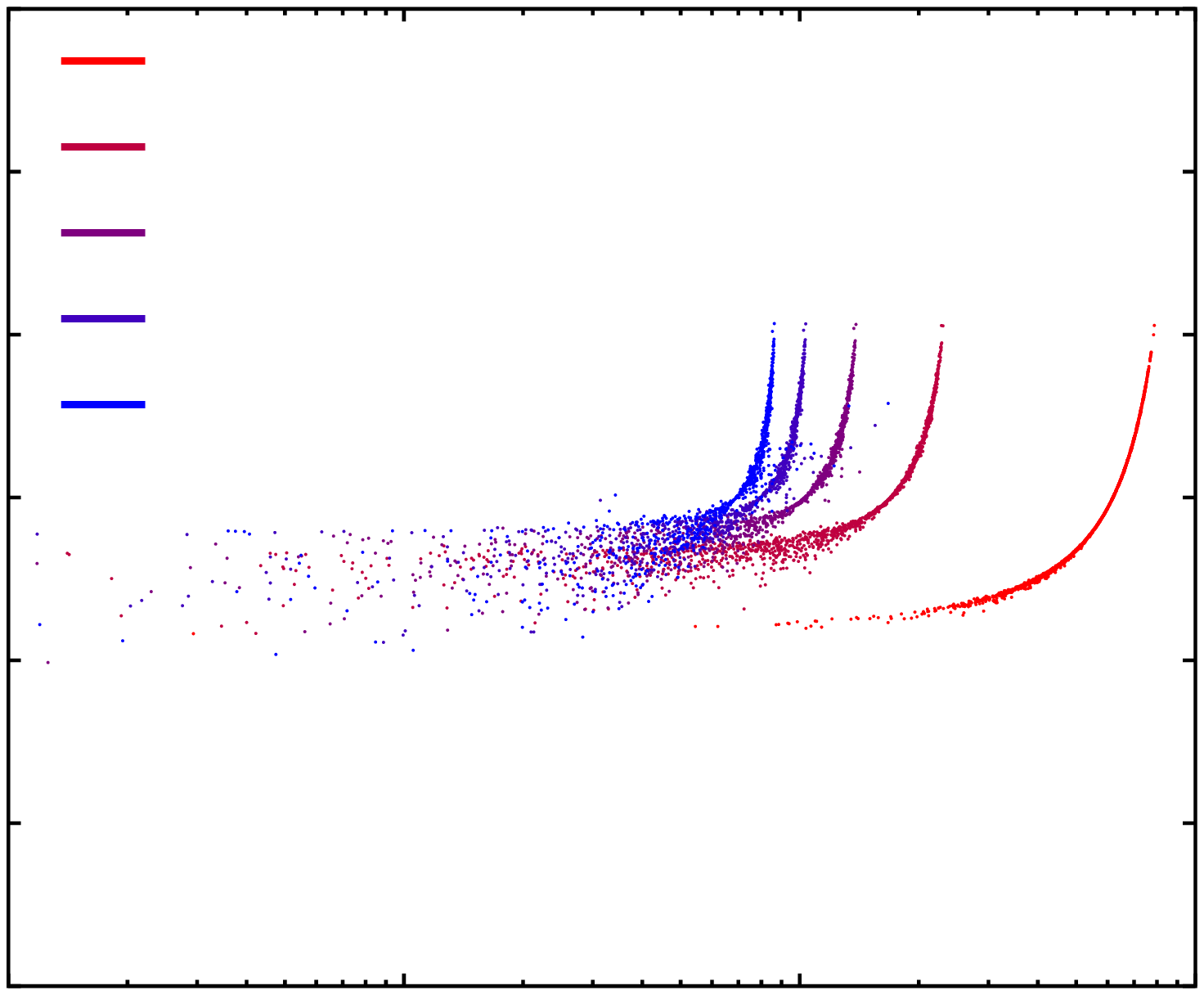}}
    \gplfronttext
  \end{picture}
\endgroup
 }\\
    \adjustbox{width=0.55\textwidth}{\begingroup
  \makeatletter
  \providecommand\color[2][]{
    \GenericError{(gnuplot) \space\space\space\@spaces}{
      Package color not loaded in conjunction with
      terminal option `colourtext'
    }{See the gnuplot documentation for explanation.
    }{Either use 'blacktext' in gnuplot or load the package
      color.sty in LaTeX.}
    \renewcommand\color[2][]{}
  }
  \providecommand\includegraphics[2][]{
    \GenericError{(gnuplot) \space\space\space\@spaces}{
      Package graphicx or graphics not loaded
    }{See the gnuplot documentation for explanation.
    }{The gnuplot epslatex terminal needs graphicx.sty or graphics.sty.}
    \renewcommand\includegraphics[2][]{}
  }
  \providecommand\rotatebox[2]{#2}
  \@ifundefined{ifGPcolor}{
    \newif\ifGPcolor
    \GPcolorfalse
  }{}
  \@ifundefined{ifGPblacktext}{
    \newif\ifGPblacktext
    \GPblacktexttrue
  }{}
  \let\gplgaddtomacro\g@addto@macro
  \gdef\gplbacktext{}
  \gdef\gplfronttext{}
  \makeatother
  \ifGPblacktext
    \def\colorrgb#1{}
    \def\colorgray#1{}
  \else
    \ifGPcolor
      \def\colorrgb#1{\color[rgb]{#1}}
      \def\colorgray#1{\color[gray]{#1}}
      \expandafter\def\csname LTw\endcsname{\color{white}}
      \expandafter\def\csname LTb\endcsname{\color{black}}
      \expandafter\def\csname LTa\endcsname{\color{black}}
      \expandafter\def\csname LT0\endcsname{\color[rgb]{1,0,0}}
      \expandafter\def\csname LT1\endcsname{\color[rgb]{0,1,0}}
      \expandafter\def\csname LT2\endcsname{\color[rgb]{0,0,1}}
      \expandafter\def\csname LT3\endcsname{\color[rgb]{1,0,1}}
      \expandafter\def\csname LT4\endcsname{\color[rgb]{0,1,1}}
      \expandafter\def\csname LT5\endcsname{\color[rgb]{1,1,0}}
      \expandafter\def\csname LT6\endcsname{\color[rgb]{0,0,0}}
      \expandafter\def\csname LT7\endcsname{\color[rgb]{1,0.3,0}}
      \expandafter\def\csname LT8\endcsname{\color[rgb]{0.5,0.5,0.5}}
    \else
      \def\colorrgb#1{\color{black}}
      \def\colorgray#1{\color[gray]{#1}}
      \expandafter\def\csname LTw\endcsname{\color{white}}
      \expandafter\def\csname LTb\endcsname{\color{black}}
      \expandafter\def\csname LTa\endcsname{\color{black}}
      \expandafter\def\csname LT0\endcsname{\color{black}}
      \expandafter\def\csname LT1\endcsname{\color{black}}
      \expandafter\def\csname LT2\endcsname{\color{black}}
      \expandafter\def\csname LT3\endcsname{\color{black}}
      \expandafter\def\csname LT4\endcsname{\color{black}}
      \expandafter\def\csname LT5\endcsname{\color{black}}
      \expandafter\def\csname LT6\endcsname{\color{black}}
      \expandafter\def\csname LT7\endcsname{\color{black}}
      \expandafter\def\csname LT8\endcsname{\color{black}}
    \fi
  \fi
    \setlength{\unitlength}{0.0200bp}
    \ifx\gptboxheight\undefined
      \newlength{\gptboxheight}
      \newlength{\gptboxwidth}
      \newsavebox{\gptboxtext}
    \fi
    \setlength{\fboxrule}{0.5pt}
    \setlength{\fboxsep}{1pt}
\begin{picture}(11520.00,8640.00)
    \gplgaddtomacro\gplbacktext{
      \colorrgb{0.00,0.00,0.00}
      \put(2200,1280){\makebox(0,0)[r]{\strut{}-3e-06}}
      \colorrgb{0.00,0.00,0.00}
      \put(2200,2427){\makebox(0,0)[r]{\strut{}-2e-06}}
      \colorrgb{0.00,0.00,0.00}
      \put(2200,3573){\makebox(0,0)[r]{\strut{}-1e-06}}
      \colorrgb{0.00,0.00,0.00}
      \put(2200,4720){\makebox(0,0)[r]{\strut{}0}}
      \colorrgb{0.00,0.00,0.00}
      \put(2200,5866){\makebox(0,0)[r]{\strut{}1e-06}}
      \colorrgb{0.00,0.00,0.00}
      \put(2200,7012){\makebox(0,0)[r]{\strut{}2e-06}}
      \colorrgb{0.00,0.00,0.00}
      \put(2200,8159){\makebox(0,0)[r]{\strut{}3e-06}}
      \colorrgb{0.00,0.00,0.00}
      \put(2440,880){\makebox(0,0){\strut{}$10^{0}$}}
      \colorrgb{0.00,0.00,0.00}
      \put(5226,880){\makebox(0,0){\strut{}$10^{1}$}}
      \colorrgb{0.00,0.00,0.00}
      \put(8013,880){\makebox(0,0){\strut{}$10^{2}$}}
      \colorrgb{0.00,0.00,0.00}
      \put(10799,880){\makebox(0,0){\strut{}$10^{3}$}}
    }
    \gplgaddtomacro\gplfronttext{
      \colorrgb{0.00,0.00,0.00}
      \put(320,4719){\rotatebox{90}{\makebox(0,0){\strut{}$\varepsilon$}}}
      \colorrgb{0.00,0.00,0.00}
      \put(6619,280){\makebox(0,0){\strut{}$H_{\CD}$ (km/s/Mpc)}}
      \colorrgb{0.00,0.00,0.00}
      \put(4855,7793){\makebox(0,0){\footnotesize $ 1.0$~Gyr}}
      \colorrgb{0.00,0.00,0.00}
      \put(4855,7188){\makebox(0,0){\footnotesize $ 4.0$~Gyr}}
      \colorrgb{0.00,0.00,0.00}
      \put(4855,6583){\makebox(0,0){\footnotesize $ 7.1$~Gyr}}
      \colorrgb{0.00,0.00,0.00}
      \put(4855,5978){\makebox(0,0){\footnotesize $ 9.9$~Gyr}}
      \colorrgb{0.00,0.00,0.00}
      \put(4855,5373){\makebox(0,0){\footnotesize $13.0$~Gyr}}
      \colorrgb{0.00,0.00,0.00}
      \put(2858,1624){\makebox(0,0)[l]{\strut{}$\LD =  10.0$~Mpc/$\hzeroeff$}}
    }
    \gplbacktext
    \put(0,0){\includegraphics[scale=0.4]{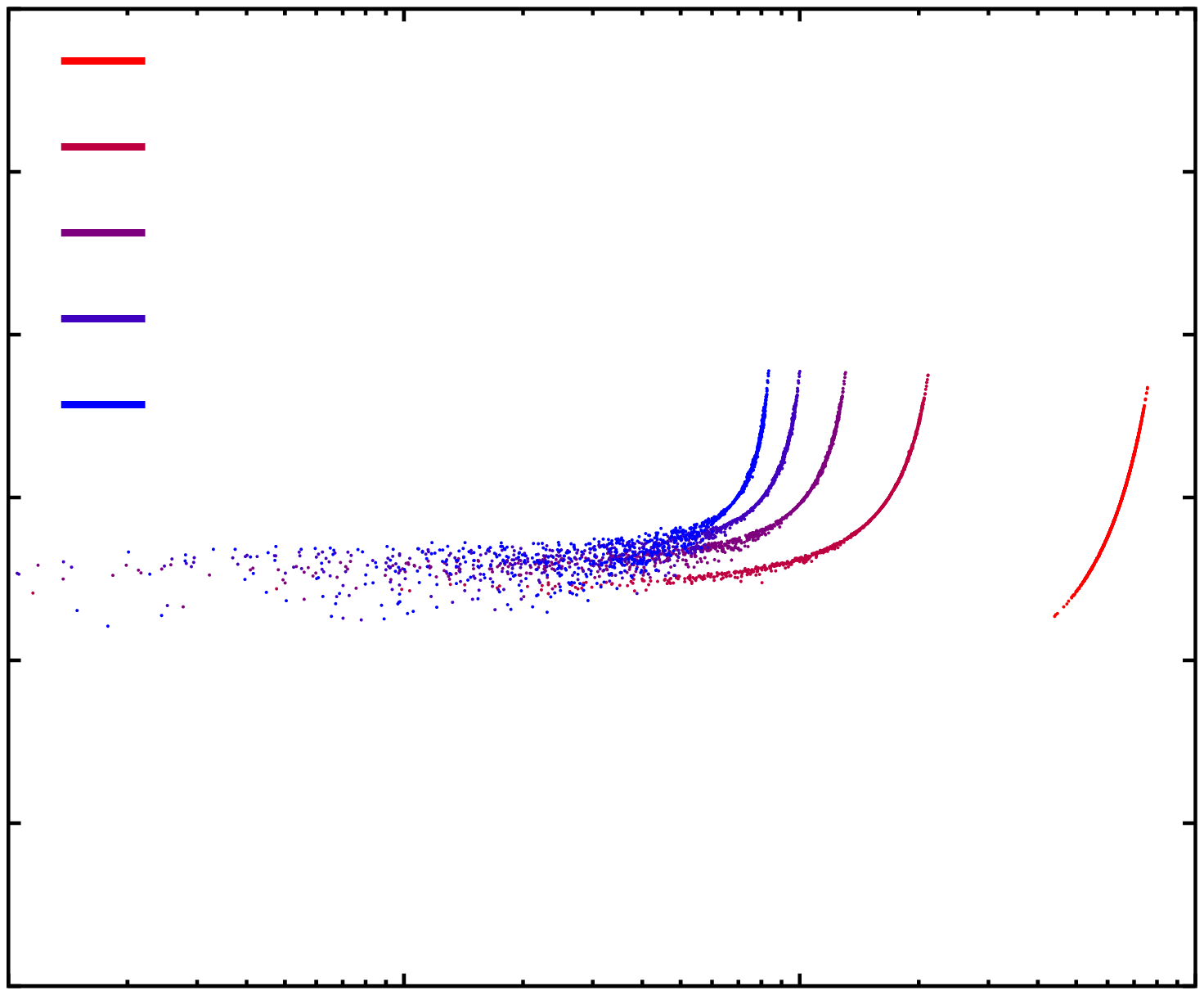}}
    \gplfronttext
  \end{picture}
\endgroup
 }\\
    \adjustbox{width=0.55\textwidth}{\begingroup
  \makeatletter
  \providecommand\color[2][]{
    \GenericError{(gnuplot) \space\space\space\@spaces}{
      Package color not loaded in conjunction with
      terminal option `colourtext'
    }{See the gnuplot documentation for explanation.
    }{Either use 'blacktext' in gnuplot or load the package
      color.sty in LaTeX.}
    \renewcommand\color[2][]{}
  }
  \providecommand\includegraphics[2][]{
    \GenericError{(gnuplot) \space\space\space\@spaces}{
      Package graphicx or graphics not loaded
    }{See the gnuplot documentation for explanation.
    }{The gnuplot epslatex terminal needs graphicx.sty or graphics.sty.}
    \renewcommand\includegraphics[2][]{}
  }
  \providecommand\rotatebox[2]{#2}
  \@ifundefined{ifGPcolor}{
    \newif\ifGPcolor
    \GPcolorfalse
  }{}
  \@ifundefined{ifGPblacktext}{
    \newif\ifGPblacktext
    \GPblacktexttrue
  }{}
  \let\gplgaddtomacro\g@addto@macro
  \gdef\gplbacktext{}
  \gdef\gplfronttext{}
  \makeatother
  \ifGPblacktext
    \def\colorrgb#1{}
    \def\colorgray#1{}
  \else
    \ifGPcolor
      \def\colorrgb#1{\color[rgb]{#1}}
      \def\colorgray#1{\color[gray]{#1}}
      \expandafter\def\csname LTw\endcsname{\color{white}}
      \expandafter\def\csname LTb\endcsname{\color{black}}
      \expandafter\def\csname LTa\endcsname{\color{black}}
      \expandafter\def\csname LT0\endcsname{\color[rgb]{1,0,0}}
      \expandafter\def\csname LT1\endcsname{\color[rgb]{0,1,0}}
      \expandafter\def\csname LT2\endcsname{\color[rgb]{0,0,1}}
      \expandafter\def\csname LT3\endcsname{\color[rgb]{1,0,1}}
      \expandafter\def\csname LT4\endcsname{\color[rgb]{0,1,1}}
      \expandafter\def\csname LT5\endcsname{\color[rgb]{1,1,0}}
      \expandafter\def\csname LT6\endcsname{\color[rgb]{0,0,0}}
      \expandafter\def\csname LT7\endcsname{\color[rgb]{1,0.3,0}}
      \expandafter\def\csname LT8\endcsname{\color[rgb]{0.5,0.5,0.5}}
    \else
      \def\colorrgb#1{\color{black}}
      \def\colorgray#1{\color[gray]{#1}}
      \expandafter\def\csname LTw\endcsname{\color{white}}
      \expandafter\def\csname LTb\endcsname{\color{black}}
      \expandafter\def\csname LTa\endcsname{\color{black}}
      \expandafter\def\csname LT0\endcsname{\color{black}}
      \expandafter\def\csname LT1\endcsname{\color{black}}
      \expandafter\def\csname LT2\endcsname{\color{black}}
      \expandafter\def\csname LT3\endcsname{\color{black}}
      \expandafter\def\csname LT4\endcsname{\color{black}}
      \expandafter\def\csname LT5\endcsname{\color{black}}
      \expandafter\def\csname LT6\endcsname{\color{black}}
      \expandafter\def\csname LT7\endcsname{\color{black}}
      \expandafter\def\csname LT8\endcsname{\color{black}}
    \fi
  \fi
    \setlength{\unitlength}{0.0200bp}
    \ifx\gptboxheight\undefined
      \newlength{\gptboxheight}
      \newlength{\gptboxwidth}
      \newsavebox{\gptboxtext}
    \fi
    \setlength{\fboxrule}{0.5pt}
    \setlength{\fboxsep}{1pt}
\begin{picture}(11520.00,8640.00)
    \gplgaddtomacro\gplbacktext{
      \colorrgb{0.00,0.00,0.00}
      \put(2200,1280){\makebox(0,0)[r]{\strut{}-1e-05}}
      \colorrgb{0.00,0.00,0.00}
      \put(2200,3000){\makebox(0,0)[r]{\strut{}-5e-06}}
      \colorrgb{0.00,0.00,0.00}
      \put(2200,4720){\makebox(0,0)[r]{\strut{}0}}
      \colorrgb{0.00,0.00,0.00}
      \put(2200,6439){\makebox(0,0)[r]{\strut{}5e-06}}
      \colorrgb{0.00,0.00,0.00}
      \put(2200,8159){\makebox(0,0)[r]{\strut{}1e-05}}
      \colorrgb{0.00,0.00,0.00}
      \put(2440,880){\makebox(0,0){\strut{}$10^{0}$}}
      \colorrgb{0.00,0.00,0.00}
      \put(5226,880){\makebox(0,0){\strut{}$10^{1}$}}
      \colorrgb{0.00,0.00,0.00}
      \put(8013,880){\makebox(0,0){\strut{}$10^{2}$}}
      \colorrgb{0.00,0.00,0.00}
      \put(10799,880){\makebox(0,0){\strut{}$10^{3}$}}
    }
    \gplgaddtomacro\gplfronttext{
      \colorrgb{0.00,0.00,0.00}
      \put(320,4719){\rotatebox{90}{\makebox(0,0){\strut{}$\varepsilon$}}}
      \colorrgb{0.00,0.00,0.00}
      \put(6619,280){\makebox(0,0){\strut{}$H_{\CD}$ (km/s/Mpc)}}
      \colorrgb{0.00,0.00,0.00}
      \put(4855,7793){\makebox(0,0){\footnotesize $ 1.0$~Gyr}}
      \colorrgb{0.00,0.00,0.00}
      \put(4855,7188){\makebox(0,0){\footnotesize $ 4.0$~Gyr}}
      \colorrgb{0.00,0.00,0.00}
      \put(4855,6583){\makebox(0,0){\footnotesize $ 7.1$~Gyr}}
      \colorrgb{0.00,0.00,0.00}
      \put(4855,5978){\makebox(0,0){\footnotesize $10.0$~Gyr}}
      \colorrgb{0.00,0.00,0.00}
      \put(4855,5373){\makebox(0,0){\footnotesize $13.0$~Gyr}}
      \colorrgb{0.00,0.00,0.00}
      \put(2858,1624){\makebox(0,0)[l]{\strut{}$\LD =  40.0$~Mpc/$\hzeroeff$}}
    }
    \gplbacktext
    \put(0,0){\includegraphics[scale=0.4]{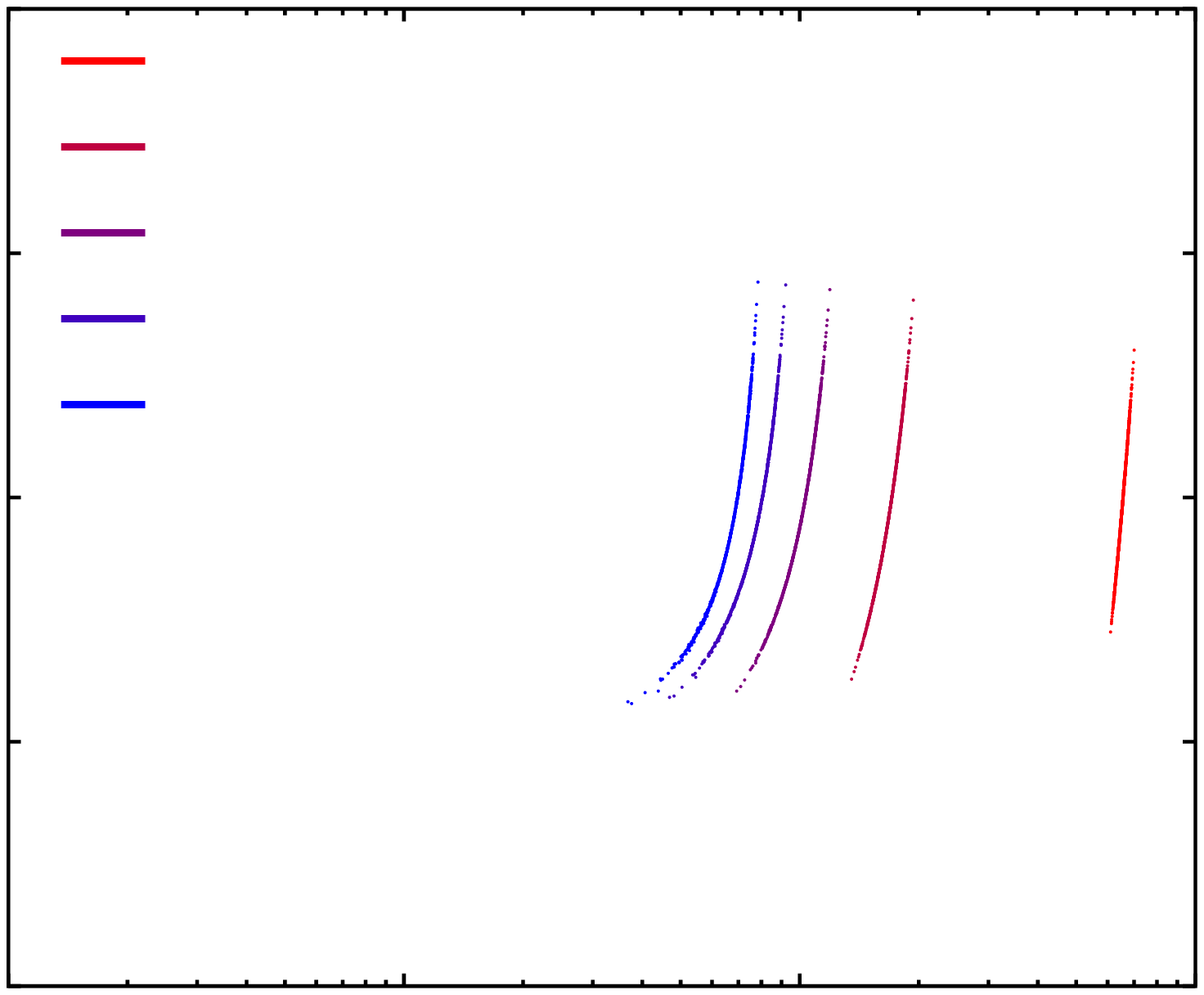}}
    \gplfronttext
  \end{picture}
\endgroup
 }
  \end{center}
  \caption{Curvature deviation parameter $\varepsilon$ as a function of
    expansion rate $H_{\CD}$, as in Fig.~\protect\ref{f-epsilonEdS}, for the almost-{\LCDM} model.
    \label{f-epsilonLCDM}}
\end{figure} 
 
Overall, it is overwhelmingly clear in Figs~\ref{f-II-OmRD-EdS} and
\ref{f-II-OmRD-LCDM} and Table~\ref{t-OmRDTA-stats} that turnaround is
almost always associated with a strong positive average curvature. We
further quantify this as follows.

Table~\ref{t-nonnegOmRDTA} lists the fractions of domains
which have negative or zero curvature. No such domains were found in any of the
cases simulated.
The total numbers of domains listed in
Table~\ref{t-nonnegOmRDTA} give an estimate of an upper limit to the frequency of
occurrence of non-positive curvature at turnaround in the almost-EdS
and almost-{\LCDM} models.
In order to increase the significance of the limit,
we performed 29 additional independent $N=256^3$ realisations,
each with the same parameters as the original. None of the domains
had $\OmRD \ge 0$ at turnaround. Thus, we estimate 99\% numerical upper limits
on the probability of a domain on these scales not having positive curvature at turnaround
as $P=0.0002$ in the almost-EdS case (23125 domains reaching turnaround)
and $P=0.0003$ in the almost-{\LCDM} case (14526 domains).
We estimate this upper limit by assuming that on average, $\bar{\mu}=5$ domains should have zero or negative
curvature at turnaround out of the full sample reaching turnaround,
but zero were detected due to random selection according to a Poisson distribution of mean
$\bar{\mu}$.
It is clearly very rare for a domain to be able to collapse in the average sense without
having positive 3-Ricci curvature.

In the generalisation from pointwise collapse to average collapse in a domain,
the expansion variance term appears in Eq.~\eqref{e-Buch00dust-4a-averaged},
and is usually combined with the shear scalar in the
kinematical backreaction (Eq.~\eqref{e-defn-Q}). To see how this could, in principle,
allow a domain to reach the turnaround epoch in an averaged sense despite being
spatially flat or negatively curved, the Hamiltonian constraint at turnaround,
Eq.~\eqref{e-hamilton-TA}, can be rewritten
\begin{align}
  \OmRD \ge 0 \, &\Leftrightarrow \,
  \left(\OmmD +  \OmQD + \OmLam^{\CD}\right) \le 0 \,.
  \label{e-hamilton-TA-Omm-OmQ}
\end{align}
Figures~\ref{f-OmmOmQ-EdS} and~\ref{f-OmmOmQ-LCDM} show the
$\OmmD$--$\OmQD$ relation for domains near turnaround. Since $\OmLam^{\CD}$ is
low compared to $\OmmD + \OmQD$ in these diagrams, a conservative bound for negative
curvature is shown in both figures. Reaching zero or negative curvature would require
a domain's position in one of these diagrams to lie just at or to the left of/below the
$\OmmD + \OmQD = 0$ line in Fig.~\ref{f-OmmOmQ-EdS}, or somewhat to the left of/below
the corresponding line in Fig.~\ref{f-OmmOmQ-LCDM} after taking into account
the value of $\OmLam^{\CD}$ for the domain.

The bands of points in Figs~\ref{f-OmmOmQ-EdS} and~\ref{f-OmmOmQ-LCDM}
give the impression that a selection criterion in $\OmmD + \OmQD$ or
$\OmRD$ was applied in selecting the points. This is true only
indirectly, in the sense that the points represent domains at
pre-turnaround epochs when $|H_{\CD}| < 1$~km/s/Mpc. In other words,
selecting for the turnaround epoch effectively makes a selection for
spatial curvature to lie in a band not too far from the EdS special
case of $\OmRD = -5\alpha^2$.

\section{Discussion} \label{s-discuss}

\subsection{Foliation, gauge and vorticity}
\label{s-gauge-weak-dependence}

Are any of the results above gauge dependent?
\postrefereeAAchanges{Proposition}~{\ELemmaFlatNoTurnaround} (\SSS\ref{e-lemma-flat-no-turnaround})
and the definitions in
\SSS\ref{s-method-GR-fluid} are given in terms of a foliation given
certain assumptions restricting the allowed spacetimes.
Defining a hypersurface orthogonal to the fluid flow provides a physical
definition, so the hypersurfaces are gauge independent.
The quantities of interest
are scalars, which are invariant (coordinate independent). Choosing a different
gauge to study the same quantities on the same spatial hypersurface would complicate
the calculations, but could not modify the results unless the use of the
new gauge change imposed additional constraints that modified the metric.
Thus, the results presented here are not gauge dependent. If the
EdS and {\LCDM} models
are interpreted as strictly FLRW models, with strictly flat spatial sections, then
the relativistic forbidding of gravitational collapse cannot be avoided by
gauge-dependence arguments. Similar reasoning applies to the plane-symmetric subcase.

In the almost-FLRW numerical QZA modelling in which curvature is allowed
to vary above and below zero on any given spatial hypersurface
and averages in Lagrangian domains are studied, gauge dependence
is again not an issue (provided, again, that no restrictions are imposed by a gauge
transformation), but foliation dependence could, in principle, be significant.
\postrefereeBBchanges{Buchert, Mourier \& Roy}
\cite*{BuchMourRoy18} argue that volume would differ by a factor of the order
of the mean Lorentz factor $\gamma$ relating the fluid rest frame to an
alternative reference frame. If the latter is that of a best-fit FLRW model
to observational data, then fluid velocities of the order of 200 km/s would yield
changes in volume or volume-based functionals such as $\OmRD$ by
about $\gamma -1 < 10^{-6}$, that is, the tilt between vectors normal to the different
foliations is negligible in the present context.
See \cite{BuchMourRoy18} for more details, including the role of vorticity,
for which in Newtonian cosmology simulations, the vorticity scalar is
generally found to be weaker
than the shear scalar \citep[e.g.][Fig.~10]{BernardeauWeygaert96}.

\tversionsnbody

\begin{figure}
  \begin{center}
    \adjustbox{width=0.55\textwidth}{\begingroup
  \makeatletter
  \providecommand\color[2][]{
    \GenericError{(gnuplot) \space\space\space\@spaces}{
      Package color not loaded in conjunction with
      terminal option `colourtext'
    }{See the gnuplot documentation for explanation.
    }{Either use 'blacktext' in gnuplot or load the package
      color.sty in LaTeX.}
    \renewcommand\color[2][]{}
  }
  \providecommand\includegraphics[2][]{
    \GenericError{(gnuplot) \space\space\space\@spaces}{
      Package graphicx or graphics not loaded
    }{See the gnuplot documentation for explanation.
    }{The gnuplot epslatex terminal needs graphicx.sty or graphics.sty.}
    \renewcommand\includegraphics[2][]{}
  }
  \providecommand\rotatebox[2]{#2}
  \@ifundefined{ifGPcolor}{
    \newif\ifGPcolor
    \GPcolorfalse
  }{}
  \@ifundefined{ifGPblacktext}{
    \newif\ifGPblacktext
    \GPblacktexttrue
  }{}
  \let\gplgaddtomacro\g@addto@macro
  \gdef\gplbacktext{}
  \gdef\gplfronttext{}
  \makeatother
  \ifGPblacktext
    \def\colorrgb#1{}
    \def\colorgray#1{}
  \else
    \ifGPcolor
      \def\colorrgb#1{\color[rgb]{#1}}
      \def\colorgray#1{\color[gray]{#1}}
      \expandafter\def\csname LTw\endcsname{\color{white}}
      \expandafter\def\csname LTb\endcsname{\color{black}}
      \expandafter\def\csname LTa\endcsname{\color{black}}
      \expandafter\def\csname LT0\endcsname{\color[rgb]{1,0,0}}
      \expandafter\def\csname LT1\endcsname{\color[rgb]{0,1,0}}
      \expandafter\def\csname LT2\endcsname{\color[rgb]{0,0,1}}
      \expandafter\def\csname LT3\endcsname{\color[rgb]{1,0,1}}
      \expandafter\def\csname LT4\endcsname{\color[rgb]{0,1,1}}
      \expandafter\def\csname LT5\endcsname{\color[rgb]{1,1,0}}
      \expandafter\def\csname LT6\endcsname{\color[rgb]{0,0,0}}
      \expandafter\def\csname LT7\endcsname{\color[rgb]{1,0.3,0}}
      \expandafter\def\csname LT8\endcsname{\color[rgb]{0.5,0.5,0.5}}
    \else
      \def\colorrgb#1{\color{black}}
      \def\colorgray#1{\color[gray]{#1}}
      \expandafter\def\csname LTw\endcsname{\color{white}}
      \expandafter\def\csname LTb\endcsname{\color{black}}
      \expandafter\def\csname LTa\endcsname{\color{black}}
      \expandafter\def\csname LT0\endcsname{\color{black}}
      \expandafter\def\csname LT1\endcsname{\color{black}}
      \expandafter\def\csname LT2\endcsname{\color{black}}
      \expandafter\def\csname LT3\endcsname{\color{black}}
      \expandafter\def\csname LT4\endcsname{\color{black}}
      \expandafter\def\csname LT5\endcsname{\color{black}}
      \expandafter\def\csname LT6\endcsname{\color{black}}
      \expandafter\def\csname LT7\endcsname{\color{black}}
      \expandafter\def\csname LT8\endcsname{\color{black}}
    \fi
  \fi
    \setlength{\unitlength}{0.0200bp}
    \ifx\gptboxheight\undefined
      \newlength{\gptboxheight}
      \newlength{\gptboxwidth}
      \newsavebox{\gptboxtext}
    \fi
    \setlength{\fboxrule}{0.5pt}
    \setlength{\fboxsep}{1pt}
\begin{picture}(11520.00,8640.00)
    \gplgaddtomacro\gplbacktext{
      \colorrgb{0.00,0.00,0.00}
      \put(1480,1280){\makebox(0,0)[r]{\strut{}-10}}
      \colorrgb{0.00,0.00,0.00}
      \put(1480,3000){\makebox(0,0)[r]{\strut{}-5}}
      \colorrgb{0.00,0.00,0.00}
      \put(1480,4720){\makebox(0,0)[r]{\strut{}0}}
      \colorrgb{0.00,0.00,0.00}
      \put(1480,6439){\makebox(0,0)[r]{\strut{}5}}
      \colorrgb{0.00,0.00,0.00}
      \put(1480,8159){\makebox(0,0)[r]{\strut{}10}}
      \colorrgb{0.00,0.00,0.00}
      \put(1720,880){\makebox(0,0){\strut{}$10^{0}$}}
      \colorrgb{0.00,0.00,0.00}
      \put(4746,880){\makebox(0,0){\strut{}$10^{1}$}}
      \colorrgb{0.00,0.00,0.00}
      \put(7773,880){\makebox(0,0){\strut{}$10^{2}$}}
      \colorrgb{0.00,0.00,0.00}
      \put(10799,880){\makebox(0,0){\strut{}$10^{3}$}}
    }
    \gplgaddtomacro\gplfronttext{
      \colorrgb{0.00,0.00,0.00}
      \put(320,4719){\rotatebox{90}{\makebox(0,0){\strut{}$\OmkD$}}}
      \colorrgb{0.00,0.00,0.00}
      \put(6259,280){\makebox(0,0){\strut{}$H_{\CD}$ (km/s/Mpc)}}
      \colorrgb{0.00,0.00,0.00}
      \put(4135,7793){\makebox(0,0){\footnotesize $ 1.1$~Gyr}}
      \colorrgb{0.00,0.00,0.00}
      \put(4135,7188){\makebox(0,0){\footnotesize $ 4.1$~Gyr}}
      \colorrgb{0.00,0.00,0.00}
      \put(4135,6583){\makebox(0,0){\footnotesize $ 6.9$~Gyr}}
      \colorrgb{0.00,0.00,0.00}
      \put(4135,5978){\makebox(0,0){\footnotesize $ 9.9$~Gyr}}
      \colorrgb{0.00,0.00,0.00}
      \put(4135,5373){\makebox(0,0){\footnotesize $13.1$~Gyr}}
      \colorrgb{0.00,0.00,0.00}
      \put(2174,1624){\makebox(0,0)[l]{\strut{}$\LD =   2.5$~Mpc/$\hzeroeff$}}
    }
    \gplbacktext
    \put(0,0){\includegraphics[scale=0.4]{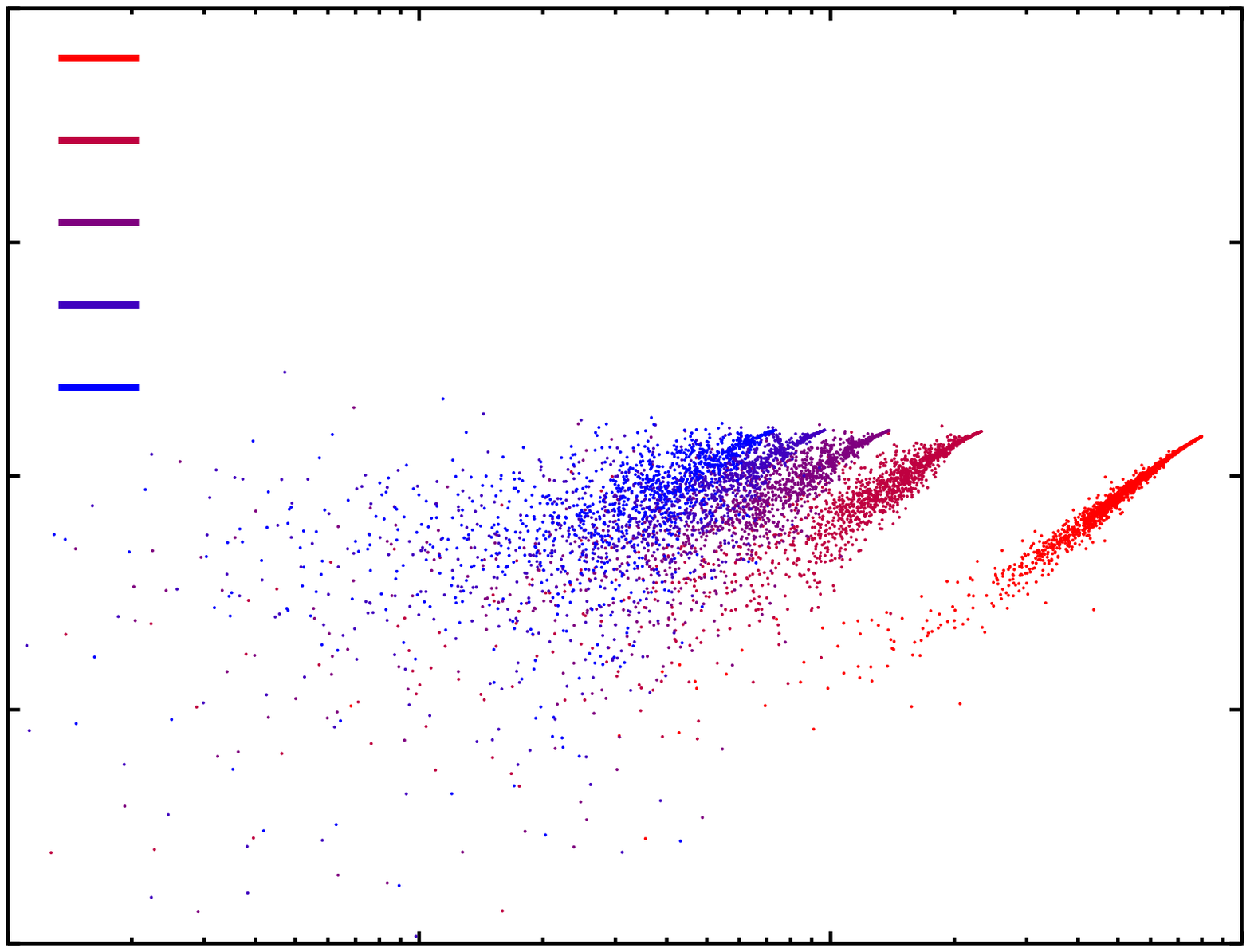}}
    \gplfronttext
  \end{picture}
\endgroup
 }\\
    \adjustbox{width=0.55\textwidth}{\begingroup
  \makeatletter
  \providecommand\color[2][]{
    \GenericError{(gnuplot) \space\space\space\@spaces}{
      Package color not loaded in conjunction with
      terminal option `colourtext'
    }{See the gnuplot documentation for explanation.
    }{Either use 'blacktext' in gnuplot or load the package
      color.sty in LaTeX.}
    \renewcommand\color[2][]{}
  }
  \providecommand\includegraphics[2][]{
    \GenericError{(gnuplot) \space\space\space\@spaces}{
      Package graphicx or graphics not loaded
    }{See the gnuplot documentation for explanation.
    }{The gnuplot epslatex terminal needs graphicx.sty or graphics.sty.}
    \renewcommand\includegraphics[2][]{}
  }
  \providecommand\rotatebox[2]{#2}
  \@ifundefined{ifGPcolor}{
    \newif\ifGPcolor
    \GPcolorfalse
  }{}
  \@ifundefined{ifGPblacktext}{
    \newif\ifGPblacktext
    \GPblacktexttrue
  }{}
  \let\gplgaddtomacro\g@addto@macro
  \gdef\gplbacktext{}
  \gdef\gplfronttext{}
  \makeatother
  \ifGPblacktext
    \def\colorrgb#1{}
    \def\colorgray#1{}
  \else
    \ifGPcolor
      \def\colorrgb#1{\color[rgb]{#1}}
      \def\colorgray#1{\color[gray]{#1}}
      \expandafter\def\csname LTw\endcsname{\color{white}}
      \expandafter\def\csname LTb\endcsname{\color{black}}
      \expandafter\def\csname LTa\endcsname{\color{black}}
      \expandafter\def\csname LT0\endcsname{\color[rgb]{1,0,0}}
      \expandafter\def\csname LT1\endcsname{\color[rgb]{0,1,0}}
      \expandafter\def\csname LT2\endcsname{\color[rgb]{0,0,1}}
      \expandafter\def\csname LT3\endcsname{\color[rgb]{1,0,1}}
      \expandafter\def\csname LT4\endcsname{\color[rgb]{0,1,1}}
      \expandafter\def\csname LT5\endcsname{\color[rgb]{1,1,0}}
      \expandafter\def\csname LT6\endcsname{\color[rgb]{0,0,0}}
      \expandafter\def\csname LT7\endcsname{\color[rgb]{1,0.3,0}}
      \expandafter\def\csname LT8\endcsname{\color[rgb]{0.5,0.5,0.5}}
    \else
      \def\colorrgb#1{\color{black}}
      \def\colorgray#1{\color[gray]{#1}}
      \expandafter\def\csname LTw\endcsname{\color{white}}
      \expandafter\def\csname LTb\endcsname{\color{black}}
      \expandafter\def\csname LTa\endcsname{\color{black}}
      \expandafter\def\csname LT0\endcsname{\color{black}}
      \expandafter\def\csname LT1\endcsname{\color{black}}
      \expandafter\def\csname LT2\endcsname{\color{black}}
      \expandafter\def\csname LT3\endcsname{\color{black}}
      \expandafter\def\csname LT4\endcsname{\color{black}}
      \expandafter\def\csname LT5\endcsname{\color{black}}
      \expandafter\def\csname LT6\endcsname{\color{black}}
      \expandafter\def\csname LT7\endcsname{\color{black}}
      \expandafter\def\csname LT8\endcsname{\color{black}}
    \fi
  \fi
    \setlength{\unitlength}{0.0200bp}
    \ifx\gptboxheight\undefined
      \newlength{\gptboxheight}
      \newlength{\gptboxwidth}
      \newsavebox{\gptboxtext}
    \fi
    \setlength{\fboxrule}{0.5pt}
    \setlength{\fboxsep}{1pt}
\begin{picture}(11520.00,8640.00)
    \gplgaddtomacro\gplbacktext{
      \colorrgb{0.00,0.00,0.00}
      \put(1480,1280){\makebox(0,0)[r]{\strut{}-10}}
      \colorrgb{0.00,0.00,0.00}
      \put(1480,3000){\makebox(0,0)[r]{\strut{}-5}}
      \colorrgb{0.00,0.00,0.00}
      \put(1480,4720){\makebox(0,0)[r]{\strut{}0}}
      \colorrgb{0.00,0.00,0.00}
      \put(1480,6439){\makebox(0,0)[r]{\strut{}5}}
      \colorrgb{0.00,0.00,0.00}
      \put(1480,8159){\makebox(0,0)[r]{\strut{}10}}
      \colorrgb{0.00,0.00,0.00}
      \put(1720,880){\makebox(0,0){\strut{}$10^{0}$}}
      \colorrgb{0.00,0.00,0.00}
      \put(4746,880){\makebox(0,0){\strut{}$10^{1}$}}
      \colorrgb{0.00,0.00,0.00}
      \put(7773,880){\makebox(0,0){\strut{}$10^{2}$}}
      \colorrgb{0.00,0.00,0.00}
      \put(10799,880){\makebox(0,0){\strut{}$10^{3}$}}
    }
    \gplgaddtomacro\gplfronttext{
      \colorrgb{0.00,0.00,0.00}
      \put(320,4719){\rotatebox{90}{\makebox(0,0){\strut{}$\OmkD$}}}
      \colorrgb{0.00,0.00,0.00}
      \put(6259,280){\makebox(0,0){\strut{}$H_{\CD}$ (km/s/Mpc)}}
      \colorrgb{0.00,0.00,0.00}
      \put(4135,7793){\makebox(0,0){\footnotesize $ 1.1$~Gyr}}
      \colorrgb{0.00,0.00,0.00}
      \put(4135,7188){\makebox(0,0){\footnotesize $ 4.1$~Gyr}}
      \colorrgb{0.00,0.00,0.00}
      \put(4135,6583){\makebox(0,0){\footnotesize $ 6.9$~Gyr}}
      \colorrgb{0.00,0.00,0.00}
      \put(4135,5978){\makebox(0,0){\footnotesize $ 9.9$~Gyr}}
      \colorrgb{0.00,0.00,0.00}
      \put(4135,5373){\makebox(0,0){\footnotesize $13.1$~Gyr}}
      \colorrgb{0.00,0.00,0.00}
      \put(2174,1624){\makebox(0,0)[l]{\strut{}$\LD =  10.0$~Mpc/$\hzeroeff$}}
    }
    \gplbacktext
    \put(0,0){\includegraphics[scale=0.4]{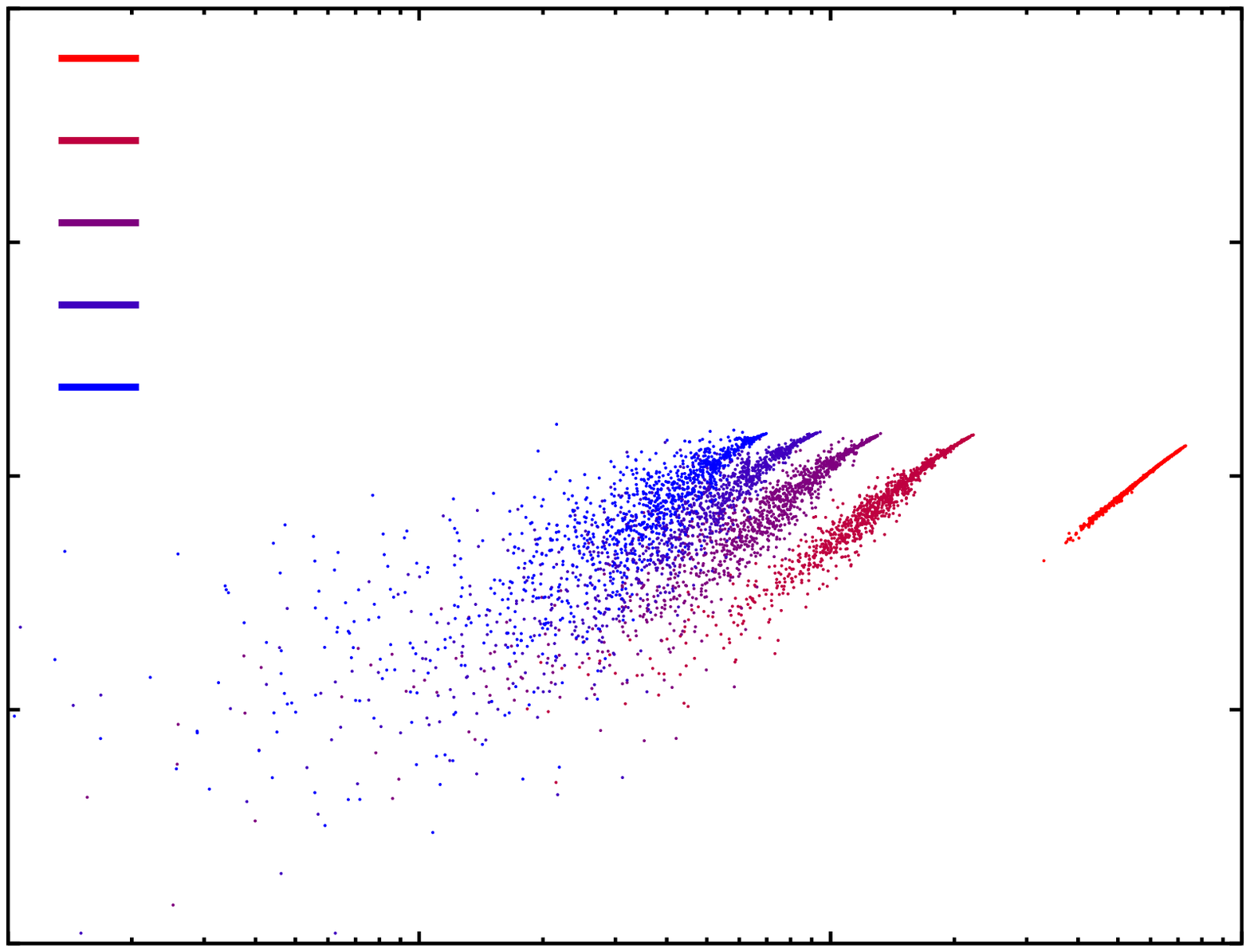}}
    \gplfronttext
  \end{picture}
\endgroup
 }\\
    \adjustbox{width=0.55\textwidth}{\begingroup
  \makeatletter
  \providecommand\color[2][]{
    \GenericError{(gnuplot) \space\space\space\@spaces}{
      Package color not loaded in conjunction with
      terminal option `colourtext'
    }{See the gnuplot documentation for explanation.
    }{Either use 'blacktext' in gnuplot or load the package
      color.sty in LaTeX.}
    \renewcommand\color[2][]{}
  }
  \providecommand\includegraphics[2][]{
    \GenericError{(gnuplot) \space\space\space\@spaces}{
      Package graphicx or graphics not loaded
    }{See the gnuplot documentation for explanation.
    }{The gnuplot epslatex terminal needs graphicx.sty or graphics.sty.}
    \renewcommand\includegraphics[2][]{}
  }
  \providecommand\rotatebox[2]{#2}
  \@ifundefined{ifGPcolor}{
    \newif\ifGPcolor
    \GPcolorfalse
  }{}
  \@ifundefined{ifGPblacktext}{
    \newif\ifGPblacktext
    \GPblacktexttrue
  }{}
  \let\gplgaddtomacro\g@addto@macro
  \gdef\gplbacktext{}
  \gdef\gplfronttext{}
  \makeatother
  \ifGPblacktext
    \def\colorrgb#1{}
    \def\colorgray#1{}
  \else
    \ifGPcolor
      \def\colorrgb#1{\color[rgb]{#1}}
      \def\colorgray#1{\color[gray]{#1}}
      \expandafter\def\csname LTw\endcsname{\color{white}}
      \expandafter\def\csname LTb\endcsname{\color{black}}
      \expandafter\def\csname LTa\endcsname{\color{black}}
      \expandafter\def\csname LT0\endcsname{\color[rgb]{1,0,0}}
      \expandafter\def\csname LT1\endcsname{\color[rgb]{0,1,0}}
      \expandafter\def\csname LT2\endcsname{\color[rgb]{0,0,1}}
      \expandafter\def\csname LT3\endcsname{\color[rgb]{1,0,1}}
      \expandafter\def\csname LT4\endcsname{\color[rgb]{0,1,1}}
      \expandafter\def\csname LT5\endcsname{\color[rgb]{1,1,0}}
      \expandafter\def\csname LT6\endcsname{\color[rgb]{0,0,0}}
      \expandafter\def\csname LT7\endcsname{\color[rgb]{1,0.3,0}}
      \expandafter\def\csname LT8\endcsname{\color[rgb]{0.5,0.5,0.5}}
    \else
      \def\colorrgb#1{\color{black}}
      \def\colorgray#1{\color[gray]{#1}}
      \expandafter\def\csname LTw\endcsname{\color{white}}
      \expandafter\def\csname LTb\endcsname{\color{black}}
      \expandafter\def\csname LTa\endcsname{\color{black}}
      \expandafter\def\csname LT0\endcsname{\color{black}}
      \expandafter\def\csname LT1\endcsname{\color{black}}
      \expandafter\def\csname LT2\endcsname{\color{black}}
      \expandafter\def\csname LT3\endcsname{\color{black}}
      \expandafter\def\csname LT4\endcsname{\color{black}}
      \expandafter\def\csname LT5\endcsname{\color{black}}
      \expandafter\def\csname LT6\endcsname{\color{black}}
      \expandafter\def\csname LT7\endcsname{\color{black}}
      \expandafter\def\csname LT8\endcsname{\color{black}}
    \fi
  \fi
    \setlength{\unitlength}{0.0200bp}
    \ifx\gptboxheight\undefined
      \newlength{\gptboxheight}
      \newlength{\gptboxwidth}
      \newsavebox{\gptboxtext}
    \fi
    \setlength{\fboxrule}{0.5pt}
    \setlength{\fboxsep}{1pt}
\begin{picture}(11520.00,8640.00)
    \gplgaddtomacro\gplbacktext{
      \colorrgb{0.00,0.00,0.00}
      \put(1480,1280){\makebox(0,0)[r]{\strut{}-10}}
      \colorrgb{0.00,0.00,0.00}
      \put(1480,3000){\makebox(0,0)[r]{\strut{}-5}}
      \colorrgb{0.00,0.00,0.00}
      \put(1480,4720){\makebox(0,0)[r]{\strut{}0}}
      \colorrgb{0.00,0.00,0.00}
      \put(1480,6439){\makebox(0,0)[r]{\strut{}5}}
      \colorrgb{0.00,0.00,0.00}
      \put(1480,8159){\makebox(0,0)[r]{\strut{}10}}
      \colorrgb{0.00,0.00,0.00}
      \put(1720,880){\makebox(0,0){\strut{}$10^{0}$}}
      \colorrgb{0.00,0.00,0.00}
      \put(4746,880){\makebox(0,0){\strut{}$10^{1}$}}
      \colorrgb{0.00,0.00,0.00}
      \put(7773,880){\makebox(0,0){\strut{}$10^{2}$}}
      \colorrgb{0.00,0.00,0.00}
      \put(10799,880){\makebox(0,0){\strut{}$10^{3}$}}
    }
    \gplgaddtomacro\gplfronttext{
      \colorrgb{0.00,0.00,0.00}
      \put(320,4719){\rotatebox{90}{\makebox(0,0){\strut{}$\OmkD$}}}
      \colorrgb{0.00,0.00,0.00}
      \put(6259,280){\makebox(0,0){\strut{}$H_{\CD}$ (km/s/Mpc)}}
      \colorrgb{0.00,0.00,0.00}
      \put(4135,7793){\makebox(0,0){\footnotesize $ 1.2$~Gyr}}
      \colorrgb{0.00,0.00,0.00}
      \put(4135,7188){\makebox(0,0){\footnotesize $ 4.2$~Gyr}}
      \colorrgb{0.00,0.00,0.00}
      \put(4135,6583){\makebox(0,0){\footnotesize $ 7.0$~Gyr}}
      \colorrgb{0.00,0.00,0.00}
      \put(4135,5978){\makebox(0,0){\footnotesize $10.1$~Gyr}}
      \colorrgb{0.00,0.00,0.00}
      \put(4135,5373){\makebox(0,0){\footnotesize $13.1$~Gyr}}
      \colorrgb{0.00,0.00,0.00}
      \put(2174,1624){\makebox(0,0)[l]{\strut{}$\LD =  40.0$~Mpc/$\hzeroeff$}}
    }
    \gplbacktext
    \put(0,0){\includegraphics[scale=0.4]{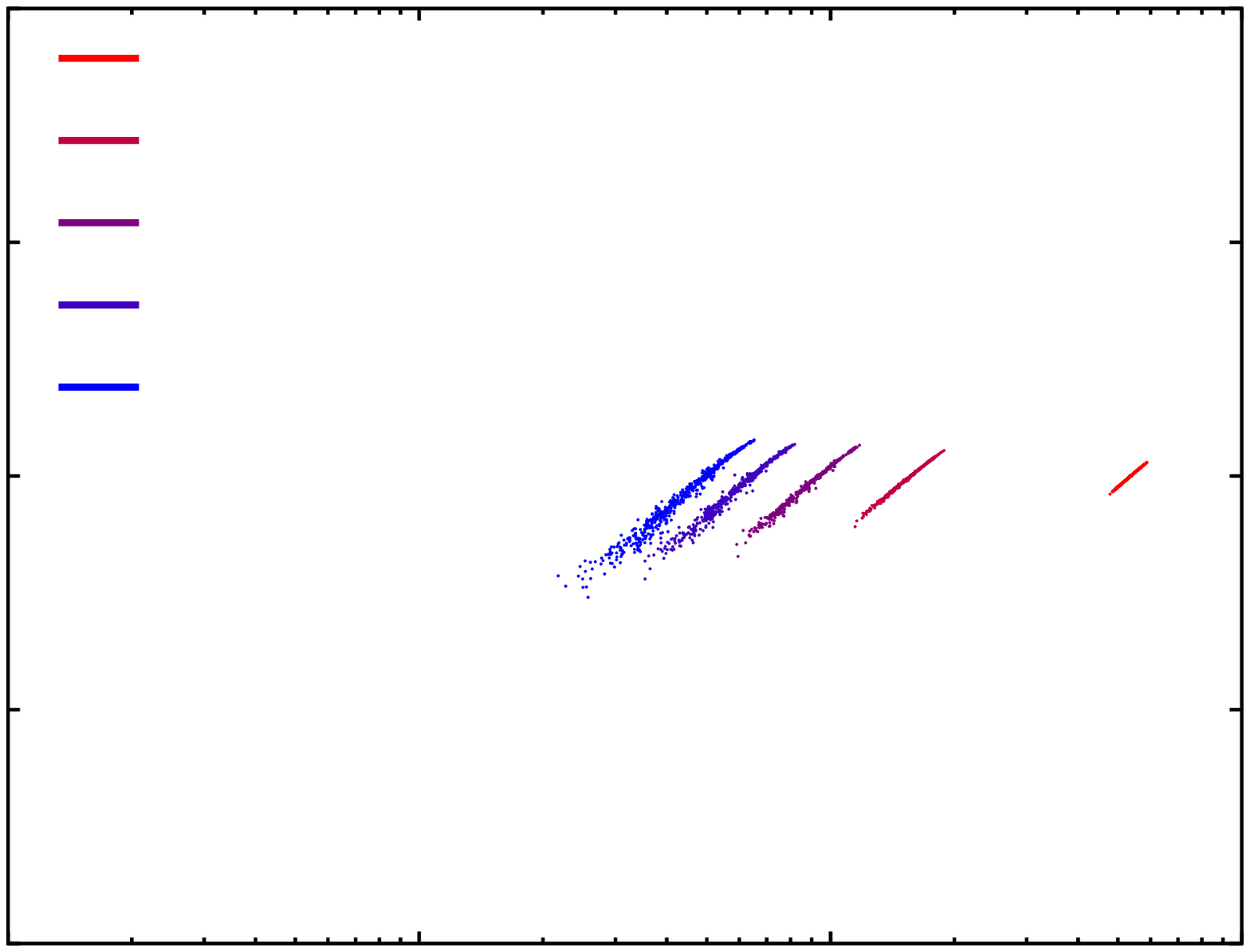}}
    \gplfronttext
  \end{picture}
\endgroup
 }
  \end{center}
  \caption{\postrefereeBBchanges{Comoving domain pseudo-curvature functional $\OmkD$ versus
    expansion rate $H_{\CD}$,
    from non-relativistic $N$-body simulations ($N$-body initial conditions)
    for
    a CMB-normalised almost-EdS model.
    {\em {\Toptobottom} panels}: averaging scales
    $\LD = 2.5, 10, 40$~Mpc/$\hzeroeff$, respectively, where
    $L_{\mathrm{box}}  = 16 \LD = 64 L_{\mathrm{DTFE}} = 128 L_N$
    and
    $N=128^3$ particles.
    Colours and ranges are as in Fig.~\protect\ref{f-OmRDEdS}.}
    \label{f-OmRDEdS-nbody}}
\end{figure} 
 
\subsection{Newtonian $N$-body equivalents} \label{s-pseudo-curvature}
\postrefereeBBstart
Adamek et al.\/ \cite{Adamek19backreaction} argue that a change in the
choice of foliation, which necessarily changes the choice of gauge, can affect
parameters calculated in relation to backreaction by 2--3 orders of magnitude.
In this discussion, we do not try to resolve the apparent disagreement between
the claims of \cite{BuchMourRoy18} and \cite{Adamek19backreaction}.

Instead, we consider the equivalent of the curvature--expansion-rate relation in a conventional
$N$-body simulation and compare it with our RZA-based curvature--expansion-rate relation.
We need a \enquote*{pseudo-curvature} parameter that can be estimated from Newtonian
$N$-body simulations and interpreted as what would constitute curvature in the relativistic
case. We replace Eq.~\eqref{e-defn-Q} by its Newtonian equivalent \cite[][Eq.~(6)]{BKS00};
we retain the equivalent of Eq.~\eqref{e-defn-OmQD}, that is,
\begin{align}
  \OmQDNewt := -\CQ_{\CD}/(6 H^2)\;,
  \label{e-defn-OmQD-Newt}
\end{align}
using the expansion rate $H$ of the reference model; and in a slight variation of the approach
in ref.~\cite[][II.D]{BKS00}, we use the Newtonian averaged Hamiltonian constraint
\cite[][Eq.~(17)]{BKS00} to define a Newtonian pseudo-curvature parameter
\begin{align}
  \OmkD := 1 - \OmmD -\OmLam^{\CD} - \OmQDNewt \;.
  \label{e-defn-OmkD}
\end{align}
(We run the simulations for an EdS reference model, so $\OmLam^{\CD} \equiv 0$ in this case.)
We use the same numerical method as above, apart from switching {\RAMSESscalav} from RZA
integration to $N$-body mode, as indicated using the software version listed in
Table~\ref{t-versions-nbody}.
This method provides an approximate comparison with the above results, with
some differences. The definitions in this section
apply to the conventional Newtonian foliation, which (i) is not fluid-orthogonal.
Two other differences with the RZA method
are that (ii) vorticity is not set to zero (so the foliation is not irrotational), and (iii)
the domains for calculating the invariants of the peculiar expansion tensor and averaging
have fixed comoving-FLRW spatial boundaries, not Lagrangian spatial
boundaries.

We performed these simulations for the EdS reference model and show the results in Fig.~\ref{f-OmRDEdS-nbody}.
Comparing these to Fig.~\ref{f-OmRDEdS} and keeping in mind the three differences in the method,
the curvature--expansion-rate relations appear to be broadly consistent between the two figures.
The use of comoving-FLRW boundaries
to define domains prevents any domain from collapsing to zero size and weakens the ability of a domain
to achieve turnaround (since turnaround in this case requires a larger spatial region than the original
Lagrangian region to have slowed down to momentary mean staticity in physical spatial coordinates).
It is thus unsurprising that the rather sharply defined
relations pointing towards $\OmRD \sim -5$ at turnaround in Fig.~\ref{f-OmRDEdS}
are less apparent in Fig.~\ref{f-OmRDEdS-nbody}.
The intermediate scale simulation, for $\LD = 10$~Mpc/$\hzeroeff$, best suggests an approach
towards $\OmRD \sim -5$ at turnaround ($H_{\CD} \rightarrow 0$~km/s/Mpc),
but the use of comoving-FLRW boundaries appears to lead to fuzzier curvature--expansion-rate relations
than for Lagrangian boundaries.
For the largest scale, $\LD = 40$~Mpc/$\hzeroeff$, turnaround is not reached for this comoving domain size.
The smaller scale, $\LD = 2.5$~Mpc/$\hzeroeff$, has a broad scatter of domains consistent with approaching
turnaround at curvatures that are mostly positive.
It would seem reasonable to attribute the other differences (such as straight versus curved relations)
to the different choices of foliation and vorticity.
Overall, the tendency to require positive average spatial cuvature to
reach turnaround remains present in these Newtonian pseudo-curvature calculations
as shown in Fig.~\ref{f-OmRDEdS-nbody}, but is less sharply defined than in the RZA calculations.
\postrefereeBBstop

\subsection{The curvature-induced deviation $\varepsilon$}

The relations in Figs~\ref{f-OmRDEdS} and \ref{f-OmRDLCDM}
indicate methods both of calibrating cosmological structure formation
simulations that claim to be fully relativistic, and, in principle, of
being measured observationally.  What possible avenues could there be
for measuring $\OmRD$ in a given spatial domain?  Here, we introduce a
dimensionless \enquote*{curvature-induced deviation} variable $\varepsilon$
defined for non-zero curvature that
depends both on the curvature and on the averaging scale.
We express the average 3-Ricci curvature for non-zero curvature
as a typical curvature radius $\RCeff$ (in physical units)
with a value that is real for positive curvature and imaginary
for negative curvature,
\begin{align}
  \RCeff := \left(\Heff\, \sqrt{-\OmRD}\right)^{-1} \,,
  \label{e-defn-RCeff}
\end{align}
We consider a length $l$ over which to estimate the deviation of
a pair of straight
lines (spatial geodesics at constant $t$ in the foliation)
that at one location are locally parallel.
We set $l = a_{\CD} \LD$, the approximate average physical size of a domain
(the cube root of the volume) at the averaging scale $\LD$,
where as above, $\LD$ is expressed in effective-model comoving units.
This approximatoin is valid to first order in $\RCeff$.
The curvature-induced deviation's functional dependence is
\begin{align}
  \varepsilon(\average{\CR}, l) &\equiv
  \varepsilon(\OmRD, \Heff, a_{\CD}, \LD) \equiv
  \varepsilon(\RCeff, a_{\CD}, \LD).
  \label{e-varepsilon-dependencies}
\end{align}
This functional is designed to measure the difference of an arc length
on the constant foliation time hypersurface, interpreted as having
constant curvature $\average{R}$, from the corresponding arc length in a corresponding
flat spatial section, at a distance
corresponding to the averaging scale $\LD$
(again in comoving effective units, as for $\LD$ throughout this work),
and normalised by that
same distance $\LD$,
\begin{align}
  \varepsilon &:=
  \frac{1}{a_{\CD}\, \LD} \left( \RCeff \sin \frac{a_{\CD}\,\LD}{\RCeff} - a_{\CD}\,\LD \right)
  \nonumber \\
  &=
  -\frac{1}{6}\, \left( a_{\CD}\, \Heff\, \LD \right)^2 \OmRD + \ldots
  \,,
  \label{e-defn-varepsilon}
\end{align}
in which the third-order Taylor expansion for $\sin$ (or $\sinh$) should
be used for numerical stability in nearly flat domains
($|\OmRD| \ll 1 $; in exactly flat domains, $\varepsilon := 0$), and
positive curvature corresponds to negative $\OmRD$ and negative $\varepsilon$.
For example, two locally parallel spatial geodesics separated by 1~Mpc
should be separated by about $(1+\varepsilon)$~Mpc after being extended by a distance
of $\LD$. Time-integrated dynamical quantities will differ from flat-space calculations
by integrals including $\varepsilon$ terms.

Figures~\ref{f-epsilonEdS} and \ref{f-epsilonLCDM} show the
curvature-induced deviation $\varepsilon$.  The amplitudes of this
effect range from about $10^{-7}$ at the $\LD=2.5$~Mpc/$\hzeroeff$
scale (top panels) up to a little above
$10^{-5}$ at the largest comoving
scale, $\LD = 40$~Mpc/$\hzeroeff$ (bottom panels).
This can be interpreted using
the rightmost expression in Eq.~\eqref{e-defn-varepsilon},
since an increase in $\LD^2$ by a factor of 16
approximately corresponds to the
increase in the typical magnitudes of $\varepsilon$
from the top to middle panels
in these two figures. The increase is weaker in shifting to
the largest scale (bottom panels). In particular, the bottom panel in
Fig.~\ref{f-epsilonLCDM} shows a scale in
the almost-{\LCDM} model at which
turnaround is not reached, so only weak positive curvature
results and the curvature-induced deviation $\varepsilon$
is correspondingly weak.

Given that the {\LCDM} model is a fair observational proxy,
the lower two panels of
Fig.~\ref{f-epsilonLCDM} indicate that
\postrefereeBBchanges{spatial (constant $t$ for our foliation)}
geodesics that in strict {\LCDM} are assumed to be parallel must, in a
relativistic almost-{\LCDM} model including structure formation,
converge or diverge at about
the $10^{-6}$ to $10^{-5}$ level after passing through a
typical turnaround domain.
\postrefereeBBchanges{Observations are performed on the past light cone,
  so the curvature-induced deviation
  $\varepsilon$ defined here in Eq.~\eqref{e-defn-varepsilon}
  on a constant-$t$ hypersurface in the fluid-orthogonal irrotational
  foliation will correspond to a somewhat different deviation of
  observers' null geodesics
  on the light cone. It is unlikely that the amplitude of deviation should
  be significantly weaker on the light cone than in the spatial hypersurface.}
Thus, observational analyses of
dense regions of the cosmic web made under
the assumption of an FLRW metric, without taking into account
spatially varying curvature, should be relativistically inaccurate at
about this level.
\postrefereeAAchanges{While weak lensing surveys take
  spatial curvature into account, as modelled by linear perturbation
  theory \citep[e.g.][]{BartelmannSchneider01},
  there are many cosmological observational analyses
  that do not,
  and instead, algebraically assume that turnaround-epoch
  spatial curvature is insignificant. For example,
  the baryon acoustic oscillation discovery papers
  \citep{Cole05BAO,Eisenstein05} do not appear to
  assume any deviation from a strict FLRW model, whether
  by weak lensing or by turnaround-epoch lensing;
  global curvature constraint estimates, such as that of
  \cite{RyanYunRatra19}, do not appear to take into
  account any non-FLRW effects;
  the main analyses of \cite{Bennett13WMAP9} and
  \cite{Planck2015cosmparam}, leaving aside
  gravitational weak lensing analyses, adopt a flat
  FLRW model.
  \prerefereeAAchanges{Here, we have shown that the inaccuracy of
    these assumptions should lie at roughly the $10^{-6}$ to $10^{-5}$
    level when passing through a single turnaround domain.}
  A potential application of the non-perturbative RZA approach would be
  to extend standard weak lensing methods beyond
  first and second order perturbation theory.
  In the new generation of surveys,
  detecting turnaround-epoch lensing effects (geodesic deviations)
  is likely to be difficult,}
but should in principle provide a test of precise, accurate cosmology.

These low values of $\varepsilon$
do not necessarily contradict the recently emerged
negative average curvature hypothesis (\cite{Rasanen06negemergcurv,
  NambuTanimoto05,KaiNambu07, Rasanen08peakmodel,
  Larena09template,Chiesa14Larenatest, WiegBuch10,BuchRas12,
  Wiltshire09timescape,DuleyWilt13,NazerW15CMB, ROB13,
  Barbosa16viscos, BolejkoC10SwissSzek, LRasSzybka13,
  Roukema17silvir, Sussman15LTBpertGIC,Chirinos16LTBav,
  Bolejko17negcurv,Bolejko17styro,
  Kras81kevolving,Kras82kevolving,Kras83kevolving,Stichel16,Stichel18,
  Coley10scalars,KasparSvitek14Cartan,RaczDobos16}) of explaining dark
energy as a misinterpretation of (non-relativistic fit to) cosmological
observations. The curvature deviation $\varepsilon$ indicates how much spatial
geodesics should converge or diverge, not how much expansion rates
should be spatially inhomogeneous. It is already known
observationally that the BAO scale is inhomogeneous
when using the {\LCDM} model as a proxy to
interpret the luminous red galaxy distribution in the
Sloan Digital Sky Survey
(\cite{RBOF15,RBFO15};
see \cite{HeinBlakeLiWilt18BAO} for interpretation in terms of an inhomogeneous
model and \cite{Neyrinck18BAOenv, BlakeABR18BAOenv} for related analyses).
The main order of magnitude observational argument supporting the emergent negative average
curvature hypothesis
is the void peculiar expansion rate, that is,
the ratio of the peculiar velocities of galaxies falling out of voids to the void
sizes, which is typically of a few tens of km/s/Mpc -- a substantial fraction
of the Hubble--Lema\^{\i}tre constant
\citep[][]{ROB13,RMBO16Hbg1}.

Non-perturbative work on the recently emerged negative average curvature hypothesis
shows in more detail how the curvature functional $\OmRD$
reveals negative curvature in a void and positive curvature in an overdensity.
Using a Szekeres model to model a local void and nearby overdensity
and domains that are spheres of radius 5~Mpc,
\cite[][Fig.~2, bottom-right panel]{Bolejko17negcurv} showed negative curvature
in the (central) void and positive curvature in the overdensity (lying at
$-40$~Mpc~$\ltapprox z \ltapprox -20$~Mpc).
The results presented in this work
are consistent with the conclusion of \cite{Bolejko17negcurv} that
for non-perturbative, non-linear calculations, curvature
associated with structure formation is highly inhomogeneous, and of an order of magnitude
at least as great as that of the FLRW density parameter $\Omm$.

\subsection{New GR test: the $\OmRD$--$H_{\CD}$ relation}
\label{s-OmRD-HCD-relation}

The turnaround-epoch positive spatial curvature requirement is
\postrefereeAAchanges{not overtly coded into} the Euclidean spatial geometry of a Newtonian
simulation.
\postrefereeAAchanges{In terms of linear perturbation theory interpretations,
  the Newtonian-gauge gravitational potential gradient matches spatial curvature to first
  order. In this sense, it could be argued that the requirement is present implicitly
  in an approximate sense.
  However, such an interpretation does not account for the fact that
  vector} arithmetic is, \postrefereeAAchanges{in principle,} no longer globally justified
\postrefereeAAchanges{in the presence of positive (or negative) spatial curvature}
-- tangent and cotangent spaces of vectors and 1-forms at individual
spacetime events have to be related to each other via a connection,
typically the covariant derivative.

\postrefereeAAchanges{Nevertheless}, \cite{Fidler15,Fidler16,Fidler17initconds,Fidler17}
have proposed the \enquote*{Newtonian motion (Nm)} gauge formalism,
which derives diffeomorphisms (a \enquote*{dictionary}) that under appropriate
conditions relate Newtonian $N$-body simulations to general-relativistic
spacetime. The authors argue that the approximations used are cosmologically
accurate. The role of spatial curvature is
implicitly described in \cite[][\SSS{}5.2]{Fidler17}.
\postrefereeAAchanges{A useful crosscheck between the Nm formalism
  and the QZA formalism (also used by \cite{VignBuch19} in the plane-symmetric case)
  would be to evaluate the $\OmRD$--$H_{\CD}$
relation on a fluid-orthogonal, irrotational foliation of a
Newtonian cosmological $N$-body simulation and check the resulting values
against the scalar averaging results found in this work.
The two approaches make differing simplifications, so consistency of the
results would suggest that the simplifications do not have strong effects.
Similarly, crosschecks against the partially relativistic cosmological simulation code
{\sc gevolution} \citep{AdamekDDK16code} and the fully relativistic
cosmological simulation packages used within the {\sc Einstein Toolkit}
\citep[][({\sc et})]{BentivegnaBruni15,Macpherson17} would be useful. Differences between these
would have to be understood. In the subcases in
which RZA is exact, such as the plane-symmetric case, the simplicity of the RZA approach
would potentially provide a good test for calibrating the
computational accuracy of {\sc gevolution} and {\sc et}.}

\postrefereeAAchanges{In principle, it should be possible to test the
  $\OmRD$--$H_{\CD}$ relation observationally, as a new test of the Einstein equation.
  In practice, this is likely to be very difficult.}

\subsection{Geometrical dark matter}

Earlier discussion of the role of exact relativistic solutions has
pointed out the difference between these and the perturbed FLRW
approach, arguing that there is effectively a \enquote*{general-relativistic
dark matter} component associated with gravitational collapse on
cosmologically relevant scales, using Tolman--Lema\^{\i}tre--Bondi
models
\citep{KrasHellaby02LTBDM,KrasHellaby04LTBDM,Bolejko06SzekeresDM}, the
quasi-spherical Szekeres model
\citep{Bolejko06SzekeresDM,Bolejko07SzekeresDM}, and Szekeres Class-II
models \citep{IshakPeel11}, though without a clear focus on the role
of pointwise or domain averaged positive curvature.  The expectation
that there is an effective form of general-relativistic dark matter
has been discussed in the more general context of scalar averaging
by, in particular, \cite{RoyFOCUS} and \cite{BuchRZA2}, but without
making calculations based on
\postrefereeAAchanges{generic realisations of}
a standard initial power spectrum of
density perturbations.  \cite{BuchRZA2} coined the term \enquote*{kinematical
dark matter}, suggesting that the shear scalar, on small scales, was
the most likely explanation to provide a kinematical dark matter
contribution to the usual observational interpretation of dark matter,
through its role in the Raychaudhuri equation, especially at the later
phases of gravitational collapse.
Equations \eqref{e-Buch00dust-4a} and
\eqref{e-Buch00dust-4a-averaged} show that at late phases, a higher
compression rate, that is, a greater value of $|H_D|$ during the
$H_D < 0$ post-turnaround phase, would also be contributed by the shear scalar
to a late-phase kinematical dark matter effect.
In this work, we focussed instead on the turnaround epoch and
made calculations based on a standard initial power spectrum of Gaussian fluctuations,
finding that at turnaround,
positive curvature much more frequently plays the dominant role in
gravitational collapse, at least near the turnaround epoch, rather than
kinematical backreaction, which is
the net effect of expansion variance and the shear scalar together.
\postrefereeAAchanges{Simultaneously to the present work,
  \cite{VignBuch19} showed that in the plane-symmetric case,
  kinematical backreaction becomes stronger in amplitude
  than curvature during the post-turnaround phase, rather than
  remaining at its turnaround value of one-fifth of the latter in
  absolute value (see \eqref{e-TA-critical-values}).}

\postrefereeAAchanges{Since turnaround-epoch positive spatial
  curvature is a geometrical phenomenon, not just dynamical, it
  should, in principle, contribute to weak lensing effects.  It could
  well play a role in modifying the usual calculations of weak lensing
  effects -- again, possibly} substituting for some of the present
role of FLRW \enquote*{perturbative} dark matter.  Since the effective
\enquote*{source} of dark matter in this sense is positive curvature
rather than kinematical backreaction, we suggest \enquote*{geometrical
  dark matter} as an appropriate term when positive curvature in the
averaged Hamiltonian constraint is the dominating dark-matter--like
relativistic effect.  \postrefereeBBchanges{Whether or not
  turnaround-epoch positive spatial curvature constitutes a new
  contribution to weak lensing that justifies the term
  \enquote*{geometrical dark matter} will be a useful question for
  further work in this field. The discussion and figure in
  \SSS\ref{s-pseudo-curvature} suggest that conventional $N$-body
  simulations would be sufficient to provide an approximate answer to
  this question, although the use of relativistic simulations and/or
  RZA calculations would provide a relativistically more accurate
  answer.}

\section{Conclusion} \label{s-conclu}

It is now clear, both from a general argument
(\SSS\ref{s-results-Hamiltonian-constraint},
\postrefereeAAchanges{Proposition}~{\ELemmaFlatNoTurnaround})
and from an exact cosmological solution close to an EdS or {\LCDM} reference
model (\SSS\ref{s-results-P-only-case}),
that the interpretation of the {\LCDM} model as
\prerefereeAAchanges{only containing literally 3-Ricci--flat
  spatial domains, rather than interpreting it as} an
almost-FLRW model with inhomogeneous curvature,
would forbid almost all formation
of dense structures.
This is because
inhomogeneities that are initially expanding in terms of physical distances
cannot sufficiently slow down their expansion (isotropically and pointwise)
to pass through the turnaround epoch
if zero spatial curvature is strictly imposed
in a fluid-orthogonal foliation.

We thus considered the more reasonable hypothesis that
relativistic constraints permit a standard initial power spectrum
of Gaussian random density fluctuations that
evolves according to the growing mode.
By using the relativistic Zel'dovich approximation,
we first showed that for null
initial average second and third peculiar-expansion tensor invariants
($\initaverageriem{\initial{\invII}}(\theta^i_j) = 0$,
$\initaverageriem{\initial{\invIII}}(\theta^i_j) = 0$)
in an almost-EdS model,
a critical value of the curvature functional
$\OmRD = -5 \alpha^2$
(where $\alpha := {H}/{\Heff}$ and $0.5 \ltapprox \alpha^2 \ltapprox 1$;
alternatively,
we can write this as $\OmRD = -5$ for normalisation using the EdS expansion
rate instead of the effective expansion rate),
corresponding
to positive spatial scalar curvature, must occur
in a domain as it passes through the turnaround epoch
(\SSS\ref{s-results-EdS-II-III-zero}).

For the more general case of standard initial conditions,
using kinematical backreaction evolution as modelled by the RZA
and implemented using the {\inhomog} library
(QZA, \SSS\ref{s-method-QZA}), we showed that
almost-EdS and almost-{\LCDM} models
give values of $\OmRD$ at turnaround corresponding to positive curvature
and lying in a range that includes this critical value
(\SSS\ref{s-results-EdS-LCDM-subsub},
Figs~\ref{f-OmRDEdS}--\ref{f-OmmOmQ-LCDM},
Tables~\ref{t-OmRDTA}--\ref{t-OmRDTA-stats}).
In the context where FLRW cosmological
parameters are believed to be approaching precision at the
one percent level, and possibly also a similar level of accuracy,
\postrefereeAAchanges{we find that neglecting strong turnaround-epoch curvature
is unlikely to lead to signficant}
inaccuracies in standard flat-space cosmological $N$-body simulations and
observational data analyses (\SSS\ref{s-discuss},
Figs~\ref{f-epsilonEdS}, \ref{f-epsilonLCDM}).
\postrefereeAAchanges{The explicit inclusion of turnaround-epoch
  positive spatial curvature in the analysis of the upcoming
  generation of major extragalactic surveys
  would nevertheless, in principle, be useful
  as an improvement beyond the methods of linear perturbation theory.}

\section*{Acknowledgments}
Thank you to
\postrefereeAAchanges{Pierre Mourier for a thorough reading of the text and equations and extensive
  comments and suggestions, to Quentin Vigneron for deriving the expressions
  in \eqref{e-aTA-tTA-subcase} and for helpful comments, and to}
Krzysztof Bolejko,
Justyna Borkowska,
Thomas Buchert,
Matteo Cinus,
Johan Comparat,
and
Marius Peper
for helpful comments and suggestions.
A part of this project was funded by the National
Science Centre, Poland, under grant 2014/13/B/ST9/00845.
Part of this work was supported by the \enquote*{A next-generation worldwide quantum sensor network with optical atomic
clocks} project, which is carried out within the TEAM IV programme of the
Foundation for Polish Science co-financed by the European Union under the
European Regional Development Fund.
This work has benefited from funding under the Polish MNiSW grant DIR/WK/2018/12.
JJO acknowledges hospitality and support by
Catalyst grant CSG--UOC1603 during his visit to the University of Canterbury and grant ANR-10-LABX-66 within the `Lyon Institute of Origins'.
A part of this project has made use
of computations made under grant 314 of the Pozna\'n
Supercomputing and Networking Center (PSNC).  This work has used the
free-licensed {\sc GNU Octave} package \citep{Eaton14}.

\subm{ \clearpage }

\bibliographystyle{\BIBAABST} 
\providecommand{\href}[2]{#2}\begingroup\raggedright\endgroup

\end{document}